\documentclass[12pt,a4paper]{article}

\usepackage[utf8]{inputenc}
\usepackage[T1]{fontenc}
\usepackage{amsmath}
\usepackage{ae}
\usepackage{color}
\usepackage[affil-it]{authblk}
\usepackage{xcolor}
\pdfoutput=1
\usepackage{graphicx}
\usepackage{longtable}
\usepackage{bbm}
\usepackage{bm}
\usepackage{xspace}
\usepackage{tabularx}
\usepackage{cite}
\usepackage[colorlinks=true, linkcolor=blue, bookmarks=true]{hyperref}

\newcommand{\rd}{\ensuremath{\mathrm{d}}}

\newcommand{\cZ}{\ensuremath{Z}}

\newcommand{\bu}{\bm{u}}
\newcommand{\bh}{\bm{h}}
\newcommand{\bx}{\bm{x}}

\newcommand{\ads}{AdS$_3$\xspace}
\newcommand{\pc}[1]{\tilde{#1}}

\numberwithin{equation}{section}
\begin{document}

\title{Collisions of massive particles, timelike thin shells and formation of black holes in three dimensions}

\author[1,2]{Jonathan Lindgren\thanks{Electronic adress: \texttt{Jonathan.Lindgren@vub.ac.be}}}
\affil[1]{Theoretische Natuurkunde, Vrije Universiteit Brussel, and the International Solvay Institutes, Pleinlaan 2, B-1050 Brussels, Belgium}
\affil[2]{Physique Th\'eorique et Math\'ematique, Universit\'e Libre de Bruxelles, Campus Plaine C.P.\ 231, B-1050 Bruxelles, Belgium}
\date{}

\maketitle

\abstract{We study collisions of massive pointlike particles in three dimensional anti-de Sitter space, generalizing the work on massless particles in \cite{Lindgren:2015fum}. We show how to construct exact solutions corresponding to the formation of either a black hole or a conical singularity from the collision of an arbitrary number of massive particles that fall in radially and collide at the origin of AdS. No restrictions on the masses or the angular and radial positions from where the particles are released, are imposed. We also consider the limit of an infinite number of particles, obtaining novel timelike thin shell spacetimes. These thin shells have an arbitrary mass distribution as well as a non-trivial embedding where the radial location of the shell depends on the angular coordinate, and we analyze these shells using the junction formalism of general relativity. We also consider the massless limit and find consistency with earlier results, as well as comment on the stress-energy tensor modes of the dual CFT.}
\newpage
\tableofcontents

\section{Introduction}
It is a remarkable fact that there exists non-trivial analytic solutions of Einstein's equations, the most well known being spherically symmetric black holes and the Schwarzschild metric. However, in time-dependent situations, particularly in the process of collapse of matter that forms a black hole, analytic solutions are very hard to find. Studying formation of black holes thus often requires us to resort to numerical methods, which can be very difficult except for problems with a lot of symmetry. Having access to analytical toy models of black hole formation can thus be of great value when trying to understand these processes.\\
\linebreak
Black hole formation was, not surprisingly, studied first in flat spacetime (with zero cosmological constant), see for instance \cite{Oppenheimer:1939ue,Goldwirth:1987nu,Choptuik:1992jv}. However, in recent years black hole formation in anti-de Sitter space (AdS) has attracted much attention due to the AdS/CFT correspondence\cite{Maldacena:1997re,Aharony:1999ti} (also called holography or gauge/gravity-duality). According to this correspondence, black holes in $(d+1)$-dimensional anti-de Sitter space are dual to thermal states in a $d$-dimensional conformal field theory, and thus the dynamical process of forming a black hole is dual to the process of {\it thermalization} in a conformal field theory. This is a very useful observation, since studying time dependent processes in strongly coupled quantum field theories using conventional techniques can be extremely difficult. Many papers have now been published using holography to study thermalizaion in conformal field theories \cite{Balasubramanian:2011ur,Balasubramanian:2013oga,Chesler:2008hg,Bhattacharyya:2009uu,Bizon:2011gg,Liu:2013iza,Buchel:2014dba,Craps:2014vaa,Craps:2014jwa,Balasubramanian:2014cja} as well as in non-conformal field theories \cite{Craps:2013iaa,Craps:2014eba,Fonda:2014ula,Alishahiha:2014cwa,Ishii:2015gia,Buchel:2015saa,Craps:2015upq,Gursoy:2016tgf}.\\
\linebreak
In the case of spherical or translational symmetry, there do exist analytic solutions corresponding to pressureless null-matter collapsing to form a black hole, the so called Vaidya spacetimes. These metrics have been used to study black hole formation in flat space, and the analog in anti-de Sitter space (the AdS-Vaidya metric) has been used extensively in the context of the AdS/CFT correspondence and holographic thermalization. The special case where the matter takes the shape of a infinitely thin shell is of particular importance, and the AdS-analog has been used in the study of thermalization after an {\it instantaneous} quench in the dual field theory\cite{Balasubramanian:2011ur,Fonda:2014ula,Ziogas:2015aja}. Spacetimes with timelike shells collapsing to form black holes (including shells with non-zero pressure) in AdS have also been studied\cite{Keranen:2015fqa,Keranen:2014zoa,Erdmenger:2012xu}. However, all of these constructions suffer from the disadvantage that the setups have spherical or translational symmetry.\\
\linebreak
Another interesting toy model of black hole formation, which only works in three dimensions, is that of colliding pointlike particles. This was first studied in \cite{Matschull:1998rv}, and the author found that when two pointlike particles collide a BTZ black hole\cite{Banados:1992wn,Banados:1992gq} can form if the energy is large enough. This process has also been studied from a holographic perspective\cite{Ageev:2015xoz,Balasubramanian:1999zv}. In \cite{Lindgren:2015fum}, more general solutions consisting of many colliding massless particles without symmetry restrictions were studied. Furthermore, a connection to thin shell spacetimes was obtained by building a shell of lightlike matter by taking the limit of an infinite number of massless pointlike particles. Interestingly, this construction resulted in generalizations of the thin shell \ads-Vaidya spacetimes, which are {\it not} rotationally symmetric.\\
\linebreak
In this paper we will extend the results of \cite{Lindgren:2015fum} to the case of massive particles. We will show, by using similar techniques, how to construct spacetimes corresponding to an arbitrary number of massive pointlike particles colliding in the center of \ads. The particles will generically have different rest masses and be released from arbitrary angles as well as from arbitrary radial positions. We will then take the limit of an infinite number of particles and obtain thin shell spacetimes corresponding to (non-rotationally symmetric) timelike shells collapsing to form a black hole (or to a conical singularity). Since the particles can now be released from different radial positions, the shells are specified by two free functions: the rest mass density as well as the maximal radial location of the shell (corresponding to the initial radial position of the particles). This should be compared to the lightlike case, where there is only one free function which specifies the energy density. This non-trivial embedding of the shells makes the computations significantly more involved compared to the lightlike shells. Spacetimes with spherically symmetric massive thin shells collapsing to form black holes have been studied before\cite{Keranen:2014zoa,Keranen:2015fqa}, but the solutions constructed in this paper are the first examples of massive thin shells which break rotational symmetry. We also comment on the massless limit and the dual CFT description, in particular reproducing some of the results in \cite{Lindgren:2015fum}.\\
\linebreak
The paper is organized as follows. In Section \ref{adssec} we review our coordinates and convention for \ads. In Section \ref{ppsec} and Section \ref{bhsec} we review how to construct pointlike particles and BTZ black holes by identifying points in \ads. In Section \ref{collisionsec} we explain how to construct spacetimes with colliding pointlike particles. In Section \ref{limitsec} we take the limit of an infinite number of particles and show how a thin shell spacetime emerges. In Section \ref{junctionsec} we analyze these thin shell spacetimes using the junction formalism, notably computing the stress-energy tensor of the shell. In Section \ref{masslesssec} we explore the massless limit. In Section \ref{numericssec} we explain some numerical algorithms used when constructing the solutions in this paper. Appendices include some lengthy derivations as well as a summary of the notation used in this paper.
\section{Three-dimensional anti-de Sitter space}\label{adssec}
Anti-de Sitter (AdS) space is the maximally symmetric solution of Einsteins equations with negative cosmological constant. It can also be constructed as an embedding in a higher dimensional Minkowski space with two time directions. For three-dimensional AdS (\ads), the ambient space is four dimensional Minkwoski space with signature $(-,-,+,+)$, and the embedding equation takes the form
\begin{equation}
x_3^2+x_0^2-x_1^2-x_2^2=\ell^2,\label{embedding_eq}
\end{equation}
where $\ell$ is the AdS radius, related to the cosmological constant by $\Lambda=-1/\ell^2$. The ambient space has the metric
\begin{equation}
ds^2=dx_1^2+dx_2^2-dx_3^2-dx_0^2,
\end{equation}
which induces a metric on the submanifold. To parametrize this manifold, one can use the coordinates $(t,\chi,\phi)$ defined by
\begin{equation}
\begin{array}{cc}
x^3=\ell\cosh \chi \cos t, & x^0=\ell\cosh \chi \sin t,\\
x^1=\ell\sinh \chi \cos \phi, & x^2=\ell\sinh \chi \sin \phi, \label{adscoord}
\end{array}
\end{equation}
where $t$ is interpreted as a time coordinate, $\chi$ a radial coordinate and $\phi$ an angular coordinate. The spacetime so defined has closed time-like curves, and to solve this issue one ``unwinds'' the time coordinate, effectively dropping the periodicity of $t$. The ranges of the coordinates are then $0\leq\chi\leq\infty$, $0\leq\phi<2\pi$ and $-\infty<t<\infty$. The boundary of AdS is located at $\chi\rightarrow\infty$ and we will refer to the point $\chi=t=0$ as the {\it origin}. The metric is 
\begin{equation}
ds^2=\ell^2\left(-\cosh^2\chi dt^2+d\chi^2+\sinh^2\chi d\phi^2\right).\label{adsmetric}
\end{equation}
From now on we will set $\ell=1$. For figures in this paper, we will use a different parametrization of the radial coordinate, namely $r=\tanh(\chi/2)$ with $0\leq r\leq1$. With this parametrization, the metric at constant time $t$ takes the form of a Poincar\'e disc.\\
\linebreak
A very useful property of AdS$_3$ is that it is locally\footnote{SL(2,R) is isomorphic to AdS$_3$ {\it before} unwinding the time coordinate. This will not be an issue for any of the computations in this paper.} isomorphic to SL(2,R), the group manifold of real $2\times2$ matrices with unit determinant (which we will for simplicity for now on refer to as just SL(2)). Following \cite{Matschull:1998rv}, we define the basis 
\begin{equation}
\gamma_0=\left(\begin{array}{ccc} 0 & 1 \\ -1 & 0 \\ \end{array}\right),\hspace{23pt}\gamma_1=\left(\begin{array}{ccc} 0 & 1 \\ 1 & 0 \\ \end{array}\right),\hspace{22pt} \gamma_2=\left(\begin{array}{ccc} 1 & 0 \\ 0 & -1 \\ \end{array}\right).
\end{equation}
We can then expand an arbitrary matrix as $\bm x=x_3\bm 1+\gamma_ax^a$, and the condition of unit determinant yields the embedding equation \eqref{embedding_eq} (note that indices are here lowered and raised by $\eta_{ab}=\textrm{diag}(-1,1,1)$). The isometries of AdS$_3$ can now be implemented by left and right multiplications as
\begin{equation}
\bx\rightarrow \bm{g}^{-1}\bx\bm{h} \hspace{21pt} \bm{g},\bm{h}\in \textrm{SL(2)}.\label{isom}
\end{equation}
We will use this technique repeatedly to generate isometries of \ads. Note that in the coordinates \eqref{adscoord}, the matrix $\bx$ takes the form
\begin{align}
\bx=&\cosh\chi\cos t+\cosh\chi\sin t\gamma_0+\sinh\chi\cos\phi\gamma_1+\sinh\chi\sin\phi\gamma_2\nonumber\\
&=\cosh\chi\omega(t)+\sinh\chi\gamma(\phi),
\end{align}
where we have defined the convenient matrices
\begin{equation}
\omega(\alpha)=\cos \alpha + \sin\alpha\gamma_0, \hspace{20pt}\gamma(\alpha)=\cos \alpha\gamma_1 + \sin\alpha\gamma_2.\label{omegagamma}
\end{equation}
\subsection{Geodesics}\label{geodesicssec}
We will now illustrate the power of the isomorphism of AdS$_3$ and the group manifold SL(2) by computing timelike and lightlike geodesics. Starting with the static geodesic located at $\chi=0$, we can apply isometries of AdS$_3$ to generate any timelike geodesic. The isometries are generated using equation \eqref{isom} with $\bm{g}=\bm{h}=\bu$, where the group element $\bu=e^{-\frac{1}{2}\zeta\gamma(\psi-\pi/2)}=\cosh\frac{1}{2}\zeta-\gamma(\psi-\pi/2)\sinh\frac{1}{2}\zeta$, and $\gamma$ is given by \eqref{omegagamma}. Computing $\pc\bx=\bu^{-1}\bx \bu$ results in the equations
\begin{align}
\begin{split}
\cos \pc{t}\cosh \pc\chi&=\cos t\cosh \chi,\\
\sin \pc t\cosh \pc\chi&=\cosh\chi\sin t \cosh \zeta - \sinh\chi\sinh\zeta\cos(\phi-\psi),\\
\sinh\pc\chi\cos\pc\phi&=-\cosh\chi\sinh\zeta\sin t\cos\psi+\sinh\chi(-\sin\psi\sin(\phi-\psi)+\cosh\zeta\cos\psi\cos(\phi-\psi)),\\
\sinh\pc\chi\sin\pc\phi&=-\cosh\chi\sin t\sinh\zeta\sin\psi+\sinh\chi(\cosh\zeta\cos(\phi-\psi)\sin\psi+\sin(\phi-\psi)\cos\psi).\label{boosteqs}
\end{split}
\end{align}
It can now be shown that the static geodesic at $\chi=0$ transforms into the oscillating timelike trajectory $\tanh\pc\chi=-\tanh\zeta\sin \pc t$ with $\pc\phi=\psi$. Of course, ($\psi$, $\zeta$) is the same transformation as ($\psi+\pi$, $-\zeta$). Note also that, for $\tanh\zeta\sin \pc t>0$, we have $\pc\chi<0$. We can thus choose either to allow for negative $\pc\chi$ or rotate the angle by $\pi$ radians whenever $\tanh\zeta\sin \pc t>0$.\\
\linebreak
Note also that in the unprimed coordinates, the proper time $\tau$ of a stationary particle at the origin $\chi=0$ is given by $\tau=t$, and thus we can use this to express the trajectory of a moving particle in terms of the proper time. By taking the ratio of the first and second equations in \eqref{boosteqs}, after setting $\chi=0$, we obtain
\begin{equation}
\tan\pc t=\tan \tau\cosh\zeta.\label{proptime1}
\end{equation}
From the third equation after setting $\pc\phi=\psi$ we obtain
\begin{equation}
\sinh\pc\chi=-\sin \tau \sinh\zeta.\label{proptime2}
\end{equation}
We can also derive the useful relation
\begin{equation}
\cosh\pc\chi=\frac{\cos \tau }{\cos \pc t}=\sqrt{\cos^2\tau +\sin^2\tau \cosh^2\zeta}.\label{proptime3}
\end{equation}
\linebreak
Massless geodesics can be obtained by taking $\zeta\rightarrow\infty$, and these take the form $\tanh\pc\chi=-\sin \pc t$.

\section{Pointlike particles in anti-de Sitter space}\label{ppsec}
\begin{figure}
\begin{center}
\includegraphics[scale=0.7]{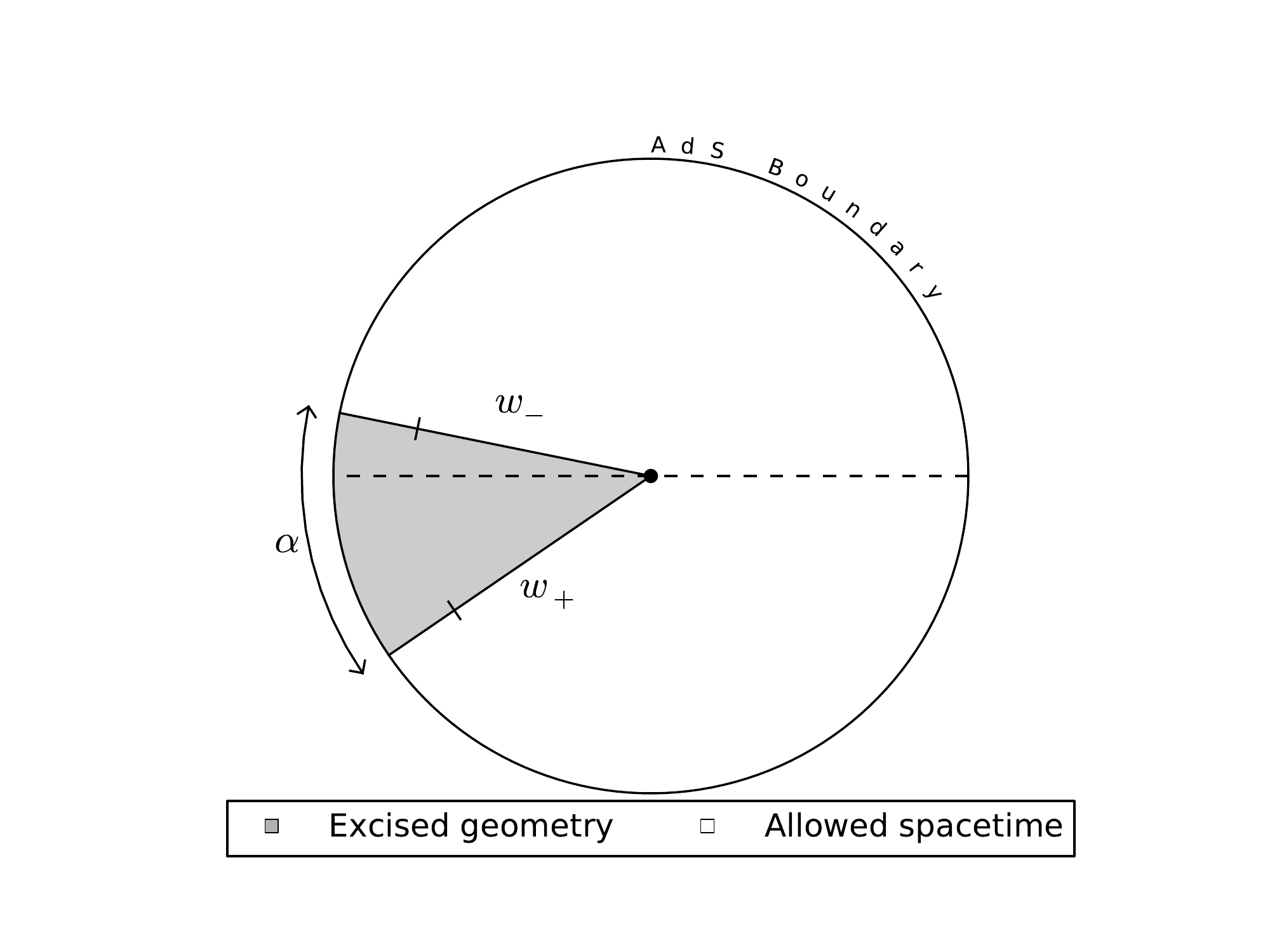}
\caption{\label{staticpp}A static particle in \ads. A wedge has been removed, and the surfaces $w_-$ and $w_+$ are identified by a rotation, resulting in a spacetime with a conical singularity with deficit angle $\alpha$. The wedge is not located symmetrically around $\phi=0$, so a boost in this direction will result in a moving pointparticle constructed by excising a wedge not located symmetrically around its trajectory (see Fig. \ref{1particle}). }
\end{center}
\end{figure}

\begin{figure}
\begin{center}
\includegraphics[scale=0.8]{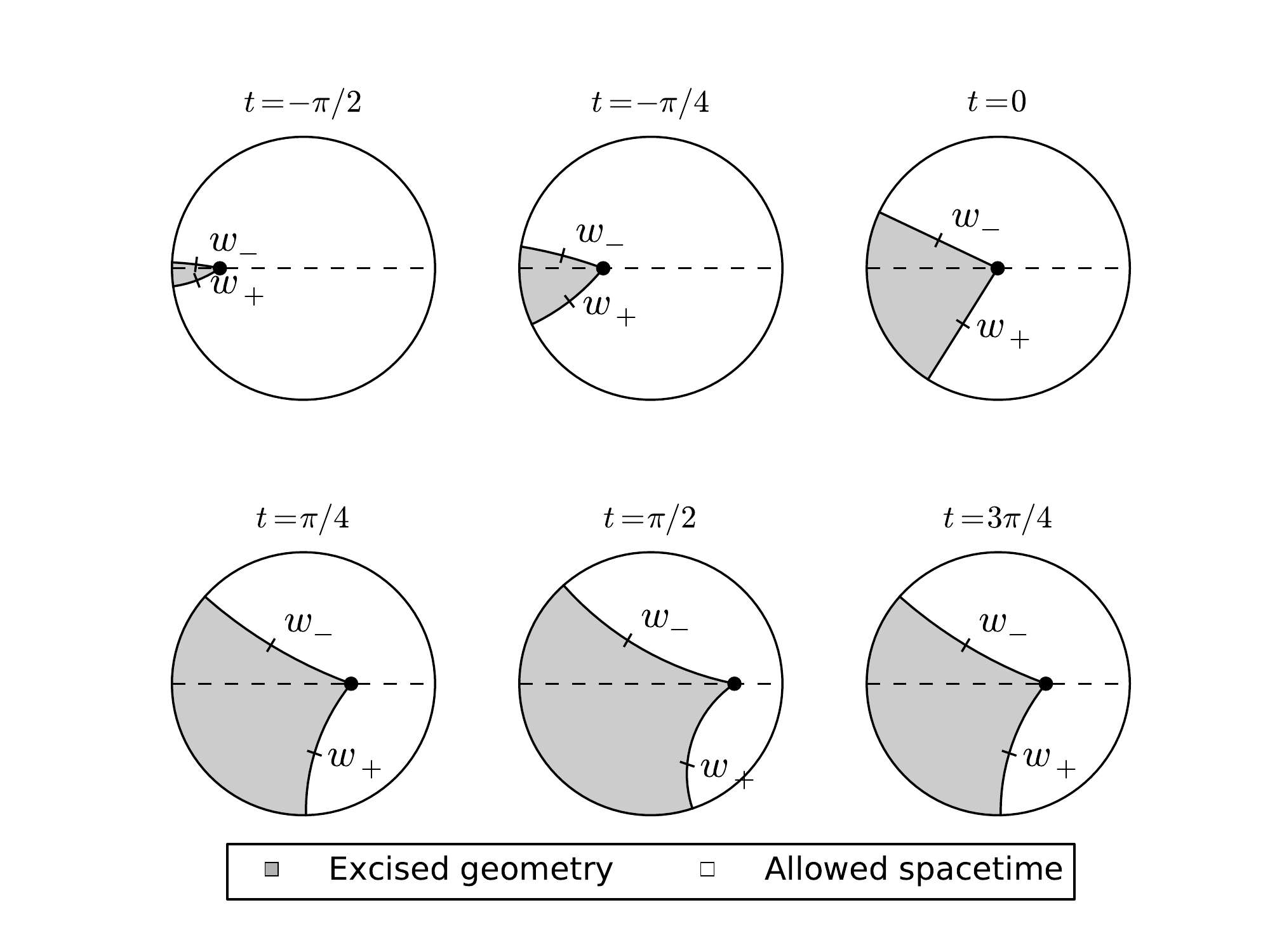}
\caption{\label{1particle} A moving massive particle, obtained by cutting out a wedge from \ads. The wedge in this example is not located symmetrically around the trajectory of the particle, meaning the parameter $p$ in equation \eqref{Gammapm} is non-zero.}
\end{center}
\end{figure}

In three-dimensional anti-de Sitter space, a static pointlike particle will take the form of a conical singularity at the origin of \ads. We can easily construct it by cutting out a piece of geometry and identifying the edges of this piece by a rotation, see Fig. \ref{staticpp} (we would like to remind the reader that all figures in this paper uses the compactification of the radial coordinate mentioned after equation \eqref{adsmetric}). More generally, if we have a particle moving along some geodesic, we can excise a piece of geometry that induces a conical singularity along the world line of this particle, and the edges of this piece are then identified by some non-trivial isometry of \ads that has the geodesic as fixed points. We will refer to all such excised pieces of geometry, that induce some conical deficit along a geodesic, as {\it wedges}, and for the special case of a static particle at the origin of \ads we will refer to them as {\it static wedges}. The element in SL(2) that induces the isometry is called the {\it holonomy} of the particle. In this section, we will construct moving pointlike particles by boosting the static particle, using the isometry \eqref{boosteqs}.\\
\linebreak
We thus start with a static particle as in Fig. \ref{staticpp}, with a wedge bordered by two surfaces $w_\pm$ at angles $\phi_\pm$. We will now boost this spacetime along a direction $\psi$ (dashed line in Fig. \ref{staticpp}). The most common construction of a moving pointlike particle would be to align the wedge such that the boost parameter $\psi$ is right in between $\phi_+$ and $\phi_-$. We will refer to this parametrization as a {\it symmetric wedge}. However, this is just an arbitrary choice of coordinates, and it is possible to have many other. In this paper we will be interested in a one-parameter family of wedges, which are obtained by boosting a static wedge along an arbitrary direction. To this end, we write $\phi_\pm=\psi\pm\nu(1\pm p)$, such that a symmetric wedge is obtained by setting $p=0$ and the deficit angle is given by $2\nu$. When applying the boost, we thus obtain a family of wedges parametrized by a continuous parameter $p$ (note that $p$ has no real physical meaning, and is analogous to a choice of coordinates). By using \eqref{boosteqs}, it can be shown that (see \cite{Lindgren:2015fum} for a more detailed derivation), after applying a boost with boost parameter $\zeta$, the surfaces $\pc w_\pm$ bordering the wedge can be parametrized by
\begin{equation}
\tanh\pc\chi\sin(-\pc\phi +\Gamma_\pm+\psi)=-\tanh\zeta \sin\Gamma_\pm \sin\pc t, \label{c1}
\end{equation}
where
\begin{equation}
\tan\Gamma_\pm =\pm\tan((1 \pm p)\nu) \cosh\zeta.\label{Gammapm}
\end{equation}
In these formulas it is again clear that $(\psi$, $\zeta)$ is equivalent to $(\psi+\pi$, $-\zeta)$. If we insist on positive radial coordinate $\chi$, the ranges of the parameters are
\begin{equation}
\pc\phi\in(\psi,\arcsin(\tanh\zeta\sin \Gamma_\pm\sin t)+\psi+\Gamma_\pm),
\end{equation}
for $\sin \pc t<0$, and
\begin{equation}
\pc\phi\in(\arcsin(\tanh\zeta\sin \Gamma_\pm\sin t)+\psi+\Gamma_\pm,\psi\pm\pi),
\end{equation}
for $\sin \pc t>0$. For $\sin \pc t=0$  we have $\pc \phi=\psi+\Gamma_\pm$. An illustration of a boosted particle with $p\neq0$ (non-symmetric parametrization), obtained by boosting Fig. \ref{staticpp} along $\psi=\pi$ with $\zeta=1.5$, is shown in Fig. \ref{1particle}. The holonomy of the particle (meaning the group element $\bh$ such that $\pc w_-=\bh^{-1}\pc w_+\bh$) can be computed as
\begin{equation}
\bh=\cos \nu+\gamma_0\cosh \zeta \sin \nu-\sinh \zeta \sin \nu \gamma(\psi).\label{hol}
\end{equation}
\linebreak
The massless limit is obtained by letting $\zeta\rightarrow\infty$ and $\nu\rightarrow0$, such that $\sinh\zeta\tan\nu\rightarrow E$. The wedges are then given by
\begin{equation}
\tanh\pc\chi\sin(-\pc\phi +\Gamma_\pm+\psi)=-\sin\Gamma_\pm \sin \pc t, \label{c1_massless}
\end{equation}
where
\begin{equation}
\tan\Gamma_\pm =(p\pm1)E.\label{Gammapm_massless}
\end{equation}
The holonomy then turns into
\begin{equation}
\bh=1+E(\gamma_0-\gamma(\psi)).\label{hol0}
\end{equation}


\subsection{The stress-energy tensor of a pointlike particle}
In this section we will compute the stress-energy tensor of a pointlike particle. We will first compute it for a static particle, and then obtain the stress-energy tensor for a moving particle by applying a boost. For a static particle, the stress energy tensor takes the form
\begin{equation}
T^{\tau\tau}=T^{tt}=m\delta^{(2)}(x^\mu).
\end{equation}
Recall that $\tau=t$ is the proper time at the origin of \ads. $\delta^{(2)}$ is the standard two-dimensional delta function at the center of \ads such that the area integral on a constant time slice is equal to one. Assuming we have a conical singularity at the origin with conical deficit $\alpha$, we want to relate $m$ and $\alpha$. The Einstein equations take the form
\begin{equation}
R_{\mu\nu}-\frac{1}{2}Rg_{\mu\nu}+\Lambda g_{\mu\nu}=8\pi G T_{\mu\nu},
\end{equation}
where $G$ is Newtons constant in three dimensions. From this we can derive
\begin{equation}
R=6\Lambda+16\pi G m\delta^{(2)} (x^\mu),
\end{equation}
by taking the trace. By now decomposing the metric as $g_{tt}=f(\chi)$, $g_{ti}=0$ and $g_{ij}=\gamma_{ij}(\chi)$ it can be shown that for the two-dimensional Ricci scalar on a constant time slice, we have $^{(2)}\!R=16\pi G m\delta^{(2)} (x^\mu)+$(finite terms). Now consider a small disc $D$ of radius $\chi$ around the origin. The geodesic curvature around the edge of the disc is $k_g=\coth \chi$. Now from the Gauss-Bonnet theorem, as $\chi\rightarrow0$, we obtain
\begin{equation}
2\pi-\alpha=\lim_{\chi\rightarrow0}\int_0^{2\pi-\alpha}k_gds=2\pi\chi(D)-\lim_{\chi\rightarrow0}\frac{1}{2}\int_D {}^{(2)}\!R=2\pi-8\pi G m,
\end{equation}
and thus $\alpha=8\pi G m$, which is the same result as in \cite{ Deser:1983tn}. Thus it may be appropriate to refer to the deficit angle $\alpha$ as the mass of the particle. Note that we also used that the Euler characteristic $\chi(D)$ of a disc is 1, as can be easily seen from Euler's formula for a triangle.\\
\newline
We have seen that a moving pointlike particle can be obtained by applying a boost, and thus we can now obtain the stress-energy tensor by applying the boost to the static stress-energy tensor. The transformation will be given by \eqref{boosteqs} and for clarity we will denote the coordinates in the boosted coordinate system by $(\pc t,\pc\chi,\pc\phi)$ to distinguish them from the static coordinates $(t,\chi,\phi)$. Recall that $t$ coincides with the proper time of the static particle. Let us first figure out what the delta function looks like, which we will denote by $\delta_{\psi,\zeta}$. The delta function will be taken to be proportional to $\delta(\pc\phi-\psi)\delta(\tanh\pc\chi+\sinh\zeta\sin\pc t)$, such that it singles out the trajectory $\tanh\pc\chi=-\sinh\zeta\sin\pc t$ and $\pc\phi=\psi$. Note that since this is a two-dimensional delta function in a three-dimensional space, the normalization must be such that $\int \delta_{\psi,\delta}=\Delta \tau$, where $\Delta \tau$ is the elapse in proper time along the geodesic that is inside the domain of the integral (the domain can be taken as a very thin tube covering a segment of the world line of the particle). It can be easily shown that
\begin{equation}
\delta_{\psi,\zeta}=\frac{\delta(\pc\phi-\psi)\delta(\tanh\pc\chi+\tanh\zeta\sin \pc t)}{\cosh\pc\chi\sinh\pc\chi\cosh\zeta},
\end{equation}
satisfies
\begin{equation}
\int_{\mathcal{D}}\sqrt{-g} \delta_{\psi,\zeta}d\pc\phi d\pc\chi d\pc t=\Delta \tau,
\end{equation}
where $\mathcal{D}$ is some region that covers $\Delta \tau$ of proper time of the particles trajectory. The stress-energy tensor is now given by $T^{\mu\nu}=T^{\tau\tau}\dot{x}^\mu\dot{x}^\nu$. For the transformation of the stress-energy tensor, we have the relations $\dot{\pc\chi}=-\sinh\zeta\cos \pc t$ and $\dot{\pc t}=\cosh\zeta/\cosh^2\pc \chi$, which gives us finally
\begin{equation}
\begin{array}{cc}
 T^{\pc t\pc t}&=m\frac{\cosh^2\zeta}{\cosh^4\pc \chi}\delta_{\psi,\zeta},\\
 T^{\pc\chi\pc t}&=-m\frac{\cosh\zeta\sinh\zeta\cos \pc t}{\cosh^2\pc\chi}\delta_{\psi,\zeta},\\
 T^{\pc\chi\pc\chi}&=m\sinh^2\zeta\cos^2\pc t\delta_{\psi,\zeta},\\\label{Tpointparticle}
\end{array}
\end{equation}
while all other components vanish.


\section{The BTZ black hole}\label{bhsec}
In this section we will briefly review the construction of the BTZ black hole\cite{Banados:1992wn,Banados:1992gq} that will be useful for our purposes. The discussion follows closely that of \cite{Lindgren:2015fum}. The BTZ black hole can be constructed by doing an identification of points in \ads, similar to that of a conical singularity. The identification will now have a spacelike geodesic as fixed points, as opposed to when constructing a conical singularity where the set of fixed points is a timelike geodesic. We will refer to this set of fixed points as the {\it singularity} of the black hole. For our purposes, the singularity will be a general radial geodesic passing through the origin of \ads, but we will start by choosing the simple radial geodesic given by $\phi=0$ and $t=0$ and then apply a set of isometries to obtain a singularity given by a more general spacelike geodesic. The isometry we will pick that leaves this geodesic invariant is given by $\bu=e^{\mu\gamma_1}=\cosh\mu+\gamma_1\sinh\mu$. We can now define a region of \ads, bordered by two surfaces $w_\pm$, and then cut out everything outside this region and make sure that these two surfaces are identified by the isometry as $\bu w_-=w_+\bu$. If we write the equations for $w_\pm$ as $w_\pm=\cosh\chi\omega(t)+\sinh\chi\gamma(\phi)$, and we assume that the two surfaces $w_\pm$ are located symmetrically around the singularity, the coordinates must satisfy
\begin{equation}
w_\pm: \tanh\chi\sin\phi=\mp\sin t\tanh\mu.\label{bhwedge1}
\end{equation}
We will also be interested in surfaces $w_\pm$ that are not symmetric around the singularity. This can be obtained by acting with an isometry of the form $e^{-\frac{1}{2}\xi\gamma_1}$ for some parameter $\xi$. This group element also has the same radial geodesic as its fixed points, and after applying this isometry, the equations for the surfaces are instead
\begin{equation}
w_\pm: \tanh\chi\sin\phi=\mp\sin t\tanh(\mu\pm\xi).\label{bhwedge2}
\end{equation}
Now we would like to apply another isometry to change the singularity to a more general geodesic. This can be obtained by applying the boost $e^{\frac{1}{2}\zeta\gamma_2}$. This is the same type of isometry as was used in Section \ref{ppsec} and will transform the singularity such that it now obeys $\tanh\chi=-\sin t\cosh \zeta$. After also applying a rotation such that the singularity is along an arbitrary angle $\psi$, it can be shown that the surfaces now take the form
\begin{equation}
w_\pm: \tanh \chi \sin(-\phi+\Gamma_\pm+\psi)=-\sin\Gamma_\pm\coth\zeta\sin t,\label{bhwedge3}
\end{equation}
where
\begin{equation}
\tan\Gamma_\pm=\mp\tanh(\mu\pm\xi)\sinh\zeta.\label{Gammapmu}
\end{equation}
Note the resemblence to the equations for the moving pointlike particle, \eqref{c1} and \eqref{Gammapm}.\\
\linebreak
Another useful parametrization of the surfaces $w_\pm$ is to move to a different set of coordinates where the metric takes the form of a black hole metric with unit mass. The piece bordered by $w_\pm$ will then take the form of a circle sector, but defined in a BTZ black hole background. These coordinates can be obtained by a different parametrization of the embedding equation \eqref{embedding_eq}. We start by parametrizing $x_0=\rho\cosh y$ and $x_2=\rho\sinh y$. It can then be shown that the isometry $\bu$ just acts as a translation by $y\rightarrow y-2\mu$, and thus the surfaces $w_\pm$, given by \eqref{bhwedge1}, are just planes at constant values of $y$, bordering a circle sector with opening angle $2\mu$. Note that this only parametrizes a subset which satisfies $|x_0|>|x_2|$, but this inequality is always obeyed by the spacetime defined by \eqref{bhwedge1}. Our embedding equation \eqref{embedding_eq} has thus turned into
\begin{equation}
 x_3^2-x_1^2=\ell^2-\rho^2.
\end{equation}
It is clear that, to continue, we must decide if $\rho$ is larger or smaller than $\ell$. For $\rho<\ell$, we have the parametrization $x_1=\sqrt{\ell^2-\rho^2}\sinh (\sigma/\ell)$ and $x_3=\sqrt{\ell^2-\rho^2}\cosh (\sigma/\ell)$. For $\rho>\ell$, we can choose the parametrization $x_1=\sqrt{\rho^2-\ell^2}\cosh(\sigma/\ell)$ and $x_3=\sqrt{\rho^2-\ell^2}\sinh(\sigma/\ell)$. They both lead to the metric
\begin{equation}
ds^2=-(-1+\frac{\rho^2}{\ell^2})d\sigma^2+\frac{d\rho^2}{-1+\frac{\rho^2}{\ell^2}}+\rho^2dy^2.\label{unitbhmetric}
\end{equation}
Thus we conclude, that the surfaces defined by \eqref{bhwedge3}, can, by a series of coordinate transformations, be mapped to static circular sectors in a BTZ black hole background with unit mass, where the circular sectors will have opening angle $2\mu$. Moreover, let us now consider the case where we have many such wedges, defined by $w_\pm^i$, where instead of having $w_-^i$ and $w_+^i$ being identified, we have that $w_-^i$ is identified with $w_+^{i-1}$. It is then clear that the whole spacetime can be transformed to a set of circular sectors in a black hole background with unit mass, where each circle sector is linked to the next one. This spacetime then has a total angle of $\alpha=\sum_i2\mu_i$. Rescaling the coordinates by $y\rightarrow\alpha y/(2\pi)$, $\sigma\rightarrow\alpha\sigma/(2\pi)$ and $\rho\rightarrow2\pi\rho/\alpha$ yields the metric\\
\begin{equation}
ds^2=-(-M+\frac{\rho^2}{\ell^2})d\sigma^2+\frac{d\rho^2}{-M+\frac{\rho^2}{\ell^2}}+\rho^2dy^2,\label{blackholemetric}
\end{equation}
where the mass $M=\alpha^2/(2\pi)^2$ and $y$ has the standard range from $0$ to $2\pi$.\\
\linebreak
Another coordinate system that we will use, can be obtained by defining $\ell\cosh\beta=\rho$. This is only defined for $\rho>\ell$ and thus only covers the region outside the event horizon. After rescaling $\sigma\rightarrow\ell\sigma$, the metric takes the form
\begin{equation}
ds^2=\ell^2(-\sinh^2\beta d\sigma^2+\beta^2+\cosh^2\beta dy^2),\label{betabhcoord}
\end{equation}
which is reminiscent of \eqref{adsmetric} which is why we will find this coordinate system useful. We will set $\ell=1$ in the remainder of this paper.

\section{Colliding massive particles}\label{collisionsec}
We will now explain how to construct a spacetime where a number of massive pointlike particles collide to form a single joint object (either a new particle or a black hole). We will assume that all particles move on radial geodesics and that they all collide at the origin of \ads, $(\chi=0, t=0)$. Before tackling the general case, we will first look at a much simpler setup where two massive particles collide head-on.\\
\subsection{Colliding two particles}
We will now consider the simplest example of two colliding particles, with deficit angles (in their respective rest frames) given by $2\nu_i$ and their boost parameters are given by $\zeta_i$, where $i=1,2$. The particles will always be assumed to be colliding head-on in the center of
AdS (meaning that the first particle comes in along an angle $\psi_1=0$ and the second particle along $\psi_2=\pi$, and the particles are both created at the boundary at time $t=-\pi/2$). It should be pointed out that in the case of massless particles, it is possible to move to a center of momentum frame such that both particles have the same energy. For massive particles however, each particle has two independent parameters (the rest mass and the boost parameter), thus it is not in general possible to move to a frame where $\nu_1=\nu_2$ and $\zeta_1=\zeta_2$. The best we can do is to reduce the number of free parameters to 3. We will later use this freedom to pick parameters such that
\begin{equation}
\tan\nu_1\sinh\zeta_1=\tan\nu_2\sinh\zeta_2\equiv E,\label{restframe}
\end{equation}
which can be interpreted as the center of momentum frame for this process and will see simplify the computations considerably. For discussion purely of the (massless) two-particle case with equal energies, we refer the reader to \cite{Matschull:1998rv} and a brief discussion of colliding massive particles can be found in \cite{Birmingham:1999yt}.\\
\linebreak
The two particles will be constructed by excising a wedge, as explained in Section \ref{ppsec}. Both wedges will be located behind and symmetrically around each particles trajectory, meaning that $p_1=p_2=0$ (see equation \eqref{c1}). This is a consequence of the reflection symmetry in the $\phi=0$ plane, and we will see that in the general case for more particles and without any symmetry restrictions, we will need to use general wedges with $p_i\neq0$. The first wedge is thus bordered by two surfaces $w_1^\pm$ given by
\begin{equation}
\tanh\chi\sin(-\phi +\Gamma^1_\pm)=-\tanh\zeta_1 \sin\Gamma^1_\pm \sin t,\label{wpm1}
\end{equation}
where
\begin{equation}
\tan\Gamma^1_\pm =\pm\tan\nu_1 \cosh\zeta_1.
\end{equation}
The wedge of the second particle is bordered by two surfaces given by
\begin{equation}
\tanh\chi\sin(-\phi +\Gamma^2_\pm)=\tanh\zeta_2 \sin\Gamma^2_\pm \sin t,\label{wpm2}
\end{equation}
where
\begin{equation}
\tan\Gamma^2_\pm =\pm\tan\nu_2 \cosh\zeta_2.
\end{equation}
Past the collision, there is a natural way to continue this spacetime such that the two particles merge and form one joint object, namely by identifying the intersection $I_{1,2}$ between $w_+^1$ and $w_-^2$ and the intersection $I_{2,1}$ between $w_+^2$ and $w_-^1$ as the new joint object (note that, due to the identifications of the wedges and the reflection symmetry in the $\phi=0$ plane, these two intersections are really the same spacetime point). Thus the spacetime after the collision is now composed of two separate patches, which are glued to each other via the isometries of the particles. Let us call these two wedges of spacetime $c_{1,2}$ and $c_{2,1}$. $c_{1,2}$ is bordered by the surfaces $w_-^{1,2}=w_+^1$ and $w_+^{1,2}=w_-^2$, while $c_{2,1}$ is bordered by $w_-^{2,1}=w_+^2$ and $w_+^{2,1}=w_-^1$. These two wedges can now be identified as being part of either a conical singularity spacetime, or a black hole spacetime, by matching the equations of these wedges to that of either \eqref{c1} and \eqref{Gammapm}, or that of \eqref{bhwedge3} and \eqref{Gammapm_massless}, respectively. The easiest way to see if we have formed a massive pointlike particle (conical singularity) or a black hole, is to see if the resulting radial geodesics, meaning the intersections $I_{1,2}$ and $I_{2,1}$, are timelike or spacelike. Let us first compute the intersection angles $\phi_{1,2}$ and $\phi_{2,1}$. From \eqref{wpm1} and \eqref{wpm2} we easily find that the intersections are given by
\begin{equation}
\tan\phi_{1,2}=-\tan\phi_{2,1}=\tan\nu_1\tan\nu_2\left[\frac{\sinh\zeta_1\cosh\zeta_2+\sinh\zeta_2\cosh\zeta_1}{\sinh\zeta_2\tan\nu_2-\sinh\zeta_1\tan\nu_1}\right],
\end{equation}
with the conventions $0\leq\phi_{1,2}\leq\pi$ and $\pi\leq\phi_{2,1}\leq2\pi$. Now let us focus on the wedge $c_{1,2}$. We can write $w_\pm^{1,2}$ in the following way
\begin{equation}
w_+^{1,2}: \tanh\chi\sin(-\phi+\phi_{1,2}+(\Gamma^2_--\phi_{1,2}))=-\frac{\tanh\zeta_2\sin\Gamma_-^2}{\sin(\phi_{1,2}-\Gamma^2_-)}\sin(\Gamma^2_--\phi_{1,2})\sin t,
\end{equation}
\begin{equation}
w_-^{1,2}: \tanh\chi\sin(-\phi+\phi_{1,2}+(\Gamma^1_+-\phi_{1,2}))=-\frac{\tanh\zeta_1\sin\Gamma_+^1}{\sin(\Gamma^1_+-\phi_{1,2})}\sin(\Gamma^1_+-\phi_{1,2})\sin t.
\end{equation}
By now comparing these equations to \eqref{Gammapm} or \eqref{bhwedge3}, we see that we can identify $\Gamma_+^{1,2}=\Gamma_-^2-\phi_{1,2}$ and $\Gamma_-^{1,2}=\Gamma_+^1-\phi_{1,2}$ and
\begin{equation}
\frac{\tanh\zeta_1\sin\Gamma_+^1}{\sin(\Gamma^1_+-\phi_{1,2})}=\frac{\tanh\zeta_2\sin\Gamma_-^2}{\sin(\phi_{1,2}-\Gamma^2_-)}\equiv\Bigg\{\begin{array}{ll}
                                                                                                                                             \tanh\zeta_{1,2},&\text{Point particle}\\
                                                                                                                                             \coth\zeta_{1,2},&\text{Black hole}                                                                                                                                             
                                                                                                                                            \end{array}\label{zetadef}
\end{equation}
The definition of $\phi_{1,2}$ ensures that the above equality holds and an analogous computation can be done for the wedge $c_{2,1}$. An example of two colliding particles is shown in Fig. \ref{2particles}.\\
\linebreak
We will now focus on the case where equation \eqref{restframe} holds, which we will interpret as the center of momentum frame of the collision process, and we thus parametrize our setup with $E$, $\zeta_1$ and $\zeta_2$. In this case $\phi_{1,2}=-\phi_{2,1}=\pi/2$ and \eqref{zetadef} is equal to $-E$.
\subsubsection{Formation of a conical singularity in the center of momentum frame}
In the case of a formation of a conical singularity, we should use equation \eqref{Gammapm} to compute the deficit angle. It states that
\begin{equation}
\tan\Gamma_\pm^{1,2}=\pm\tan(\nu_{1,2}(1\pm p_{1,2}))\cosh\zeta_{1,2}.
\end{equation}
The total deficit angle of the resulting conical singularity in the restframe is then given by $2\pi-2\nu_{1,2}-2\nu_{2,1}=2\pi-4\nu_{1,2}$. It is then straight forward to compute $\nu_{1,2}$ as being given by
\begin{equation}
\tan(2\nu_{1,2})=\tan(\nu_{1,2}(1+p_{1,2})+\nu_{1,2}(1-p_{1,2}))=E\sqrt{1-E^2}\left(\frac{\tanh\zeta_1+\tanh\zeta_2}{E^2+(E^2-1)\tanh\zeta_1\tanh\zeta_2}\right).
\end{equation}
For $E>1$ this result is no longer valid, and instead a black hole will form.
\subsubsection{Formation of a BTZ black hole in the center of momentum frame}
\begin{figure}
\begin{center}
\includegraphics[scale=0.8]{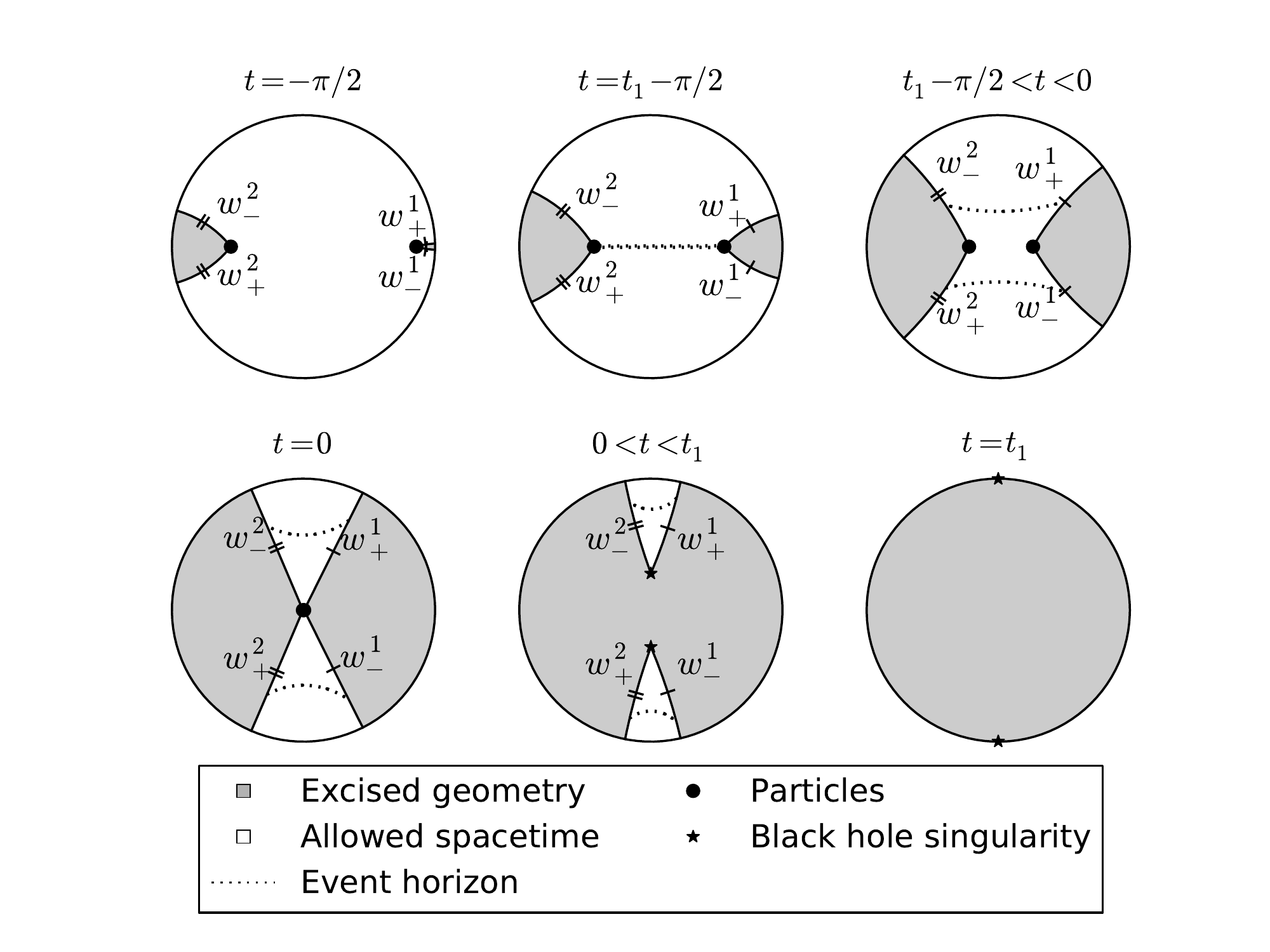}
\caption{\label{2particles} Collision of two particles with different rest masses in the center of momentum frame, forming a BTZ black hole.}
\end{center}
\end{figure}

In the case of a formation of a black hole, we should use equation \eqref{Gammapmu} to compute the mass of the black hole. It states that
\begin{equation}
\tan\Gamma_\pm^{1,2}=\mp\tanh(\mu_{1,2}\pm \xi_{1,2})\sinh\zeta_{1,2}.
\end{equation}
$\mu_{1,2}$ will determine the mass of the black hole (see Section \ref{bhsec}) and we then obtain
\begin{equation}
\tanh(2\mu_{1,2})=\tanh(\mu_{1,2}+\xi_{1,2}+\mu_{1,2}-\xi_{1,2})=E\sqrt{E^2-1}\left(\frac{\tanh\zeta_1+\tanh\zeta_2}{E^2+(E^2-1)\tanh\zeta_1\tanh\zeta_2}\right),
\end{equation}
which is only defined for $E>1$. Another way to see that a black hole has really formed is to directly compute the event horizon, which is also marked in Fig. \ref{2particles}. It coincides with the backwards lightcone of the last points of the wedges $c_{1,2}$ and $c_{2,1}$ and will be discussed in more details in Section \ref{eventsec}.

\subsection{$N$ colliding particles}
We will now go on and treat the general case and show how to construct a spacetime corresponding to $N$ colliding massive particles, without any symmetry restrictions (meaning that all particles can have arbitrary rest masses, located along arbitrary angles and be released from arbitrary radial positions). The procedure is similar to that in the massless case carried out in \cite{Lindgren:2015fum}.\\
\linebreak
We thus assume that we have $N$ particles at angles $\psi_i$, with boost parameters $\zeta_i$ and angular deficits (in the rest frame) of $2\nu_i$. All particles will then collide at $(t=0,\chi=0)$, and after the collision the resulting spacetime will be given by a set of wedges, which we will refer to the {\it allowed} geometry. We will denote the wedge bordered by $w_+^{i}$ and $w_-^{i+1}$ by $c_{i,i+1}$. The tip of this wedge, which is the intersection between $w_+^{i}$ and $w_-^{i+1}$, will be denoted by $I_{i,i+1}$ and will move along an angle $\phi_{i,i+1}$. We will refer to these wedges as {\it final wedges} while the wedges attached to the particles will be the {\it initial wedges}. However, in comparison to the two-particle case, we can no longer attach symmetric wedges to our pointlike particles. The reason is that generically, the intersections $I_{i,i+1}$ will {\it not} be mapped to each other by the isometries associated to the particles (in the two-particle case this happened due to the extra symmetries of the problem). To solve this issue, we will attach non-symmetric wedges to each particle and thus associate a parameter $p_i$ to each one (see Section \ref{ppsec} and especially equations \eqref{c1} and \eqref{Gammapm}). By tuning these parameters, it turns out that we can make sure that all the $N$ intersections $I_{i,i+1}$ are identified as one and the same geodesic in spacetime, and can thus be interpreted as the resulting single object formed in the collision. Note that this gives us $N$ conditions for our $N$ parameters $p_i$, which would then naively have a unique solution. However, in practice, we will have $2N$ unknowns and $2N$ equations. The other $N$ parameters will be the intersections $\phi_{i,i+1}$ and the other $N$ equations are just the definitions of the angles $\phi_{i,i+1}$ as being the angle of the intersections $I_{i,i+1}$.\\
\linebreak
The wedges are thus bordered by surfaces $w_\pm^i$, determined by the equations
\begin{equation}
\tanh\chi\sin(-\phi +\Gamma^i_\pm+\psi_i)=-\tanh\zeta_i \sin\Gamma^i_\pm \sin t, \label{c1i}
\end{equation}
where
\begin{equation}
\tan\Gamma_\pm^i =\pm\tan((1 \pm p_i)\nu_i) \cosh\zeta_i.\label{Gammapmi}
\end{equation}
Thus the intersection $I_{i,i+1}$ is a radial geodesic at an angle $\phi_{i,i+1}$ given by the equation
\begin{equation}
\frac{\tanh\zeta_{i}\sin \Gamma_+^i}{\sin(\psi_i+\Gamma_+^i-\phi_{i,i+1})}=\frac{\tanh\zeta_{i+1}\sin \Gamma_-^{i+1}}{\sin(\psi_{i+1}+\Gamma_-^{i+1}-\phi_{i,i+1})}.\label{intersectionangle}
\end{equation}
The parameters $p_i$ are determined by enforcing that the intersections $I_{i-1,i}$ and $I_{i,i+1}$ are mapped to each other by the isometry $\bu_i$ associated to particle $i$, which is given by equation \eqref{hol}. This computation is a bit more complicated, but results in the equation (see Appendix \ref{peqapp})
\begin{equation}
\tan(p_i \nu_i )=\frac{\tan(\phi_{i,i+1}-\psi_i)+\tan(\phi_{i-1,i}-\psi_i)}{-2\cosh \zeta_i\tan(\phi_{i,i+1}-\psi_i)\tan(\phi_{i-1,i}-\psi_i)+\cot\nu_i(\tan(\phi_{i,i+1}-\psi_i)-\tan(\phi_{i-1,i}-\psi_i))}.\label{peq}
\end{equation}
We can reformulate \eqref{intersectionangle} in terms of $p_i$ and $\nu_i$ by using \eqref{Gammapmi}, which gives the equation
\begin{align}
\frac{-\sin(\phi_{i,i+1}-\psi_i)}{\sinh\zeta_i\tan((1+p_i)\nu_i)}+\frac{\cos(\phi_{i,i+1}-\psi_i)}{\tanh\zeta_i}&=\frac{\sin(\phi_{i,i+1}-\psi_{i+1})}{\sinh\zeta_{i+1}\tan((1-p_{i+1})\nu_{i+1})}\nonumber\\
&+\frac{\cos(\phi_{i,i+1}-\psi_{i+1})}{\tanh\zeta_{i+1}}\label{nueq}
\end{align}
We now have $2N$ equations for our $2N$ parameters $p_i$ and $\phi_{i,i+1}$, and in practice it seems that this system of equations can always be solved. In Section \ref{numericssec} we will explain in detail how one can go about solving these equations in practice. Note also that when taking the massless limit $\nu_i\rightarrow0$, $\zeta_i\rightarrow\infty$ while $\tan\nu_i\sinh\zeta_i\rightarrow E_i$ goes to a constant, we reproduce the expressions in \cite{Lindgren:2015fum}. \\
\linebreak
The geometry after the collision will now either be a black hole or a pointlike particle, depending on if the intersections $I_{i,i+1}$ move on spacelike or timelike geodesics, respectively. We will now define parameters $\Gamma_\pm^{i,i+1}$ and $\zeta_{i,i+1}$ such that the wedge $c_{i,i+1}$ can be mapped either to the form \eqref{c1} and \eqref{Gammapm} or to \eqref{bhwedge3} and \eqref{Gammapmu}. By writing
\begin{align}
&\tanh\chi\sin(-\phi+\phi_{i,i+1}+(\Gamma^{i+1}_-+\psi_{i+1}-\phi_{i,i+1}))=\nonumber\\
&=-\frac{\tanh\zeta_{i+1}\sin\Gamma_-^{i+1}}{\sin(\Gamma^{i+1}_-+\psi_{i+1}-\phi_{i,i+1})}\sin(\Gamma^{i+1}_-+\psi_{i+1}-\phi_{i,i+1})\sin t,\nonumber\\
\end{align}
for $w_+^{i,i+1}=w_-^{i+1}$, and 
\begin{equation}
\tanh\chi\sin(-\phi+\phi_{i,i+1}+(\Gamma^{i}_++\psi_{i}-\phi_{i,i+1}))=-\frac{\tanh\zeta_{i}\sin\Gamma_+^{i}}{\sin(\Gamma^{i}_++\psi_i-\phi_{i,i+1})}\sin(\Gamma^{i}_++\psi_i-\phi_{i,i+1})\sin t,
\end{equation}
for $w_-^{i,i+1}=w_+^i$, we can determine the new parameters $\Gamma_\pm^{i,i+1}$ by
\begin{equation}
\Gamma_-^{i,i+1}=\psi_i+\Gamma_+^i-\phi_{i,i+1},
\end{equation}
\begin{equation}
\Gamma_+^{i,i+1}=\psi_{i+1}+\Gamma_-^{i+1}-\phi_{i,i+1},
\end{equation}
and the parameter $\zeta_{i,i+1}$ is determined by
\begin{equation}
\frac{\tanh\zeta_i\sin\Gamma_+^i}{\sin(\Gamma^i_++\psi_i-\phi_{i,i+1})}=\frac{\tanh\zeta_{i+1}\sin\Gamma_-^{i+1}}{\sin(\Gamma^{i+1}_-+\psi_{i+1}-\phi_{i,i+1})}=\Bigg\{\begin{array}{ll}
                                                                                                                                             \tanh\zeta_{i,i+1},&\text{Point particle}\\
                                                                                                                                             \coth\zeta_{i,i+1},&\text{Black hole}                                                                                                                                             
                                                                                                                                            \end{array}\label{tanhzcothz}
\end{equation}
depending on if the (absolute value) of the above ratio is smaller or larger than 1. The above equality is consistent due to the definition of $\phi_{i,i+1}$. We are now interested in going to the rest frame of each final wedge $c_{i,i+1}$, such that they take the form of static wedges. In the case of a formation of a massive particle, these static wedges will be circle sectors in AdS, while in the case of a formation of a black hole the static wedges will be circle sectors in the BTZ black hole background with unit mass. The parameters for these wedges are obtained from \eqref{Gammapm} in the conical singularity case, and from \eqref{Gammapmu} in the case of the formation of a black hole. These circle sectors are then glued together to form the whole spacetime, and the total angle of these circle sectors will determine the mass of the resulting object.
\begin{figure}
\begin{center}
\includegraphics[scale=0.8]{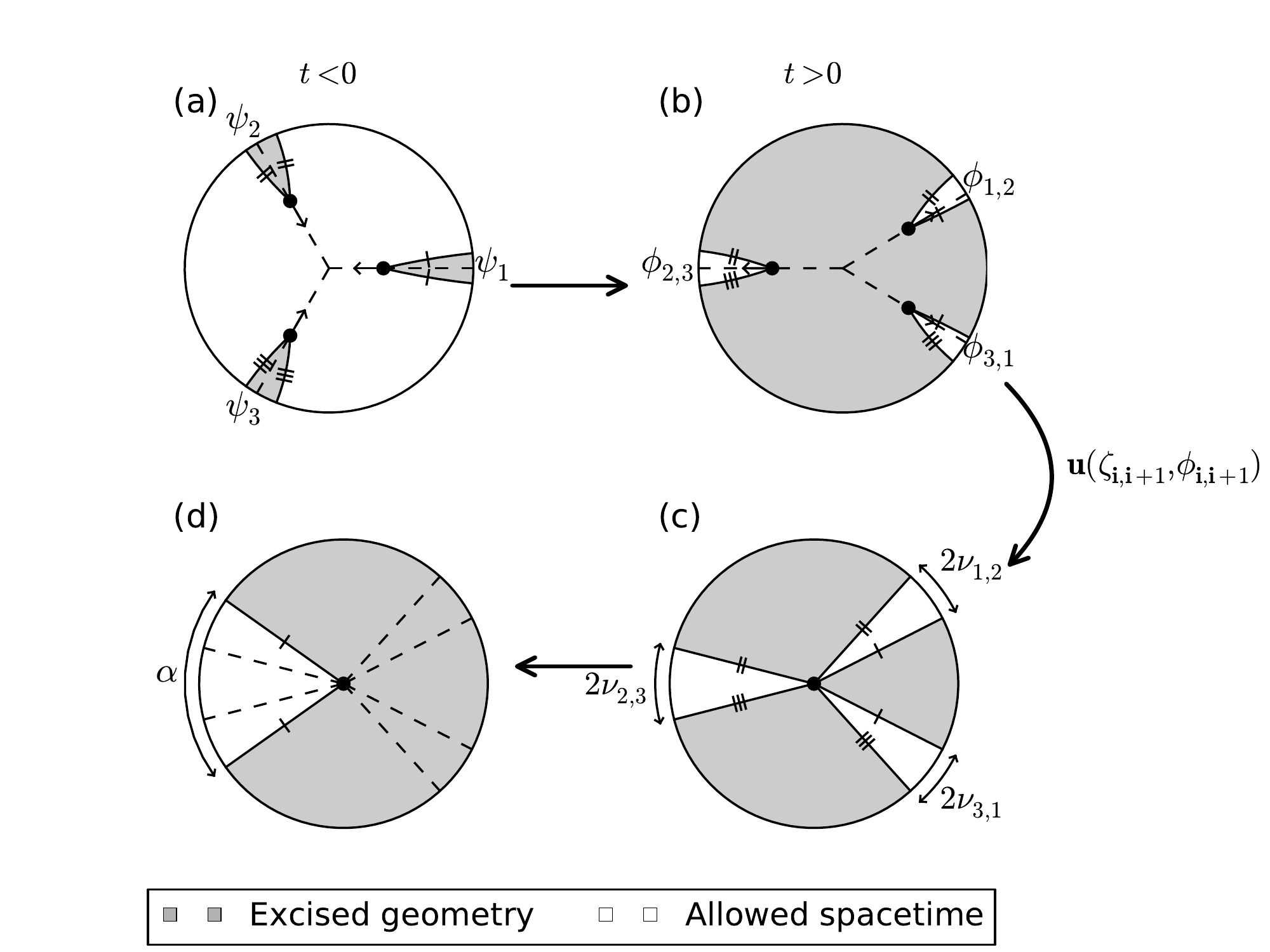}
\caption{\label{3pillustrcs} An illustration of how the transformation of the final geometry to static coordinates is carried out. The parameters are the same as in Fig. \ref{3particlessym_cs}. (a): The spacetime before the collision takes place. (b): The spacetime after the collision has taken place, showing three final wedges of allowed geometry. (c): To go from (b) to (c), we transform each wedge with the isometry $\bu_{i,i+1}$ discussed in Section \ref{ppsec}, with parameters $\zeta=\zeta_{i,i+1}$ and $\psi=\phi_{i,i+1}$. This transforms each moving wedge to static wedges with opening angle $2\nu_{i,i+1}$. (d): We finally push the wedges together to form a more common parametrization of a conical singularity, and we can then define a continuous angle $\hat\phi$ covering the whole spacetime which takes values in the range $(0,\alpha)$ where $\alpha=\sum_i2\nu_{i,i+1}$. The coordinates can then be rescaled such that the metric takes the form \eqref{blackholemetric}. Note that we have the same picture when a black hole forms, except that there is an extra coordinate transformation between panel (b) and (c) and in panel (c) and (d) the metric is a BTZ black hole metric with unit mass instead of \ads.}
\end{center}
\end{figure}

\subsubsection{Formation of a conical singularity}
\begin{figure}
\begin{center}
\includegraphics[scale=0.8]{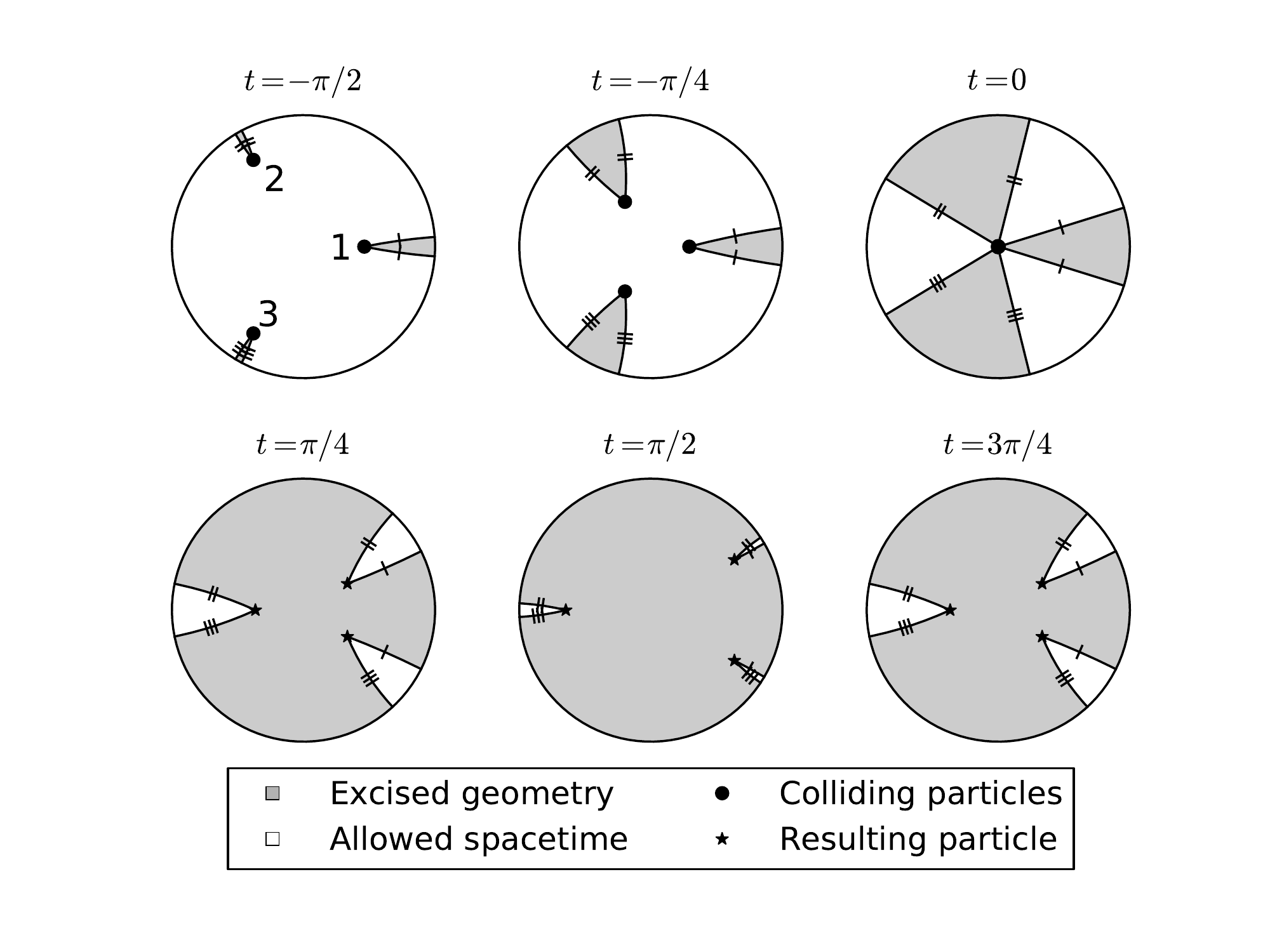}
\caption{\label{3particlessym_cs}Three colliding massive particles forming a new massive particle. The particles have the same rest mass but particle 1 is released from a different radial position compared to the other two (or in other words, it has a different boost parameter). The final wedges after the collision together form a conical singularity spacetime, but where the wedges have been boosted, and they will keep oscillating forever. Note also that, even though we have marked which surfaces are identified via the isometries, there will generically be an additional time shift under the identifications (so curves on the same time slice are generically {\it not} mapped to each other). The parameters $p_i$ which determine the orientation of the wedges have been obtained by solving equations \eqref{peq} and \eqref{nueq}.}
\end{center}
\end{figure}

In the case of the formation of a conical singularity, we want to map each final wedge $c_{i,i+1}$ to a static wedge with parameters $\nu_{i,i+1}$ and $p_{i,i+1}$. From \eqref{Gammapm}, we have the relation
\begin{equation}
\tan \Gamma^{i,i+1}_\pm=\pm \tan((1\pm p_{i,i+1})\nu_{i,i+1})\cosh\zeta_{i,i+1},\label{Gammapnu}
\end{equation}
from which we can determine $p_{i,i+1}$ and $\nu_{i,i+1}$. $2\nu_{i,i+1}$ is now the total angle of this static wedge (instead of being the angular deficit). All these $N$ wedges are then glued together by the identifications, forming a conical singularity with total angle $\alpha=\sum_{i}2\nu_{i,i+1}$. By rescaling the coordinates, the spacetime can be written in the form \eqref{blackholemetric}, and we obtain that the parameter $M$ is given by $M=-\alpha^2/(2\pi)^2$. The mapping from the moving final wedges, to the static wedges, is illustrated in Fig. \ref{3pillustrcs}, and the full solution of three colliding particles forming a new pointlike particle is shown in Fig. \ref{3particlessym_cs}.

\subsubsection{Formation of a black hole}
\begin{figure}
\begin{center}
\includegraphics[scale=0.8]{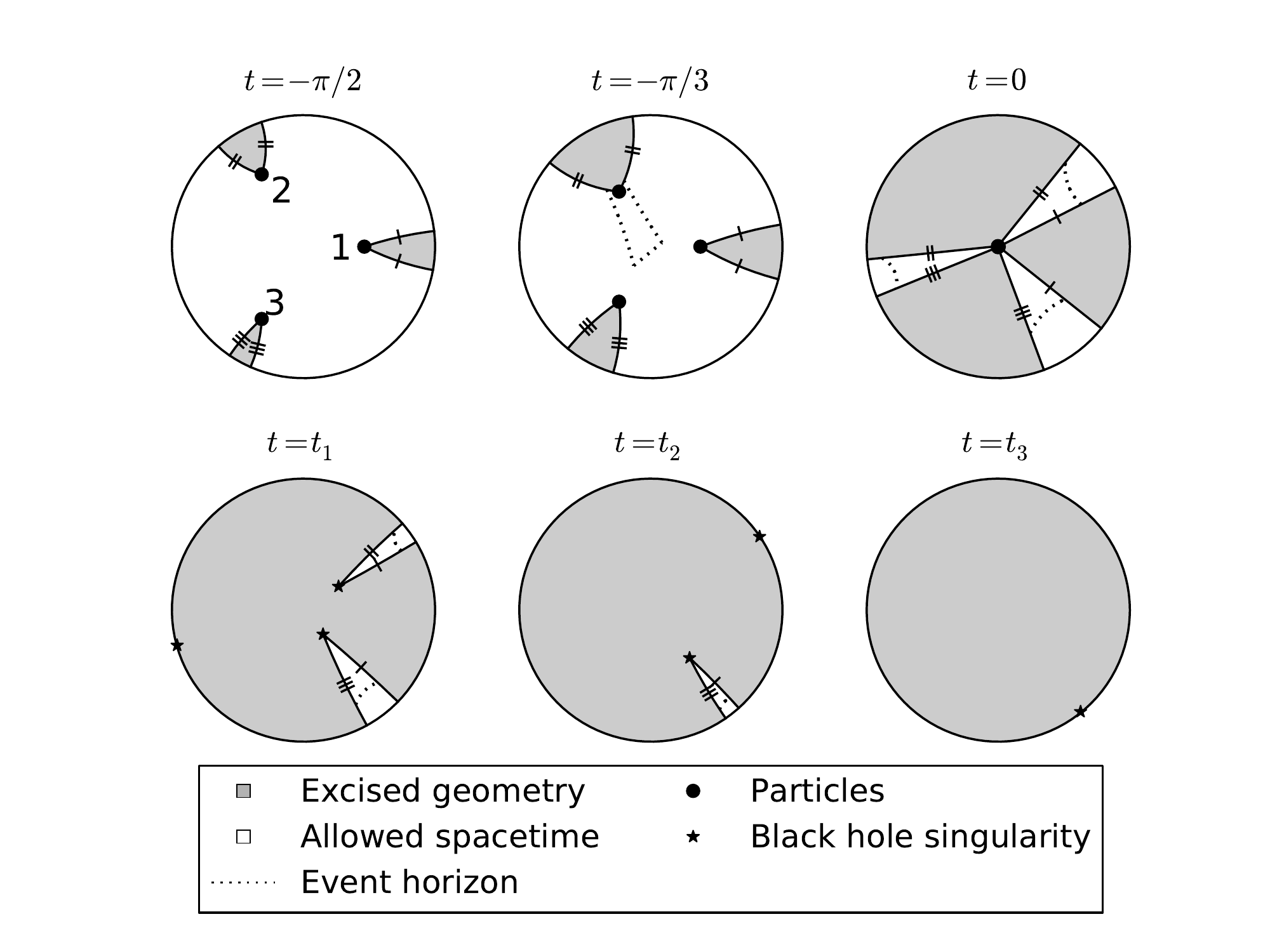}
\caption{\label{3particlesgen}Three colliding massive particles forming a BTZ black hole. In this example, there are no symmetry restrictions, and we have $\zeta_i=(1,1.5,1.5)$ and $\nu_i=(0.4,0.8,0.4)$. The final three wedges after the collision all disappear at different times $t_1$, $t_2$ and $t_3$, but note that all these final points are identified by the holonomies of the particles. The parameters $p_i$ which determine the orientation of the wedges have been obtained by solving equations \eqref{peq} and \eqref{nueq}.}
\end{center}
\end{figure}

\begin{figure}
\begin{center}
\includegraphics[scale=0.8]{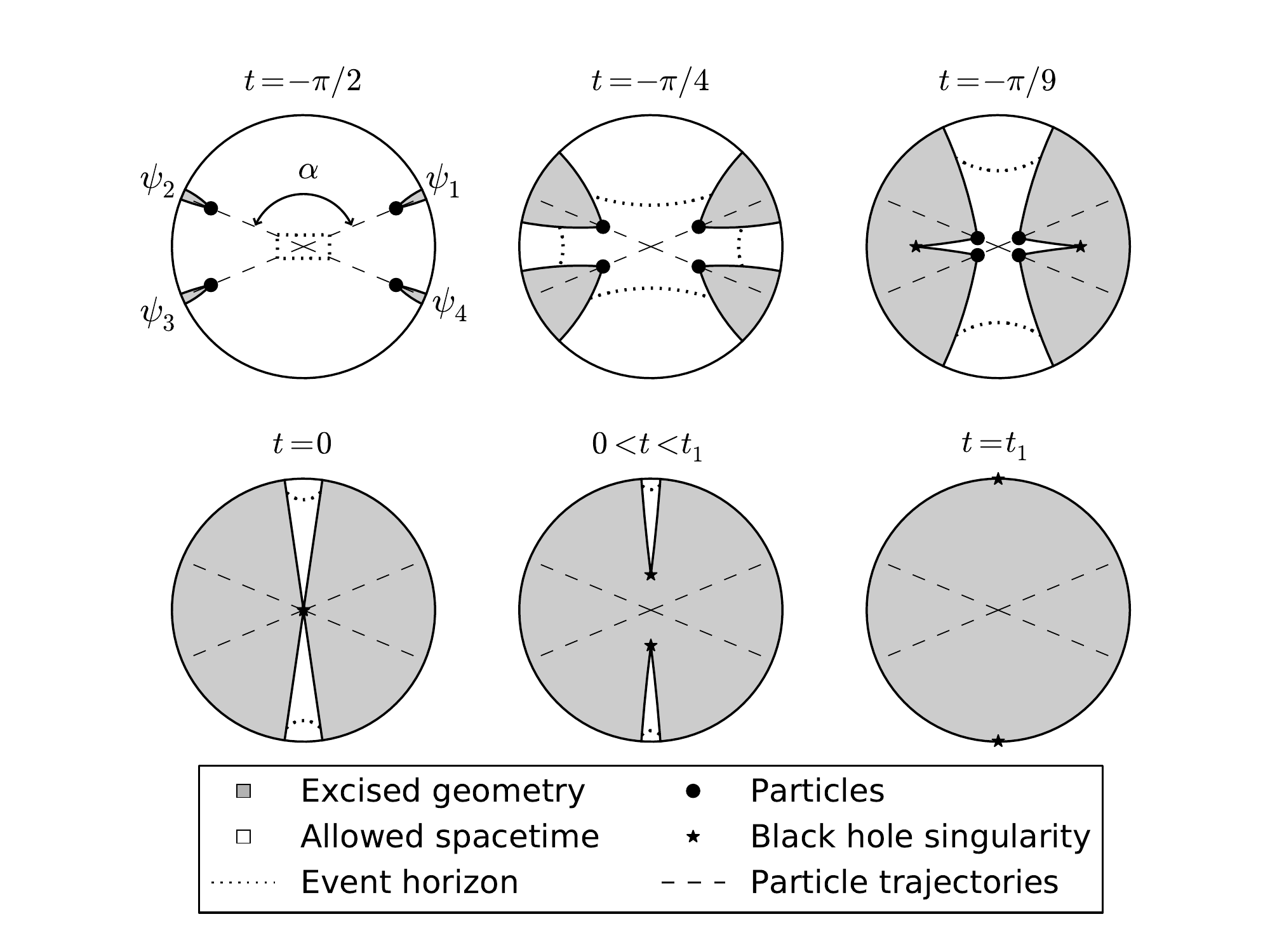}
\caption{\label{4particles} An example of four colliding particles, where they all have the same rest masses and are released from the same radial position, but the rotational symmetry is broken by the angles $\psi_i$. Specifically, we can parametrize the solution in terms of an angle $\alpha\equiv\psi_2-\psi_1=\psi_4-\psi_3$, and then $\psi_3-\psi_2=\psi_1-\psi_4=\pi-\alpha$. In this particular example, we have $\alpha=3\pi/4$. It is clear that the wedges are not symmetric around the trajectories of the particles, and they ``repel'' each other as the angle $\alpha$ increases. Note that in this example, two of the final spacelike geodesics connected to the collision point go ``backwards in time'', but this has no physical significance whatsoever. The parameters $p_i$ which determine the orientation of the wedges have been obtained by solving equations \eqref{peq} and \eqref{nueq}.}
\end{center}
\end{figure}

In the case of a formation of a black hole, we want to map each final wedge $c_{i,i+1}$ to a circle sector in the black hole geometry, with parameters $\xi_{i,i+1}$ and $\mu_{i,i+1}$. From equation \eqref{Gammapmu}, we have the relation
\begin{equation}
\tan \Gamma^{i,i+1}_\pm=\mp \tanh(\mu_{i,i+1}\pm\xi_{i,i+1})\sinh\zeta_{i,i+1},\label{Gammapmui}
\end{equation}
from which we can determine $\xi_{i,i+1}$ and $\mu_{i,i+1}$. The total angle of such a circle sector is given by $2\mu_{i,i+1}$, but it is a wedge cut out from a black hole spacetime with metric \eqref{unitbhmetric} instead of the \ads metric. These $N$ wedges are now glued together, forming a wedge with total angle $\alpha=\sum_i2\mu_{i,i+1}$. By rescaling the coordinates, we can write the spacetime in the form \eqref{blackholemetric}, and the parameter $M$ is obtained as $M=\alpha^2/(2\pi)^2$. Figures \ref{3particlesgen} and \ref{4particles} show the formation of a black hole from a collision of three and four particles, respectively.


\subsubsection{The event horizon}\label{eventsec}
We will now discuss how to compute the event horizon when a black hole forms. First let us call the last point, before the spacetime disappears, the {\it last boundary point}, and we will denote the \ads manifold where we are cutting out the wedges by $\mathcal{M}$. Now, in $\mathcal{M}$, this last boundary point is represented by $N$ points, which are identified by the holonomies of the particles. We will label these points by $P_{i,i+1}$, and are thus the end points of the final wedges $c_{i,i+1}$ of allowed spacetime (and also the endpoints of the intersections $I_{i,i+1}$). Now consider the backwards lightcones $\mathcal{L}_{i,i+1}$ in $\mathcal{M}$, of these $N$ last points. We will now show that the restriction of these lightcones to the allowed spacetime can be used to construct the event horizon. It is clear, that in the final wedge of allowed space $c_{i,i+1}$, points outside the lightcone $\mathcal{L}_{i,i+1}$ are causually disconnected from the last point $P_{i,i+1}$. They are also causually disconnected from the whole boundary in wedge $c_{i,i+1}$. One might ask the question, if lightrays from these points can somehow cross the bordering surfaces $w_-^{i,i+1}=w_+^i$ and $w_+^{i,i+1}=w_-^{i+1}$ of the wedge $c_{i,i+1}$, enter another wedge, and then reach the boundary. However, this is not possible, as we will now show. Let us denote the intersection between the surface $w_-^{i,i+1}$ and $\mathcal{L}_{i,i+1}$ by $I_{i,i+1}^-$. We will now show that the intersection $I_{i,i+1}^-$ is mapped to the intersection $I_{i-1,i}^+$ via the holonomy of particle $i$, and thus when crossing a wedge's bordering surface from inside (outside) a wedge's lightcone, will always result in again ending up inside (outside) the neighbouring wedge's lightcone. This can be seen from the following three facts:
\begin{itemize}
 \item The lightcone $\mathcal{L}_{i,i+1}$ is, by definition, composed of all lightlike geodesics ending on the point $P_{i,i+1}$.
 \item The intersection of $\mathcal{L}_{i,i+1}$ with $w_-^{i,i+1}$ or $w_+^{i,i+1}$, are geodesics, since all these surfaces are total geodesic surfaces.
 \item Since both the wedge $c_{i,i+1}$, and the lightcone $\mathcal{L}_{i,i+1}$, end on the point $P_{i,i+1}$, the intersections will also end on this point.
\end{itemize}
Thus, since the intersection $I_{i,i+1}^-$ is a geodesic, is located on $\mathcal{L}_{i,i+1}$, as well as end on $P_{i,i+1}$, it must be a lightlike geodesic. Furthermore, since it is located on $w_+^{i}$, it is {\it the} unique lightlike geodesics that lies on this two-dimensional surface and ends on $P_{i,i+1}$. A similar argument can be applied to show that also $I_{i,i+1}^+$, the intersection between $w_+^{i,i+1}=w_-^{i+1}$ and $\mathcal{L}_{i,i+1}$, is {\it the} unique lightlike geodesic on the two-dimensional surface $w_+^{i,i+1}$ and ends on $P_{i,i+1}$. Therefore, since the point $P_{i,i+1}$ is identified with point $P_{i-1,i}$ and $P_{i+1,i+2}$, and the surface $w_+^{i}$ ($w_-^{i+1}$) is identified with $w_-^{i}$ ($w_+^{i+1}$), the intersections $I_{i,i+1}^-$ ($I_{i,i+1}^+$), being the unique lightlike geodesics ending at $P_{i,i+1}$ which lies on $w_+^i$ ($w_-^{i+1}$), must be mapped to $I_{i-1,i}^+$ ($I_{i+1,i+2}^-$), the unique lightlike geodesics lying on $w_-^{i}$ ($w_+^{i+1}$) and ending on $P_{i-1,i}$ ($P_{i+1,i+2}$). This gives a complete characterization of the event horizon in the final wedges of the geometry (which would, if taking the thin shell $N\rightarrow\infty$ limit, correspond to the geometry after the shell). When the event horizon is outside the final wedges, it will be composed of piecewise segments of $\mathcal{L}_{i,i+1}$ such that the points inside the event horizon are outside {\it all} the $N$ lightcones $\mathcal{L}_{i,i+1}$. It seems difficult to find a nice expression for the horizon in a general spacetime, let alone in the $N\rightarrow\infty$ limit, and we will not pursue that further in this paper.



\subsection{The problem of colliding spinning particles}\label{spinsec}
One interesting generalization would be to collide spinning particles, with the aim to form rotating black holes. In this section we will attempt to construct such solutions, and show that the construction in \cite{Matschull:1998rv} does not generalize trivially. A spinning particle is not defined by a single group element $\bu$, but by two elements $\bu_L$ and $\bu_R$. The spacetime can then again be constructed by cutting out a wedge like in the non-spinning case, but the surfaces bordering the wedge $w_\pm$ are now identified by $w_-=\bu_L^{-1}w_+\bu_R$. The trajectory of the particle is now not a set of fixed points of the isometry. Instead, only the whole geodesic will be invariant, but points on the geodesic will be mapped to other points on the geodesic. In other words, for a massive particle, when moving around the geodesic (or equivalently when crossing the wedge), there will be a shift in the proper time along the geodesic, while for massless particles there will be a shift in the affine parameter. This means that points on the geodesic with a particular proper time (or affine parameter) will not be well defined points in the spacetime, since different points will be identified with one another via the isometry. Note also that there will be closed timelike curves close to the trajectory.\\
\linebreak
Now it is very easy to see that collision processes will be problematic. Since different points on the trajectory of each particle will be identified via the isometry, it is impossible to have a well defined collision point. Furthermore, if we try to merge two particles in a collision process, and thus abruptly change the trajectories, there will be points that are mapped to parts of the geodesic that have been removed. In other words, when crossing the wedge close to the collision point, since there will be a shift in proper time along the geodesic of the particle, one will be mapped to a part of spacetime that has been removed (that is inside the wedge of the other particle). Such a construction would thus not represent a consistent solution of Einstein's equations. We do not claim that it is impossible to construct spacetimes with colliding spinning pointlike particles, but at least the standard construction for non-rotating particles does not immediately work.

\section{The $N\rightarrow\infty$ limit and emerging thin shell spacetimes}\label{limitsec}
When taking the limit of an infinite number of particles, we will have $\nu_i\rightarrow0$, while $\zeta_i$ and $p_i\nu_i$ will go to constants. It is convenient to introduce continuous interpolating functions $T(\psi_i)=p_i\nu_i$, $\cZ(\psi_i)=\zeta_i$ and also $\Phi(\psi_i)=\phi_{i,i+1}-\psi_i$ which all approach some finite continuous functions in the limit. We will also assume that the density of particles remains constant, namely that $2\nu_i=\rd \phi \rho(\psi_i)$, where $d\phi=\psi_{i+1}-\psi_i$ (for simplicity we will assume that the angles $\psi_i$ are distributed homogeneously around the circle). Remember that $2\nu_i$, the deficit angle in the restframe, is equal to the mass of the particle (in units where $8\pi G=1$). We will thus refer to $\rho$ as the rest mass density. A straightforward calculation shows that the discrete equations \eqref{peq} and \eqref{nueq} then reduce to the following differential equations
\begin{equation}
\tan T (\tan \Phi)'=\rho\tan\Phi+\rho\cosh\cZ \tan^2\Phi\tan T-\tan T\tan^2\Phi-\tan T,\label{odeP}
\end{equation}
\begin{align}
(\tan T)'\sinh Z\tan\Phi&+\tan T\cosh Z Z'\tan\Phi-\tan^2T Z'=\nonumber\\
&\frac{\rho\sinh Z}{\cos^2 T}\tan\Phi-\sinh Z \tan T-\sinh Z\cosh Z\tan^2T\tan\Phi,\label{odeT}
\end{align}
where $'$ denotes derivative with respect to $\phi$. In the massless limit, where $\rho\rightarrow0$ and $Z\rightarrow\infty$, we should let $\tan T\cosh\cZ\rightarrow P$ and $\rho\cosh\cZ\rightarrow2\rho_0$, which reduces the expressions to those in \cite{Lindgren:2015fum}.\\
\linebreak
We will later be interested in computing $\nu_{i,i+1}$ from \eqref{Gammapnu} or $\mu_{i,i+1}$ from \eqref{Gammapmui}, for large $N$. This will require us to compute the difference $\tan\Gamma_+^{i,i+1}-\tan\Gamma_-^{i,i+1}$ to order $1/N$ (note that it vanishes at zeroth order). A straightforward calculation shows that
\begin{align}
&\tan\Gamma_-^{i,i+1}=\frac{\tan T \cosh \cZ - \tan \Phi}{1+\tan\Phi\tan T\cosh\cZ}+\frac{(1+\tan^2\Phi)\cosh\cZ \rho d\phi}{2\cos^2T(1+\tan\Phi\tan T\cosh\cZ)^2}+O(\frac{1}{N^2}),\nonumber\\
&\nonumber\\
&\tan\Gamma_+^{i,i+1}=\frac{\tan T \cosh \cZ - \tan \Phi}{1+\tan\Phi\tan T\cosh \cZ}+\nonumber\\
&+\frac{(1+\tan^2\Phi)(1+\tan^2T\cosh^2\cZ+\tan T\sinh\cZ\cZ'+\cosh\cZ(1+\tan^2T)(T'-\frac{\rho}{2}))}{(1+\tan T\cosh\cZ\tan\Phi)^2}d\phi\nonumber\\
&+O(\frac{1}{N^2}).\label{gamma}
\end{align}
Thus we obtain
\begin{align}
\tan\Gamma_+^{i,i+1}-\tan\Gamma_-^{i,i+1}=&\frac{\cos^2T(\tan T\cosh\cZ)'-\cosh\cZ\rho+\cos^2T+\sin^2T\cosh^2\cZ}{(\cos T\cos\Phi+\sin T\cosh\cZ\sin\Phi)^2}d\phi\nonumber\\
&+O(\frac{1}{N^2}).\label{gammadiff}
\end{align}
\linebreak
We will now compute explicitly the metric in the resulting thin shell spacetimes. The spacetime will consist of two patches, separated by a timelike shell of matter denoted by $\mathcal{L}$. The spacetime outside the shell will either be that of a conical singularity or that of a BTZ black hole, and the inside will just consist of empty \ads. This shell will have a non-trivial embedding in the two patches, specified by the function $\cZ$ and the mass density $\rho$. This is different from the massless shells studied in \cite{Lindgren:2015fum} where these was only one free function (the energy density) that specified the properties of the shell. We will write the spacetimes inside and outside $\mathcal{L}$ in rotationally symmetric coordinates, even though the total spacetime is not rotationally symmetric. The coordinate systems will then be discontinuous when crossing $\mathcal{L}$ with a non-trivial and angle dependent mapping, and finding the map from the coordinates inside to the coordinates outside is the main goal of this section, as well as computing the induced metric on $\mathcal{L}$. We will separate the two calculations into that of formation of a conical singularity and that of formation of a black hole. Conceptually these two calculations are different, as they will rely on applying timelike respectively spacelike isometries of \ads, but the end result will essentially be the same but with different signs of the mass. Although the case of a formation of a conical singularity is not very physical, the computations are easier to understand and more intuitively visualized, thus it is recommended to understand it first before doing the black hole computation. We will work with a finite number of particles first, and then take the limit after we have transformed the final geometry to a static coordinate system.

\subsection{Formation of a conical singularity}
In the patch inside the shell, the embedding of the surface $\mathcal{L}$ will be determined by the equation $\tan\chi=-\tanh\cZ(\phi)\sin t$. The geometry outside the shell is consisting of several moving wedges, and before we take the $n\rightarrow\infty$ limit, we will have to transform these wedges to a static geometry. The coordinates after doing this transformation will be denoted by $(\pc t,\pc\chi,\pc\phi)$, and consists of several disconnected (static) wedges, which are glued together via the isometries, and the metric is still given by \eqref{adsmetric}. We will then ``push the wedges together'' and define a new continuous angular coordinate $\hat\phi$, see Fig. \ref{3pillustrcs} for an illustration. The metric is now still given by \eqref{adsmetric}, but the angular variable takes values in the range $(0,\alpha)$ (so that the angular deficit is $2\pi-\alpha$). In the static coordinates, the wedge labeled by $(i,i+1)$, is a circle sector we will denote $c^{\text{static}}_{i,i+1}$, which has opening angle $2\nu_{i,i+1}$. Thus when passing $c^{\text{static}}_{i,i+1}$, $\hat\phi$ will increase by $2\nu_{i,i+1}$. This means that, in the $n\rightarrow\infty$ limit, we can write $\hat\phi=\hat{\phi}_0+\sum_{0\leq j\leq i} 2\nu_{j,j+1}+O(1/N)$ when $\hat\phi\in c^{\text{static}}_{i,i+1}$, where $\hat{\phi}_0$ is an overall angular shift (the approximate value of $\hat\phi$ when $\hat\phi\in c^{\text{static}}_{0,1}$). $\alpha$ is of course given by $\alpha=\sum 2\nu_{i,i+1}$. \\
\linebreak
Let us define $\hat \cZ$ and $\hat T$ to be the continuous interpolating functions corresponding to $\zeta_{i,i+1}$ and $p_{i,i+1}\nu_{i,i+1}$. Then, by taking the limit in equation \eqref{tanhzcothz} we obtain
\begin{equation}
\tanh\hat \cZ=\frac{\tan T\sinh\cZ}{\tan T\cosh\cZ\cos\Phi-\sin\Phi},\label{tanhhatZ}
\end{equation}
and from \eqref{Gammapnu} and \eqref{gamma}, we obtain
\begin{equation}
\tan \hat T\cosh \hat \cZ=\frac{\tan T\cosh Z-\tan\Phi}{1+\tan\Phi\tan T\cosh Z}. \label{tanhatTcoshhatZ}
\end{equation}
From this we can also derive the useful relation (see Appendix \ref{usefulrel})
\begin{equation}
\cos\hat T=\cos\Phi\cos T+\sin\Phi\sin T\cosh Z,\label{coshatT}
\end{equation}
and using this relation we can immediately derive from \eqref{tanhhatZ} and \eqref{tanhatTcoshhatZ} that
\begin{equation}
\sin\hat T\cosh\hat Z=\sin T\cosh Z\cos\Phi-\sin\Phi\cos T,\label{sinhatTcoshhatZ}
\end{equation}
\begin{equation}
\sin\hat T\sinh\hat Z=\sin T\sinh Z.\label{sinhatTsinhhatZ}
\end{equation}
From \eqref{Gammapnu}, we can also compute
\begin{equation}
\tan \Gamma_+^{i,i+1}-\tan \Gamma_-^{i,i+1}=\frac{2\cosh\hat Z}{\cos^2\hat T} \nu_{i,i+1}+O(\frac{1}{n^2}).\label{gammadiffcs}
\end{equation}
Now using \eqref{gammadiff}, it can be proven that in the limit we have (see Appendix \ref{anglemap})
\begin{equation}
\hat\phi=\hat{\phi}_0-\int_0^\phi\frac{\sin \hat T}{\sin \Phi}\left(\cos T-\frac{\sin T}{\sinh\cZ}\partial_\phi \cZ\right)d\phi.\label{cs_anglemap}
\end{equation}

\begin{figure}
\begin{center}
\includegraphics[scale=0.8]{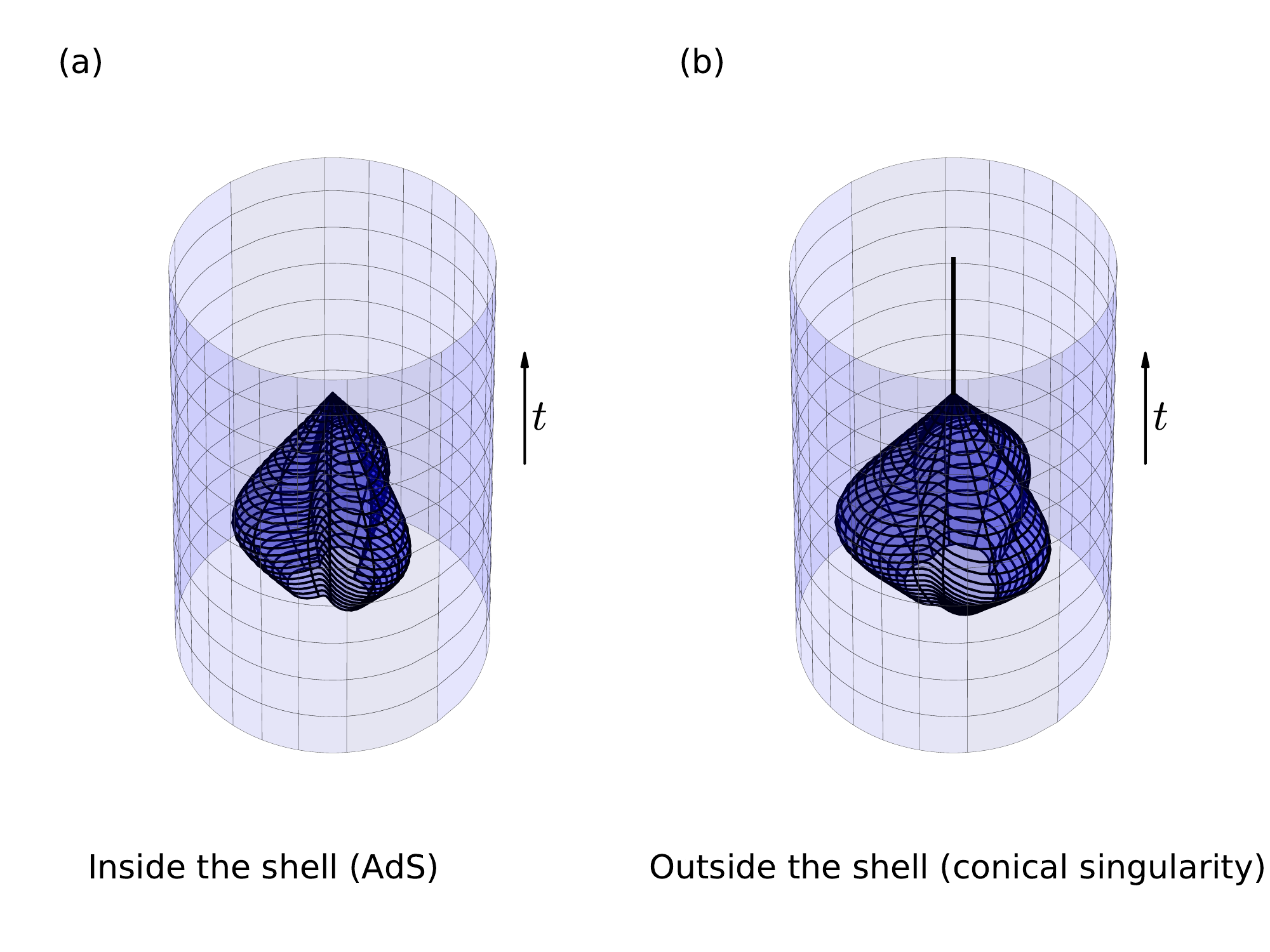}
\caption{\label{shellcsfig} An illustration of a massive thin shell spacetime forming a conical singularity. Panel (a) shows the embedding of the shell in a spacetime with the AdS metric, as seen from the inside part of the shell. The allowed part of spacetime is inside the plotted surface, while everything else should be discarded. The surface in (a) is then glued to the surface in (b) via a non-trivial coordinate transformation. The surface in (b) is embedded in a spacetime with a conical singularity at the origin, which is shown as the thick black line. The allowed part of the spacetime is outside the plotted surface in (b) while the inside part should be discarded. The particular form of the embedding inside the shell has been taken to be $Z(\phi)=1+\frac{1}{2}\cos(3\phi)$. To make the illustration possible we are again using the compactification of the radial coordinate mentioned after equation \eqref{adsmetric}}
\end{center}
\end{figure}

\noindent We will also be interested in the shape of the shell in the $(\pc\chi,\pc t,\pc\phi)$ coordinates, which will be specified by a function $\pc Z$. Note that $\pc Z$ will depend non-trivially on $Z$ and $\rho$, and this is what makes the computations much more involved compared to the lightlike case (for the lightlike shells, since the lightlike geodesics are invariant under the coordinate transformations that bring us to the static coordinate system, the shape of the shell is the same in both coordinate systems). To determine this, let us first see how generic massive geodesics are mapped under the coordinate transformation that brings us to the static coordinates. It will be useful to work in the static coordinates $(\pc t,\pc\chi,\pc\phi)$ before we push the wedges together. Let us assume that some arbitrary massive geodesic is given by the equation $\tanh\chi=-\tanh\xi\sin t$. The coordinate transformation to go from the static coordinates is given by \eqref{boosteqs} with $\psi=\phi_{i,i+1}$ and $\zeta=\zeta_{i,i+1}$. This is mapped to a new geodesic given by $\tanh\pc\chi=-\tanh\pc\xi\sin\pc t$ at some angle $\pc\phi$. From \eqref{boosteqs} we obtain then that the radial coordinates are related as
\begin{equation}
\sinh\chi=\sinh\pc\chi\tanh\xi\left(\frac{\cosh\zeta_{i,i+1}}{\tanh\pc\xi}+\sinh\zeta_{i,i+1}\cos(\phi_{i,i+1}-\pc\phi)\right).
\end{equation}
To see how this coordinate transformation acts on the massive geodesics that the particles follow, we can set $\xi=\zeta_i$ and $\pc\xi=\pc\zeta_i$, and in the continuous limit these become $Z$ respectively $\pc Z$. The embedding equation of the shell will then be $\tanh\chi=-\tanh Z\sin t$ inside the shell, and $\tanh\pc\chi=-\tanh \pc Z\sin\pc t$ outside the shell. We also note that in the limit, we have $\pc\phi-\phi_{i,i+1}\rightarrow \hat{T}$, thus
\begin{equation}
\sinh\chi=\sinh\pc\chi\tanh Z\left(\frac{\cosh\hat Z}{\tanh \pc Z}+\sinh\hat Z\cos\hat T\right).
\end{equation}
To determine $\pc Z$, we can use the relations \eqref{proptime2} between the proper time and $\chi$ and $\pc\chi$, which read
\begin{equation}
\begin{array}{ll}
\sinh\chi=-\sinh Z\sin \tau, &\quad \sinh\pc\chi=-\sinh \pc Z\sin \tau.\\\label{ZZpproptime}
\end{array}
\end{equation}
This yields
\begin{equation}
\cosh Z=\cosh \pc Z\cosh\hat Z+\sinh \pc Z\sinh \hat Z\cos\hat T.\label{Zprel1}
\end{equation}
Analogously, it is also possible to obtain a similar relation by looking at the inverse transformation, from which we instead obtain
\begin{equation}
\cosh \pc Z=\cosh Z\cosh\hat Z-\sinh Z\sinh\hat Z\cos\Phi.\label{Zprel2}
\end{equation}
From the above equations, we can eliminate $\cosh \pc Z$, and then after using \eqref{coshatT}, \eqref{sinhatTcoshhatZ} and \eqref{sinhatTsinhhatZ} we can derive the very simple relation
\begin{equation}
\frac{\sinh \pc Z}{\sinh Z}=-\frac{\sin \Phi}{\sin\hat T}.\label{ZpZ}
\end{equation}
This together with \eqref{ZZpproptime}, implies a relation between $\pc\chi$ and $\chi$ when crossing the shell, namely
\begin{equation}
\sinh\chi=-\frac{\sin\hat T}{\sin \Phi}\sinh\pc\chi=\frac{\partial_\phi \hat \phi}{\cos T-\frac{\sin T}{\sinh Z}\partial_\phi Z}\sinh\pc\chi.\label{chimap}
\end{equation}
This equation is useful when comparing with the massless limit. We now know exactly how the coordinates are related when crossing the shell, as well as the shape of the shell in both patches, but it will also be of interest to compute the induced metric. A convenient time coordinate to use for the intrinsic geometry on the shell is the proper time $\tau$ of the pointlike particles. As is shown in Appendix \ref{app_ind_metric}, the induced metric as seen from the patch inside the shell can be simplified to the form

\begin{equation}
ds^2=-d\tau^2+\sin^2\tau(\sinh^2Z+(\partial_\phi Z)^2)d\phi^2.\label{indmetric}
\end{equation}
Analogously, the induced metric outside the shell can be computed as
\begin{equation}
ds^2=-d\tau^2+\sin^2\tau(\sinh^2\pc Z+(\partial_{\hat\phi} \pc Z)^2)d\hat\phi^2.
\end{equation}
Note that $\hat\phi$ still takes values in the range $(0,\alpha)$.\\
\linebreak
We will now move to the final coordinate system that is most suitable for thin shell spacetimes and for applying junction conditions, namely the coordinate system where the metric inside the shell takes the form
\begin{equation}
ds^2=-f(r)dt^2+\frac{dr^2}{f(r)}+r^2d\phi^2,\label{rtpmetric}
\end{equation}
while the metric outside the shell is 
\begin{equation}
d\bar s^2=-\bar{f}(\bar r)d\bar t^2+\frac{d\bar r^2}{\bar{f}(\bar r)}+\bar r^2d\bar\phi^2,\label{rtpbarmetric}
\end{equation}
where we have defined $f(r)=1+r^2$ and $\bar{f}(\bar r)=-M+\bar{r}^2$. These coordinates are convenient since they take the same form for both the conical singularity case and the black hole case (note that for the conical singularity spacetimes we currently consider, we have $M<0$). Both angular coordinates now take values in $(0,2\pi)$, and the coordinate transformations that brings the metrics to these forms are
\begin{equation}
\begin{array}{ll}
r=\sinh\chi, &\quad \bar r=\sqrt{-M}\sinh\pc\chi,\\
\bar\phi=\frac{\hat \phi}{\sqrt{-M}}, &\quad \bar t=\frac{\pc t}{\sqrt{-M}},\label{bartransf_cs}\\
\end{array}
\end{equation}
while $t$ and $\phi$ remain the same and $M$ is given by
\begin{equation}
M=-\left(\frac{\alpha}{2\pi}\right)^2,
\end{equation}
where we recall that $(0,\alpha)$ is the range of the variable $\hat\phi$. The embedding of the shell is now given by 
\begin{equation}
\frac{r}{\sqrt{f(r)}}=-\frac{R}{\sqrt{f(R)}}\sin t,
\end{equation}
inside, and 
\begin{equation}
\frac{\bar r}{\sqrt{\bar{f}(\bar r)}}=-\frac{\bar R}{\sqrt{\bar{f}(\bar R)}}\sin (\sqrt{-M}\bar t),
\end{equation}
outside, where we have defined $\sinh Z\equiv R$ and $\sqrt{-M}\sinh \pc Z\equiv\bar R$. In terms of the proper time, the embedding inside the shell is given by
\begin{equation}
\begin{array}{ll}
r=-R\sin\tau, &\quad \tan t=\sqrt{f(R)}\tan \tau,\\
\end{array}
\end{equation}
and 
\begin{equation}
\begin{array}{ll}
\bar r=-\bar R\sin\tau, &\quad \tan (\sqrt{-M} \bar t)=\frac{\sqrt{\bar{f}(\bar R)}}{\sqrt{-M}}\tan \tau,\\
\end{array}
\end{equation}
outside. The induced metric is now given by
\begin{equation}
ds^2=-d\tau^2+\sin^2\tau h^2d\phi^2=-d\tau^2+\sin^2\tau\bar h^2d\bar \phi^2,
\end{equation}
where we have defined $h^2\equiv R^2+\frac{(\partial_\phi R)^2}{f(R)}$ and $\bar h^2\equiv \bar{R}^2+\frac{(\partial_{\bar \phi} \bar R)^2}{\bar{f}(\bar R)}$. It can be proven explicitly that the induced metric is the same in both coordinate systems, or in other words that
\begin{equation}
\left(\frac{d\bar\phi}{d\phi}\right)^2=\frac{h^2}{\bar{h}^2}.\label{indmetric_consistency}
\end{equation}
This is proved in Appendix \ref{app_cons_cond}. This is a necessary condition that must be satisfied to have a consistent geometry. An illustration of this spacetime is shown in Fig. \ref{shellcsfig}.\\
\linebreak
Defining $R$ and $\bar R$ (or equivalently $h$ and $\bar h$, or $Z$ and $\pc Z$) will determine the whole spacetime. This will turn out to be a useful parametrization when analysing the geometry using the junction conditions where we will take the viewpoint that our thin shell spacetime is {\it defined} by the two free functions $R$ and $\bar R$, from which the embedding of the shell and the energy density can be derived. An interpretation in terms of pointlike particles is not necessary from this point of view.

\subsection{Formation of a black hole}
When a black hole forms we can no longer map the resulting wedges to circle sectors of \ads, essentially since the tip of a resulting wedge will now be a spacelike geodesic. Instead we must map them to static patches of a BTZ black hole, as specified by equation \eqref{unitbhmetric}. As explained in Section \ref{bhsec}, we do this in two steps. We firs go to a coordinate system ($\pc t,\pc\chi,\pc\phi)$ where the final geodesics are mapped to ``straight'' spacelike geodesics in a constant timeslice, such that the wedges take the form \eqref{bhwedge1}. These wedges are still defined in a spacetime with the standard \ads metric \eqref{adsmetric}. We will then do another coordinate transformation to a coordinate system ($\sigma,\rho,y)$ where the wedges are normal circle sectors, but where the metric takes the form \eqref{unitbhmetric}, a BTZ black hole with unit mass. We will then again ``push the wedges together'' and define a continuous angular coordinate $\hat{y}$. Since these final static wedges (denoted by $c^{\text{static}}_{i,i+1}$) will have an opening angle of $2\mu_{i,i+1}$, the coordinate $\hat{y}$ will increase by $2\mu_{i,i+1}$ when crossing one such wedge. Thus $\hat{y}=\hat{y}_0+\sum_{0\leq j\leq i}2\mu_{j,j+1}+O(1/N)$ when $\hat{y}\in c^{\text{static}}_{i,i+1}$, where $\hat{y}_0$ is an unimportant overall shift. Note that the spacetime still takes the form \eqref{unitbhmetric}, a black hole metric with unit mass, since $\hat{y}$ takes values in $(0,\alpha)$ for some value $\alpha$. The correct mass is obtained by rescaling the angular coordinate to the standard range $(0,2\pi)$.\\
\linebreak
Since each final wedge is specified by two parameters $\zeta_{i,i+1}$ and $\xi_{i,i+1}$ (analogous to the $\zeta_{i,i+1}$ and $p_{i,i+1}\nu_{i,i+1}$ in the conical singularity computation), we will define continuous interpolating functions $\hat Z$ and $\hat X$ corresponding to these quantities. By then taking the limit in equation \eqref{tanhzcothz}, we obtain
\begin{equation}
\coth\hat Z=\frac{\tan T\sinh Z}{\tan T\cosh Z\cos\Phi-\sin\Phi},
\end{equation}
and from \eqref{Gammapmui} and \eqref{gamma} we obtain
\begin{equation}
\tanh\hat X\sinh \hat Z=\frac{\tan\Phi-\tan T\cosh Z}{1+\tan T\cosh Z\tan \Phi}.\label{tanhXsinhhatZ}
\end{equation}
From this the following useful relation can be derived (see Appendix \ref{usefulrel}):
\begin{equation}
\cosh\hat X=\cos\Phi\cos T+\sin\Phi\sin T\cosh Z.\label{coshX}
\end{equation}
From the above three relations it then follows that
\begin{equation}
\sinh\hat X\sinh\hat Z=-\sin T\cosh Z\cos\Phi+\sin\Phi\cos T,\label{sinhhatXsinhhatZ}
\end{equation}
\begin{equation}
\sinh\hat X\cosh\hat Z=-\sin T\sinh Z.\label{sinhhatXcoshhatZ}
\end{equation}
From \eqref{Gammapmui} we have
\begin{equation}
\tan\Gamma^{i,i+1}_\pm=\mp\frac{\sinh\hat Z}{\cosh^2\hat X}\mu_{i,i+1}+O(\frac{1}{n^2}).\label{gammadiffbh}
\end{equation}
The angular coordinate is now given by $\hat y=\hat y_0+\sum 2\mu_{i,i+1}+O(1/N)$. By using \eqref{gammadiff}, it can be shown (see Appendix \ref{anglemap}) that in the limit we obtain
\begin{equation}
\hat y=\hat{y}_0-\int_0^\phi\frac{\sinh \hat X}{\sin \Phi}\left(\cos T-\frac{\sin T}{\sinh\cZ}\partial_\phi \cZ\right)d\phi.\label{bh_anglemap}
\end{equation}
We will now obtain the relation between the radial coordinates as well as the embedding of the shell. This is a bit more involved than obtaining $\hat{y}$, and requires a more thorough investigation of the two mappings involved to bring us to the static wedges. This is the same set of coordinate transformations that can be induced by going from \eqref{bhwedge3}, to \eqref{bhwedge2} and then to a static circle sector in the spacetime with metric \eqref{unitbhmetric} as explained in Section \ref{bhsec}. We will thus first make a coordinate transformation to bring the final wedges to the form \eqref{bhwedge2}, namely given by
\begin{equation}
\sin\pc\phi\tanh\pc\chi=\mp\sin \pc t\tanh(\mu_{i,i+1}\pm\xi_{i,i+1}).\label{muwedge}
\end{equation}
This is done by using a transformation of the form \eqref{boosteqs} with $\zeta=-\zeta_{i,i+1}$ and $\psi=\phi_{i,i+1}$, as well as a convenient rotation to bring the spacelike geodesic to the angle $\pc\phi=0$. We will now bring the wedges to circle sectors in a black hole background with unit mass. However, since we will only be interested in the spacetime outside the horizon, we will use the coordinates $(\sigma,\beta,y)$ with metric \eqref{betabhcoord}. Here,  $\sigma$ is a time coordinate, $\beta$ is the radial coordinate, $y$ is an angular coordinate, and the circle sector has opening angle $2\mu_{i,i+1}$. Note that this only covers the spacetime {\it outside} the horizon which is located at $\beta=0$. As explained in Section \ref{bhsec}, this metric can be obtained by the embedding
\begin{equation}
\begin{array}{cc}
x^0=-\cosh\beta\cosh y, & x^2=\cosh\beta\sinh y,\\
x^1=\sinh \beta \cosh\sigma, & x^3=\sinh \beta \sinh \sigma, \label{bhcoord}
\end{array}
\end{equation}
and we will compute the coordinate transformation by comparing this embedding with \eqref{embedding_eq}.\\
\linebreak
To determine the embedding of the shell in the patch outside the shell, we will have to see how the massive geodesics, where the particles move, transform through the two coordinate transformations. The massive geodesics are given by $\tanh\chi=-\tanh\zeta_i\sin t$, and are first mapped to a geodesic on the form $\tanh\pc\chi=-\tanh\pc\zeta_i\sin\pc t$. Just as in the conical singularity situation, we can obtain a relation between these two from \eqref{boosteqs}. This results in
\begin{equation}
\sinh\pc\chi=\sinh\chi\tanh\pc\zeta_i\left(\frac{\cosh\zeta_{i,i+1}}{\tanh\zeta_i}-\sinh\zeta_{i,i+1}\cos(\phi_{i,i+1}-\psi_i)\right).
\end{equation}
In the limit, where we replace $\zeta_i$ by $Z$, $\pc\zeta_i$ by $\pc Z$ and $\zeta_{i,i+1}$ by $\hat Z$, and by using $\sinh\chi=-\sinh Z\sin \tau$ and $\sinh\pc\chi=-\sinh\pc Z\sin \tau$, this becomes
\begin{equation}
\cosh\pc Z=\cosh Z\cosh\hat Z-\sinh Z\sinh\hat Z\cos\Phi.\label{bhcoshZp}
\end{equation}
We also note that the inverse transformation can be used to obtain
\begin{equation}
\cosh Z=\cosh \pc Z\cosh\hat Z+\sinh\pc Z\sinh\hat Z\cos\pc\Phi,\label{bhcoshZ}
\end{equation}
where $\pc\Phi$ is the continuous version of the angular location of the geodesic in the $(\pc t,\pc\chi,\pc\phi)$ coordinates. By now using \eqref{coshX}, \eqref{sinhhatXsinhhatZ} and \eqref{sinhhatXcoshhatZ} we obtain the simple relation
\begin{equation}
\frac{\cosh \pc Z}{\sinh Z}=-\sin\Phi\coth\hat X.\label{bhzzp}
\end{equation}
However, we are interested in the relation between the radial coordinate $\beta$ in the $(\sigma,\beta,y)$ coordinate system, thus we will have to compare the  embedding \eqref{bhcoord} with \eqref{adscoord}. Note first, that equation \eqref{muwedge}, after taking the limit, together with $\tanh\pc\chi=-\tanh \pc Z\sin\pc t$, implies the following relation between $\pc Z$ and $\hat X$
\begin{equation}
\sin\pc\Phi=\frac{\tanh\hat X}{\tanh \pc Z}.
\end{equation}
Now note that the relation $\tanh\pc\chi=-\tanh \pc Z\sin \pc t$, which defines the trajectory of the geodesics, is equivalent to $x_1^2+x_2^2=\tanh^2\pc Z x_0^2$ by comparing with the embedding \eqref{adscoord}. By using $x_0=\cosh\beta\cosh y$ and $x_2=\cosh\beta\sinh y$, as well as $x_1=\sinh\pc\chi\cos\pc\phi$ and $x_2=\sinh\pc\chi\sin\pc\phi$, we obtain for the trajectory of the particles in the $(\sigma,\beta,y)$ coordinates that
\begin{align}
\tanh^2\pc Z\cosh^2\beta&=x_1^2+\frac{x_2^2}{\cosh^2\pc Z}=(\cos^2\pc\Phi+\frac{\sin^2\pc\Phi}{\cosh^2\pc Z})\sinh^2\pc\chi\nonumber\\
&=\frac{\sinh^2\pc\chi}{\cosh^2\hat X}=\frac{\sinh^2\chi\sinh^2\pc Z}{\sinh^2Z\cosh^2\hat X}\nonumber\\
&\Rightarrow \cosh\beta=-\frac{\sin\Phi}{\sinh\hat X}\sinh\chi,\label{betamap}
\end{align}
where we also used \eqref{bhzzp} in the last equality. Thus if we define the ``boost parameter'' $B$ such that $\cosh\beta=-\cosh B\sin\tau$, which specifies the embedding of the shell in the outside patch, we obtain the relation
\begin{equation}
\frac{\cosh B}{\sinh Z}=-\frac{\sin\Phi}{\sinh\hat X}.\label{BZ}
\end{equation}
which is similar to what was obtained in the conical singularity case. We also want to know how the time coordinate $\sigma$ is related to $\beta$ on the geodesic. By again using \eqref{bhcoord} in the relation $x_1^2+x_2^2=\tanh^2\pc Z x_0^2$, it is easy to show that
\begin{equation}
\coth\beta=\coth B\cosh\sigma.
\end{equation}
Thus we see for instance, that the maximal distance $\beta=B$ corresponds to $\sigma=0$, and that $\sigma\rightarrow\infty$ when $\beta$ approaches the horizon which is located at $\beta=0$. This is expected to happen; shells collapsing to a black hole, in ``Schwarzschild like`` coordinates (which we are using here), will not cross the horizon and these coordinates do not describe the whole spacetime outside the shell (only the part that is outside the horizon). From this we can also obtain the relation between $\sigma$ and the proper time $\tau$, which is
\begin{equation}
\tanh\sigma=-\frac{\cot\tau}{\sinh B},
\end{equation}
from which we also see that $\sigma\rightarrow\infty$ already at some finite value of $\tau<0$, as expected.\\
\linebreak
The induced metric can now also be derived (see Appendix \ref{app_ind_metric}). The induced metric seen from the coordinate patch inside the shell will be the same as computed in the conical singularity case, namely given by
\begin{equation}
ds^2=-d\tau^2+\sin^2\tau(\sinh^2Z+(\partial_\phi Z)^2)d\phi^2.
\end{equation}
Seen from the patch outside the shell, the induced metric is
\begin{equation}
ds^2=-d\tau^2+\sin^2\tau(\cosh^2B+(\partial_{\hat{y}}B)^2)d\hat{y}^2.\label{indmetric_bh}
\end{equation}
\linebreak
We will now, just like in the conical singularity case, move to more conventional metrics decribing a black hole, and which are more convenient for using the junction formalism. The metric inside the shell takes the form
\begin{equation}
ds^2=-f(r)dt^2+\frac{dr^2}{f(r)}+r^2d\phi^2,
\end{equation}
while the metric outside the shell is
\begin{equation}
d\bar s^2=-\bar{f}(\bar r)d\bar t^2+\frac{d\bar r^2}{\bar{f}(\bar r)}+\bar r^2d\bar\phi^2,
\end{equation}
where we now have $M>0$, and we have again defined $f(r)=1+r^2$ and $\bar{f}(\bar r)=-M+\bar r^2$. The coordinate transformation to go to these coordinates is now
\begin{equation}
\begin{array}{ll}
r=\sinh\chi, &\quad \bar r=\sqrt{M}\cosh\beta,\\
\bar\phi=\frac{\hat{y}}{\sqrt{M}}, &\quad \bar t=\frac{\sigma}{\sqrt{M}}.\label{bartransf_bh}\\
\end{array}
\end{equation}
while $t$ and $\phi$ remain the same and $M$ is given by
\begin{equation}
M=\left(\frac{\alpha}{2\pi}\right)^2,\label{Malpha}
\end{equation}
where $(0,\alpha)$ was the range of the variable $\hat y$. The embedding of the shell is given by 
\begin{equation}
\frac{r}{\sqrt{f(r)}}=-\frac{R}{\sqrt{f(R)}}\sin t,
\end{equation}
inside, and 
\begin{equation}
\frac{\bar r}{\sqrt{\bar{f}(\bar r)}}=\frac{\bar R}{\sqrt{\bar{f}(\bar R)}} \cosh (\sqrt{M}\bar t),
\end{equation}
outside, where we have defined $\sinh Z\equiv R$ and $\sqrt{M}\cosh B\equiv\bar R$. In terms of the proper time, the embedding inside the shell is again given by
\begin{equation}
\begin{array}{ll}
r=-R\sin\tau, &\quad \tan t=\sqrt{f(R)}\tan \tau,\\
\end{array}
\end{equation}
and 
\begin{equation}
\begin{array}{ll}
\bar r=-\bar{R} \sin\tau, &\quad \coth (\sqrt{M} \bar t)=-\frac{\sqrt{\bar f(\bar R)}}{\sqrt{M}}\tan \tau ,\\
\end{array}\label{bh_barcoord}
\end{equation}
outside. Note that the embedding of the $t$ coordinate takes a different form compared to the conical singularity case, and thus we must be careful when applying the junction formalism. The induced metric is now given by
\begin{equation}
ds^2=-d\tau^2+\sin^2\tau h^2d\phi^2=-d\tau^2+\sin^2\tau\bar h^2d\bar \phi^2,
\end{equation}
where we have defined $h^2\equiv\sinh^2Z+(\partial_\phi Z)^2=R^2+\frac{(\partial_\phi R)^2}{f(R)}$ and $\bar h^2\equiv M\cosh^2B+(\partial_{\bar \phi} B)^2=\bar{R}^2+\frac{(\partial_{\bar\phi} \bar R)^2}{\bar{f}(\bar{R})}$. Again we have the consistency condition $d\bar{\phi}/d\phi=h/\bar{h}$ which says that the induced metric is the same in both patches (see Appendix \ref{app_cons_cond}). In Fig. \ref{shellfig} we show an illustration of what these thin shell spacetimes look like.
\begin{figure}
\begin{center}
\includegraphics[scale=0.8]{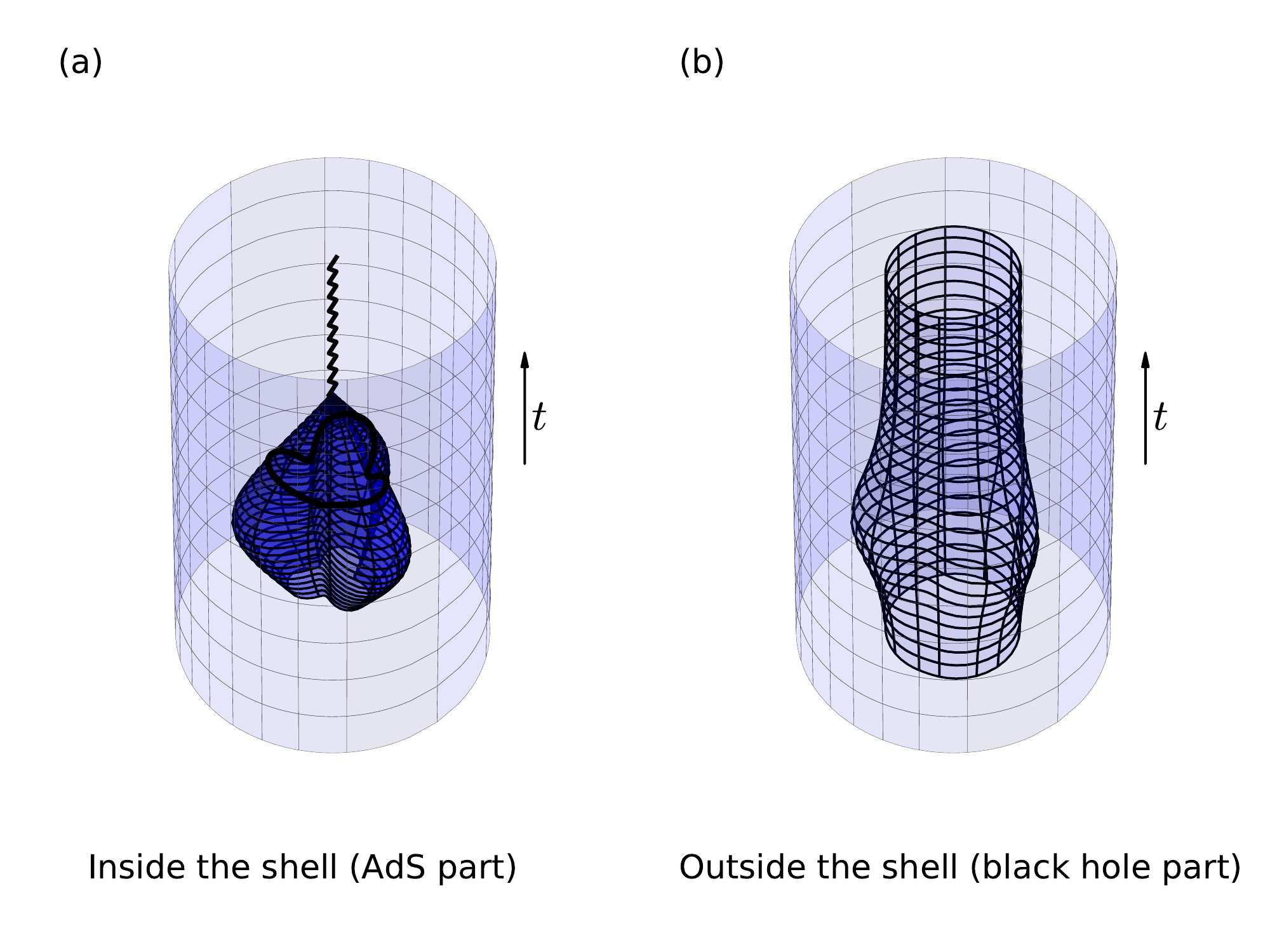}
\caption{\label{shellfig} Illustration of the massive thin shell spacetimes with an angle dependent embedding. Panel (a) shows the embedding of the shell in a spacetime with the AdS metric, as seen from the inside part of the shell. The allowed part of spacetime is inside the plotted surface, while everything else should be removed. The surface in (a) is then glued to the surface in (b) via a non-trivial coordinate transformation. The surface in (b) is embedded in a spacetime with the BTZ black hole metric, and the allowed spacetime is now the part outside the plotted surface, while the inside should be discarded. Note that in (b) we are using coordinates that only cover the region outside the horizon, which is why the shell gets stuck at the horizon for late times. However, in (a) we also cover parts of the spacetime inside the event horizon, and in particular the collision point of the shell when it forms the singularity. The intersection of the horizon with the shell is seen as the thick line in (a), which is mapped to infinite time in the spacetime in (b). When crossing the shell inside the black line one is thus mapped to the inside of the black hole outside the shell, which is not covered by the coordinates we are using. The particular form of the embedding inside the shell has been taken to be $Z(\phi)=1+\frac{1}{2}\cos(3\phi)$. To make this illustration possible, the radial coordinates have been compactified. For panel (a) it is the same compactification used in the rest of the paper (as specified after equation \eqref{adsmetric}), and for the black hole spacetime in panel (b) the radial coordinate $r$ is defined in terms of $\beta$ by $r=\tanh(s/2)$ where $\sinh s=\cosh \beta$.}
\end{center}
\end{figure}
\section{Stress-energy tensor and the junction formalism}\label{junctionsec}
In this section we will use the Israel junction formalism for timelike shells, as outlined in \cite{Israel:1966rt} and \cite{Musgrave:1995ka}, to compute the stress-energy tensor of the thin shell spacetimes that we have built from the pointlike particles.\\
\linebreak
The starting point is the two metrics
\begin{equation}
ds^2=-f(r)dt^2+\frac{dr^2}{f(r)}+r^2d\phi^2,\label{metricads}
\end{equation}
and
\begin{equation}
d\bar s^2=-\bar f(\bar r)d\bar t^2+\frac{d\bar r^2}{\bar f(\bar r)}+\bar r^2d\bar\phi^2,\label{metricbh}
\end{equation}
where the barred quantities are outside the shell and non-barred quantities inside the shell. We have also defined $f(r)=1+r^2$ and $\bar f(\bar r)=-M+\bar r^2$. For the formation of a conical singularity (black hole), we have $M<0$ ($M>0$). The embedding of the shell, as was obtained in the pointlike particle construction or can be obtained by just assuming that each point on the shell follows a radial geodesic, is given inside the shell by
\begin{equation}
\frac{r}{\sqrt{f(r)}}=-\frac{R}{\sqrt{f(R)}}\sin t,\label{trajinside}
\end{equation}
or in terms of the proper time $\tau$ by
\begin{equation}
\begin{array}{ll}
r=-R\sin\tau, &\quad \tan t=\sqrt{f(R)}\tan \tau.\\
\end{array}\label{proptimeinside}
\end{equation}
Outside the shell it is given by
\begin{equation}
\frac{\bar r}{\sqrt{\bar{f}(\bar r)}}=\frac{\bar R}{\sqrt{\bar{f}(\bar R)}} \times\Big\{\begin{array}{cc} \cosh (\sqrt{M}\bar t),& M>0 \\ \sin(\sqrt{-M}\bar t), & M<0\\ \end{array}\label{trajoutside}
\end{equation}
or in terms of the proper time $\tau$ by
\begin{equation}
\begin{array}{ll}
\bar r=-\bar{R} \sin\tau, &\quad   \Bigg\{\begin{array}{cc} \coth (\sqrt{M} \bar t)=-\frac{\sqrt{\bar{f}(\bar R)}}{\sqrt{M}}\tan \tau,& M>0 \\ \tan (\sqrt{-M} \bar t)=\frac{\sqrt{\bar{f}(\bar R)}}{\sqrt{-M}}\tan \tau, & M<0\\ \end{array}\\
\end{array}\label{proptimeoutside}
\end{equation}
outside, for two arbitrary functions $R$ and $\bar R$, which we will here use to {\it define} these spacetimes (so at this point, no reference to the other quantities that were defined or computed in the pointlike particle construction is needed). These two functions, together with how the coordinates are related when crossing the shell, completely specify the spacetime. The induced metric is
\begin{equation}
ds^2=-d\tau^2+\sin^2\tau h^2d\phi^2=-d\tau^2+\sin^2\tau\bar h^2d\bar \phi^2,\label{indmetricjunction}
\end{equation}
where $h^2=R^2+\frac{(\partial_{\phi} R)^2}{f(R)}$ and $\bar h^2=\bar{R}^2+\frac{(\partial_{\bar\phi} \bar R)^2}{\bar{f}(\bar{R})}$, and the angular coordinates when crossing the shell are related by
\begin{equation}
 \frac{d\bar\phi}{d\phi}=\frac{h}{\bar h},
\end{equation}
which must hold to make sure that the induced metric is the same from both sides (this is a necessary condition that must be satisfied for the junction formalism to be applicable, and we also showed that this follows in the pointparticle construction). It might be convenient to specify $\bar R$ as a function of $\phi$ instead as of as a function of $\bar \phi$. In that case we can obtain $\bar h$ in the following way:
\begin{align}
\bar h^2=&\bar{R}^2+\frac{(\partial_{\bar\phi} \bar R)^2}{\bar{f}(\bar{R})}=\bar{R}^2+\frac{(\partial_\phi \bar R)^2}{\bar{f}(\bar{R})}\frac{\bar h^2}{h^2}\nonumber\\
&\Rightarrow \bar h=\frac{\sqrt{\bar{f}(\bar{R})}h\bar R}{\sqrt{h^2\bar{f}(\bar R)-(\partial_{\phi} \bar R)^2}}
\end{align}
We will later in this section show that the stress-energy tensor, as computed from the junction formalism, has no pressure, and that when computing $R$ and $\bar R$ from the pointlike particle construction, the energy density coincides with the density of the pointlike particles, as expected.\\
\linebreak
By using $\phi$ as the angular coordinate in the intrinsic geometry of the shell, a basis for the tangent vectors of the shell are
\begin{equation}
\begin{array}{ll}
e^\mu_\tau=u^\mu\equiv \dot{x}^\mu=(\dot t,\dot r,0), &\quad\quad \bar e^\mu_\tau=\bar u^\mu\equiv \dot{\bar{x}}^\mu=(\dot{ \bar{ t} },\dot{ \bar{ r}},0),\\
\end{array}
\end{equation}
\begin{equation}
\begin{array}{ll}
e^\mu_\phi\equiv\frac{\partial x^\mu}{\partial \phi}=(\partial_\phi t,\partial_\phi r,1), &\quad\quad \bar e^\mu_{\phi}\equiv\frac{\partial \bar{x}^\mu}{\partial \phi}=(\partial_{\phi} \bar t,\partial_{\phi}\bar r,\frac{h}{\bar h}),\\
\end{array}
\end{equation}
where $\dot{\color{white}{t}}$ means derivative with respect to the proper time. Note that $e_\tau$ and $\bar{e}_\tau$ also coincide with the velocities of the shell. We can also use $\bar \phi$ as the angular coordinate on the shell, and in that case we have
\begin{equation}
\begin{array}{ll}
e^\mu_{\bar \phi}=(\partial_{\bar \phi} t,\partial_{\bar \phi} r,\frac{\bar h}{h})=\frac{\bar h}{h} e^\mu_{\phi}, &\quad\quad \bar e^\mu_{\bar \phi}=(\partial_{\bar \phi} \bar t,\partial_{\bar \phi} \bar r,1)=\frac{\bar h}{h} \bar e^\mu_{\phi}.\\
\end{array}
\end{equation}
From now on we will let $'$ denote derivative with respect to $\phi$. For completeness, we will list all first and second derivatives of $r$, $t$, $\bar r$ and $\bar t$ with respect to $\phi$ and $\tau$. These are obtained from equations \eqref{trajinside}-\eqref{proptimeoutside} and are as follows: 
\begin{subequations}
\begin{align}
\dot{r}=-R\cos\tau,&\quad\quad \dot{\bar{r}}=\bar{R}\cos\tau,\label{dotrdotbarr}\\
r'=-R'\sin\tau,&\quad\quad \bar{r}'=-\bar{R}'\sin\tau,\\
\dot{t}=\frac{\sqrt{f(R)}}{f(r)},&\quad\quad \dot{\bar{t}}=\frac{\sqrt{\bar{f}(\bar R)}}{\bar{f}(\bar r)},\label{dottdotbart}\\
t'=\frac{\sin\tau\cos\tau RR'}{f(r)\sqrt{f(R)}},&\quad\quad \bar{t}'=\frac{\sin\tau\cos\tau \bar{R}\bar{R}'}{\bar{f}(\bar r)\sqrt{\bar{f}(\bar{R})}},\\
\ddot{r}=R\sin\tau,&\quad\quad \ddot{\bar{r}}=\bar{R}\sin\tau,\\
r''=-R''\sin\tau,&\quad\quad \bar{r}''=-\bar{R}''\sin\tau,\\
\ddot{t}=-\frac{\sqrt{f(R)}R^2\sin(2\tau)}{f(r)^2},&\quad\quad \ddot{\bar{t}}=-\frac{\sqrt{\bar{f}(\bar{R})}\bar{R}^2\sin(2\tau)}{\bar{f}(\bar{r})^2},\\
\dot{r}'=-R'\cos\tau,&\quad\quad t''=\frac{\sin\tau\cos\tau(-(RR')^2+(RR''+(R')^2)f(R))}{f(r)\sqrt{f(R)}^3},\\
\dot{\bar{r}}'=-\bar{R}'\cos\tau,&\quad\quad \bar{t}''=\frac{\sin\tau\cos\tau(-(\bar{R}\bar{R}')^2+(\bar{R}\bar{R}''+(\bar{R}')^2)\bar{f}(\bar{R}))}{\bar{f}(\bar{r})\sqrt{\bar{f}(\bar{R})}^3},\\
\dot{t}'=\frac{RR'}{f(r)\sqrt{f(R)}},&\quad\quad \dot{\bar{t}}'=\frac{\bar{R}\bar{R}'}{\bar{f}(\bar{r})\sqrt{\bar{f}(\bar{R})}},
\end{align}
\end{subequations}
Note that these equations are only defined if $\bar{R}^2\geq M$. This will always be the case when a conical singularity forms. When a black hole forms, it only holds if $\bar{R}$, which is the maximal radial position of the shell, is outside the horizon (which we will always assume). The normal vectors $n$ and $\bar{n}$, which will be spacelike, are defined by the normalization $n\cdot n=\bar n\cdot \bar n=1$ and that they are orthogonal to the tangent vectors of the shell. We will also adopt the standard convention that the normal vectors point from the inside (the \ads part) to the outside.  These conditions results in the vectors
\begin{equation}
n_\mu=\frac{R}{\sqrt{R^2+\frac{(R')^2}{f(R)}}}\left(R\cos \tau,\frac{\sqrt{f(R)}}{f(r)},\frac{R'}{\sqrt{f(R)}}\sin\tau\right),
\end{equation}
and
\begin{equation}
\bar{n}_\mu=\sqrt{\frac{R^2-\frac{(\bar{R}')^2}{\bar{f}(\bar{R})}+\frac{(R')^2}{f(R)}}{R^2+\frac{(R')^2}{f(R)}}}\left(\bar{R}\cos \tau,\frac{\sqrt{\bar{f}(\bar{R})}}{\bar{f}(\bar{r})},\frac{\bar{R}'}{\sqrt{\bar{f}(\bar{R})}}\sin\tau\right).
\end{equation}
We will also need the extrinsic curvatures, which are defined by
\begin{equation}
K_{ij}\equiv n_\mu\left(\frac{\partial^2x^\mu}{\partial\xi^i\partial\xi^j}+\Gamma^\mu_{\nu\rho}\frac{\partial x^\nu}{\partial \xi^i}\frac{\partial x^\rho}{\partial \xi^j}\right).\label{Kij}
\end{equation}
The intrinsic stress-energy tensor $S_{ij}$ of the shell can now be computed from the junction conditions\cite{Misner:1974qy,Israel:1966rt,Musgrave:1995ka} as
\begin{equation}
[K_{ij}-\gamma_{ij}K]=8\pi G S_{ij},\label{junctioncond}
\end{equation}
where $[X]=\bar{X}-X$, $\gamma_{ij}$ is the induced metric and $K$ is the trace of the extrinsic curvature. For simplicity we will restrict to using the coordinate $\phi$ for describing the intrinsic geometry on the shell, but it is possible to translate the results as a function of $\bar{\phi}$ by using the relation between $\bar\phi$ and $\phi$. Using equation \eqref{junctioncond} and our explicit form of the induced metric, equation \eqref{indmetricjunction}, the stress-energy tensor is then expressed in terms of the extrinsic curvatures as
\begin{align}
8\pi G S_{\tau\tau}&=\frac{\bar{K}_{\phi\phi}-K_{\phi\phi}}{\sin^2\tau(R^2+\frac{(R')^2}{f(R)})},\label{Stautau}\\
8\pi G S_{\tau\phi}&=8\pi S_{\phi\tau}=\bar{K}_{\tau\phi}-K_{\tau\phi},\label{Stauphi}\\
8\pi G S_{\phi\phi}&=\sin^2\tau(R^2+\frac{(R')^2}{f(R)})(\bar{K}_{\tau\tau}-K_{\tau\tau}).\label{Sphiphi}
\end{align}
\linebreak
The Christoffel symbols computed from a metric of the form \eqref{metricads} are
\begin{align}
\begin{array}{lll}
 \Gamma^t_{tr}=\frac{f'(r)}{2f(r)},&\Gamma^r_{tt}=\frac{f'(r)f(r)}{2},&\Gamma^r_{rr}=-\frac{f'(r)}{2f(r)},\\
 \Gamma^r_{\phi\phi}=-rf(r),&\Gamma^\phi_{r\phi}=\frac{1}{r},&
\end{array}
\end{align}
and with analogous expressions outside the shell. From this it is straightforward to compute the extrinsic curvatures, and we obtain that
\begin{equation}
K_{\tau\phi}=\bar{K}_{\tau\phi}=K_{\tau\tau}=\bar{K}_{\tau\tau}=0,
\end{equation}
which implies that the only non-zero component of the stress-energy tensor is $S_{\tau\tau}$ (the energy density). In other words, the stress-energy tensor is diagonal (no momentum flux in the angular direction) and has no pressure. This is consistent with the interpretation of the shell as composed of pointlike particles falling in radially. The only non-trivial extrinsic curvatures are $K_{\phi\phi}$ and $\bar{K}_{\phi\phi}$, which are given by
\begin{equation}
\frac{K_{\phi\phi}}{\sin \tau}\equiv\frac{R^2\sqrt{f}-R \left(\frac{R''}{\sqrt{f}}-\frac{R}{\sqrt{f}^3}(R')^2\right)+2\frac{(R')^2}{\sqrt{f}}}{(R^2+\frac{(R')^2}{f})^\frac{1}{2}},\label{Kphiphi}
\end{equation}
and 
\begin{align}
\frac{\bar{K}_{\phi\phi}}{\sin\tau}\equiv&\Bigg[\left(R^2+\frac{(R')^2}{f}\right)^2\frac{\sqrt{\bar{f}}}{\bar R}-\left(R^2+\frac{(R')^2}{f}\right)\frac{(\bar R')^2}{\sqrt{\bar f}\bar{R}}+RR' \frac{\bar R'}{\sqrt{\bar f}}\nonumber\\
&-(R^2+\frac{(R')^2}{f})\left(\frac{\bar R''}{\sqrt{\bar f}}-\frac{\bar R}{\sqrt{\bar f}^3}(\bar R')^2\right)+\left(\frac{R''}{\sqrt{f}}-\frac{R}{\sqrt{f}^3}(R')^2\right)\frac{R'}{\sqrt{f}}\frac{\bar R'}{\sqrt{\bar f}}\Bigg]\nonumber\\
&\times \frac{1}{(R^2+\frac{(R')^2}{f})^\frac{1}{2}(R^2+\frac{(R')^2}{f}-\frac{(\bar R')^2}{\bar f})^{\frac{1}{2}}}.\label{barKphiphi}
\end{align}
Here we are using a shorthand notation where $f=f(R)$ and $\bar{f}=\bar{f}(\bar R)$. We have written them in terms of $\bar R$ and $R$ to be consistent with the rest of the notation in this section, but it should be pointed out the the expressions take a simpler form when written in terms of the quantities $Z$ and $\bar Z$ or $B$ which are used in Section \ref{limitsec}. An explicit expression where we have made the substitution $R=\sinh Z$ can be found in Appendix \ref{app_shellT}. The calculation of these curvatures are a bit lengthy but in principle straightforward, and can be done using a symbolic manipulation software. The only non-zero componenet of the stress-energy tensor, $S_{\tau\tau}$, is then obtained from \eqref{Stautau}.\\
\linebreak
%
%
So far we have not assumed that this spacetime is composed of pointlike particles, but instead we took the metrics and the embedding of the shell, given by the functions $R$ and $\bar R$, as the definition. If we thus want a spacetime with a particular energy density and particular starting position of the shell (which are the two physically reasonable quantities to parametrize such a spacetime with), we would have to tune $\bar{R}$ such that $S_{\tau\tau}$ equals to a particular desired profile as a function of $\phi$, which would requiring solving a relatively complicated second order ordinary differential equation with periodic boundary bonditions. However, if we assume that the spacetime arises due to the point particle construction developed in this paper, $\bar{R}$ can in principle be computed in terms of $\rho$. It can be proven that (see Appendix \ref{app_shellT})
\begin{equation}
\frac{\sqrt{\bar{f}(\bar{R})}}{\bar R}=\frac{1}{R}\left(\cosh Z\cos T-\cot \Phi\sin T\right),\label{barRrel2}
\end{equation}
where $T$ and $\Phi$ are the quantities introduced in Section \ref{limitsec}. Thus to obtain $\bar R$ in terms of $\rho$, we first solve the differential equations \eqref{odeP} and \eqref{odeT} with periodic boundary conditions to obtain $T$ and $\Phi$ in terms of $\rho$, and then use equation \eqref{barRrel2}. This looks quite involved and nothing indicates that it is possible to express $\bar R$ analytically in terms of $\rho$, but it can be proven (see Appendix \ref{app_shellT}) that the stress-energy tensor simplifies to the form
\begin{equation}
8\pi G S_{\tau\tau}^{\mathrm{point-particles}}=-\frac{\rho}{\sin\tau\sqrt{R^2+\frac{(R')^2}{f(R)}}}.\label{Stautaupp}
\end{equation}
This result makes sense intuitively, since the denominator is just the length element on the shell, and this thus represents the rest mass per unit length on the shell. However, we need to be careful with such naive interpretations, since to really know what we are talking about we must understand how $S_{ij}$ is related to the stress-energy tensor that shows up in the right hand side in Einsteins equations. We can then compare the result to the stress-energy tensor of a pointlike particle, equation \eqref{Tpointparticle}, which we will do this in Section \ref{ppcompsec}. Note also that the result is positive, since $\tau<0$. This point-particle limit can be used as a useful tool for finding thinshell solutions with a particular energy profile, which seems to be easier than solving the differential equation for $\bar R$ arising from  \eqref{Stautau} directly. We also want to stress that even if $\rho$ is more relevant when specifying initial data, knowing $\bar R$ is still crucial if we want to understand how the spacetime after the shell is linked to the spacetime before the shell and exactly how the shell is embedded.\\
\linebreak
Now the Einstein equations look like
\begin{equation}
R^{\mu\nu}-\frac{1}{2}Rg_{\mu\nu}+\Lambda g_{\mu\nu}=8\pi G T^{\mu\nu}=8\pi G\alpha \delta(F(x^\mu))S^{\mu\nu},
\end{equation}
where $F=0$ determines the embedding of the shell and $\alpha$ is a function to be determined. The tensor $S^{\mu\nu}$ is defined by the relation
\begin{equation}
S^{\mu\nu}=e_i^\mu e_j^\nu S^{ij}.
\end{equation}
This ensures that $S_{ij}$ is the induced tensor of $S_{\mu\nu}$ on the shell and that $S_{\mu\nu}n^\mu=0$. $\alpha$ should be determined such that $\alpha\int \delta(F)dn=1$, meaning that $\alpha\delta(F)$ has the standard delta-function normalization when we integrate along the direction which is normal to the surface. By expanding $F$ in the normal direction around the shell, we thus see that $\alpha=|\partial_\mu F n^\mu|$. $F$ will be chosen as
\begin{equation}
F=t+\arcsin\left(\frac{r\sqrt{1+R^2}}{R\sqrt{1+r^2}}\right),\label{Feq}
\end{equation}
by virtue of equation \eqref{trajinside}. This has the advantage that, in the lightlike limit where $R\rightarrow\infty$, we have $F\rightarrow t+\arctan(r)=v$, where $v$ is the standard infalling coordinate that is commonly used when describing \ads-Vaidya type spacetimes. Note that $\partial_\mu F\propto n_\mu$, and a straightforward computation shows that 
\begin{equation}
\alpha=\frac{\sqrt{R^2+\frac{(R')^2}{f(R)}}}{R^2\cos \tau}.\label{alphaeq}
\end{equation}
Since $S^{\tau\tau}$ is the only non-zero component, the stress-energy tensor in the $(t,r,\phi)$ coordinates is given by $S^{\mu\nu}=e_\tau^\mu e_\tau^\nu S^{\tau\tau}$. By using the relations \eqref{proptimeinside}, \eqref{dotrdotbarr} and \eqref{dottdotbart} we obtain
\begin{equation}
\begin{array}{cc}
 8\pi GT^{tt}&=\alpha \frac{f(R)}{f(r)^2}\delta(F(x^\mu))S^{\tau\tau},\\
8\pi G T^{rt}&=-\alpha \frac{R\cos\tau\sqrt{f(R)}}{f(r)} \delta(F(x^\mu))S^{\tau\tau},\\
 8\pi GT^{rr}&=\alpha R^2\cos^2\tau\delta(F(x^\mu))S^{\tau\tau}.\\\label{Tthinshell}
\end{array}
\end{equation}
If we use the relation \eqref{Stautaupp} to relate $S^{\tau\tau}$ to $\rho$, we obtain
\begin{equation}
\begin{array}{cc}
 8\pi GT^{tt}_{\mathrm{point-particles}}&=\rho\frac{f(R)}{\sqrt{f(r)}^5R r\cos t}\delta(F(x^\mu)),\\
 8\pi GT^{rt}_{\mathrm{point-particles}}&=-\rho \frac{\sqrt{f(R)}}{f(r)r} \delta(F(x^\mu)),\\
 8\pi GT^{rr}_{\mathrm{point-particles}}&=\rho \frac{\cos t\sqrt{f(r)} R}{r}\delta(F(x^\mu)),\\\label{Tthinshell2}
\end{array}
\end{equation}
where we also used the relations $\cos\tau=\sqrt{f(r)}\cos t$ and $r=-R\sin\tau$. \\
\subsection{Comparing to the stress-energy tensor of a pointlike particle}\label{ppcompsec}
We could also compute this stress-energy tensor by adding up the stress-energy tensor contributions of an infinite number of pointlike particles, by using the expression \eqref{Tpointparticle}. To build a thin shell, we would again thus place $N$ particles with such a stress-energy tensor at angles $\psi_i=2\pi i/N$, with masses and boost parameter given by $8\pi G m=\rho(\psi_i) d\phi$ and $\zeta_i=Z(\psi_i)$, where $d\phi=2\pi/N$. To translate to the quantities used in this section, we use the relations $\sinh\chi=r$, $\cosh\chi=\sqrt{f(r)}$, $\sinh Z=R$ and $\cosh Z=\sqrt{f(R)}$ to eliminate $Z$ and $\chi$. To transform the delta function in \eqref{Tpointparticle} to the one used here, we use
\begin{equation}
\delta(\tanh\chi+\tanh \zeta_i\sin t)=\frac{\sqrt{1+R^2}\delta(F(x^\mu))}{R\cos t}.
\end{equation}
We thus obtain that such a particle at angle $\phi=\psi_i$, with mass $m=\rho d\phi/8\pi G$, has stress-energy tensor
\begin{equation}
\begin{array}{cc}
8\pi G T^{tt}_i&=\rho d\phi\delta(\phi-\psi_i)\delta(F(x^\mu))\frac{f(R)}{\sqrt{f(r)}^5R r\cos t},\\
 8\pi GT^{\chi t}_i&=-\rho d\phi\delta(\phi-\psi_i) \delta(F(x^\mu))\frac{\sqrt{f(R)}}{\sqrt{f(r)}^3r},\\
 8\pi GT^{\chi\chi}_i&=\rho d\phi\delta(\phi-\psi_i)\delta(F(x^\mu))\frac{\cos t R}{r\sqrt{f(r)}},\\\label{Tthinshellpp}
\end{array}
\end{equation}
When taking the limit, we have $\sum_id\phi\delta(\psi_i-\phi)\rightarrow1$, and after transforming the tensor indices by using $\partial r/\partial\chi=\sqrt{f(r)}$, we again obtain \eqref{Tthinshell2}. This is a very non-trivial consistency check on the computations carried out in this paper.

%

\section{The massless limit revisited}\label{masslesssec}
In this section we will take the massless limit and in particular recover some of the results in \cite{Lindgren:2015fum} for massless shells. Note that, although the limit will result in a lightlike shell that is bouncing off the boundary, it is trivial to modify it such that the shell is instead created at the boundary. This is done by just removing the part of the shells worldsheet that exists before it hits the boundary and we will always assume that we are considering this modification. From an AdS/CFT point of view, it can be argued that these shells are more interesting than the bouncing shells (or the massive shells considered in this paper), since they correspond to an instantaneous energy injection in the boundary field theory. The interpretation in AdS/CFT of a massive shell in the bulk is less clear.\\
\linebreak
In the massless limit, where $R,\bar{R}\rightarrow\infty$, it is convenient to introduce the infalling coordinate $v=t+\arctan(r)$ instead of $t$, and the spacetime then takes the form
\begin{align}
ds^2=-fdv^2+2dvdr+r^2d\phi^2,&\hspace{10pt} v<0,\nonumber\\
ds^2=-\bar{f}dv^2+2dvd\bar{r}+\bar{r}^2d\bar{\phi}^2,&\hspace{10pt} v>0,
\end{align}
where again $f=1+r^2$ and $\bar{f}=-M+\bar{r}^2$. This would be the standard thin shell \ads-Vaidya spacetime, except that we still have a non-trivial coordinate transformation when crossing the shell. Specifically, when crossing the shell at $v=0$, the coordinates $(r,\phi)$ and $(\bar{r},\bar{\phi})$ are in the limit related by
\begin{equation}
\frac{R}{\bar R}=\frac{r}{\bar r}= H'(\phi),\quad \bar{\phi}(\phi)= H(\phi),\label{RoverbarR}
\end{equation}
for some function $H(\phi)$, and we use $'$ to denote derivative with respect to $\phi$. This relation follows from \eqref{cs_anglemap} and \eqref{chimap} in the conical singularity derivation, and \eqref{bh_anglemap} and \eqref{betamap} in the black hole derivation, and ensures that the induced metric is well defined. $H$ is thus defined as the mapping between $\bar{\phi}$ and $\phi$ when crossing the shell.

\subsection{Stress-energy tensor in the massless limit}
We will in this section compute the stress energy tensor in the massless limit. As was also explained in Section \ref{junctionsec}, the stress-energy tensor that shows up in the Einstein equations takes the form
\begin{equation}
8\pi G T^{\mu\nu}=8\pi G\alpha \delta(F(x^\mu))e^\mu_i e^\nu_j S^{ij}.\label{Tmunu}
\end{equation}
From \eqref{alphaeq} and \eqref{Feq} it is clear that in the limit $R\rightarrow\infty$, we have $\alpha=1/R+O(1/R^2)$ and $F\rightarrow v$. Since the only non-zero component of $S$ is $S^{\tau\tau}$, and $\dot v\rightarrow0$, we obtain in the limit that the only component of the stress-energy tensor $T$ is
\begin{equation}
T^{rr}=\delta(v)\lim_{R\rightarrow\infty}\frac{\dot{r}^2 S^{\tau\tau}}{R},
\end{equation}
where $S^{\tau\tau}$ is given by
\begin{equation}
8\pi G S^{\tau\tau}=\frac{\bar{K}_{\phi\phi}-K_{\phi\phi}}{\sin^2\tau(R^2+\frac{(R')^2}{f(R)})}.
\end{equation}
$K_{\phi\phi}$ and $\bar{K}_{\phi\phi}$ are given by \eqref{Kphiphi} and \eqref{barKphiphi}
Now we want to extract the leading behaviour of $S^{\tau\tau}$ when $R\rightarrow\infty$. Starting with the expressions for the extrinsic curvatures given by \eqref{Kphiphi} and \eqref{barKphiphi}, we eliminate $\tau$ via the relation $\sin\tau=-r/R$ (see equation \eqref{proptimeinside}) as well as express $\bar{R}$ and its derivatives in terms of $R$ and $H'$ using \eqref{RoverbarR}, a straightforward calculation results in
\begin{equation}
8\pi GS^{\tau\tau}=\frac{1}{Rr}\left(\frac{1}{2}(1+M(H')^2)+\frac{3}{2}\left(\frac{H''}{H'}\right)^2-\frac{H'''}{H'}\right)+O(\frac{1}{R^2}).
\end{equation}
Note that the leading behaviour in $\bar{K}_{\phi\phi}$ and $K_{\phi\phi}$ will cancel out, and thus it is the first subleading quantities that are relevant. Now, since $\dot{r}=-R\cos\tau$, the stress-energy tensor in the Einstein equations, given by \eqref{Tmunu}, reads
\begin{align}
8\pi GT^{rr}=&\delta(v)\frac{1}{r}\left(\frac{1}{2}(1+M(H')^2)+\frac{3}{2}\left(\frac{H''}{H'}\right)^2-\frac{H'''}{H'}\right)\nonumber\\
=&\delta(v)\frac{1}{r}\left(\frac{1}{2}(1+M(H')^2)-\{H,\phi\}\right).\label{Trrmassless}
\end{align}
where $\{F(x),x\}$ denotes the Schwarzian derivative. This is the same result as what was found in \cite{Lindgren:2015fum} using the junction formalism for lightlike shells directly\footnote{Note that the result in \cite{Lindgren:2015fum} is written in terms of the function $G\equiv1/H'$, such that $\bar{r}=Gr$.}.\\




\subsection{CFT stress-energy tensor and continuous coordinate systems}\label{contcoord}
In the formulation of the thin shell spacetimes that we have presented so far, where the coordinates are discontinuous across the shell, lightlike shells that are created at the boundary will thus also have discontinuous boundary coordinates. Inspired by this, we will construct a new coordinate system that is continuous at the boundary. This is achieved by applying a {\it large} diffeomorphism on the final spacetime, which changes the physical state and the geometry is no longer interpreted as a pure BTZ black hole. Instead the spacetime after the shell will manifestly break rotational symmetry, and the breaking of rotational symmetry is then no longer purely encoded in the discontinuities across the shell. We will only compute the first few terms of this diffeomorphism, in an expansion close to the boundary, since obtaining the full coordinate transformation is quite difficult. This is enough to compute the energy density modes (that are dual to the stress-energy tensor in the dual CFT) in the resulting spacetime. Such a formulation is more natural from an AdS/CFT point of view and much more suitable when computing AdS/CFT observables, which we plan to investigate in future work. This is a new result, that was not obtained in \cite{Lindgren:2015fum}. It is also very likely that such ``natural coordinates'' also exist for the massive shells. This is however expected to be more difficult since we do not have the discontinuity of the coordinates at the boundary to guide us, and we will not pursue this venue in this paper.\\
\linebreak
To be able to read off the stress-energy tensor modes easily, we will try to find a coordinate transformation that makes the coordinates continuous at the boundary as well as brings the metric for $v>0$ to the form
\begin{equation}
ds^2=dR^2+T_+(y_+)dy_{+}^2+T_-(y_-)dy_{-}^2-(e^{2\rho}+T_+(y_+)T_-(y_-)e^{-2\rho})dy_-dy_+.\label{TTmetric}
\end{equation}
As was pointed out in e.g. \cite{Banados:1998gg}, all metrics of the form \eqref{TTmetric} solve Einsteins equations for arbitrary $T_{\pm}$. According to the standard $AdS_3/CFT_2$ correspondence, the functions $T_\pm/8\pi G$ are identified with the stress-energy tensor modes of the dual CFT (see for instance \cite{Kraus:2006wn}).
Let us first try to get some intuition of what we expect from such a coordinate transformation. The relations between the radial coordinates and angular coordinates when crossing the shell are $\bar r=r/H'(\phi)$ and $\bar\phi=H(\phi)$, so this must be the boundary condition of the coordinate transformation on the shell $v=0$ when $\bar r\rightarrow\infty$. In other words, if we would try to define new coordinates $\tilde r$ and $\tilde \phi$ that makes the coordinates continuous at the boundary, they must satisfy $\bar\phi=H(\tilde\phi)$ and $\bar r=\tilde r/H'(\tilde\phi)$ when $\bar r\rightarrow\infty$ and $v=0$. Now if we require a flat boundary metric for all times $v>0$, it can be easily shown that this will imply that the coordinate transformation takes the form
\begin{equation}
\bar\phi\pm v=H(\tilde\phi\pm v),\label{phibc}
\end{equation}
\begin{equation}
 \bar r=\tilde r/\sqrt{H'(v+\bar{\phi})H'(v-\bar{\phi})}, \label{rbc}
\end{equation}
when $\bar r\rightarrow\infty$. This will thus be our boundary conditions on our coordinate transformation for all times $v>0$.\\
\linebreak
To construct the full coordinate transformation that brings the metric to the form \eqref{TTmetric}, we will first do a change of coordinates such that the metric for $v>0$ takes the form
\begin{equation}
ds^2=d\rho_1^2+\frac{M}{4}(dx_{+}^2+dx_{-}^2)-(e^{2\rho_1}+\frac{M^2}{16}e^{-2\rho_1})dx_-dx_+.
\end{equation}
As can be easily verified, this is achieved by the following coordinate transformation.
\begin{equation}
\bar t=v+\frac{1}{\sqrt{M}}\text{arccoth}\left(\frac{1}{\sqrt{M}}e^{\rho_1}+\frac{\sqrt{M}}{4}e^{-\rho_1}\right),\hspace{10pt}\bar{r}=e^{\rho_1}+\frac{M}{4}e^{-\rho_1},
\end{equation}
and where $x_\pm=\bar t\pm\bar{\phi}$. For the region inside the horizon, arccoth is replaced by arctanh, and for the case of formation of a conical singularity, we replace it by arctan and $\sqrt{M}$ by $\sqrt{-M}$. This step is partially just going from our infalling coordinates back to the global time coordinates used in the rest of the paper. Now we will construct the rest of the coordinate transformation enforcing the boundary conditions \eqref{phibc} and \eqref{rbc} that brings the metric to the form \eqref{TTmetric}. By expanding the coordinate transformation in an expansion close to the boundary, we obtain
\begin{align}
x_\pm=&H(y_\pm)+e^{-2\rho}\frac{H''(y_\mp)H'(y_\pm)}{2H'(y_\mp)}+e^{-4\rho}\left[\frac{(H''(y_\mp))^2H''(y_\pm)}{8(H'(y_\mp))^2}+M\frac{(H'(y_\mp))^2H''(y_\pm)}{8}\right]+O(e^{-6\rho}),\nonumber\\
e^{\rho_1}=&\frac{e^\rho}{\sqrt{H'(y_-)H'(y_+)}}-e^{-\rho}\frac{H''(y_-)H''(y_+)}{4\sqrt{H'(y_-)H'(y_+)}^3}+O(e^{-3\rho}),\label{coordtransf}
\end{align}
which is such that the metric takes the form
\begin{equation}
ds^2=d\rho^2+T(y_+)dy_{+}^2+T(y_-)dy_{-}^2-(e^{2\rho}+T(y_+)T(y_-)e^{-2\rho})dy_-dy_+,\label{TTmetric2}
\end{equation}
which is a special case of \eqref{TTmetric} where $T_+(x)=T_-(x)\equiv T(x)$. Of course, we only obtain \eqref{TTmetric2} up to higher orders in $e^{-\rho}$, but since we know that \eqref{TTmetric2} is a solution of Einsteins equations, we also know that the full coordinate transformation exists and can in principle be computed to any order. Finite forms of such coordinate transformations have been obtained in the literature (see for instance \cite{Barnich:2016lyg} and references therein), but for our purposes this is enough, since we are mainly interested in determining $T_\pm$ (however, to determine the full shape of the shell in the new coordinates as well as the map between the coordinates when crossing the shell, knowledge of the full transformation is required). The coordinate transformation \eqref{coordtransf} now implies that
\begin{equation}
T(x)=\frac{M}{4}(H'(x))^2-\frac{H'''(x)}{2H'(x)}+\frac{3}{4}\frac{(H''(x))^2}{(H'(x))^2}=\frac{M}{4}(H'(x))^2-\frac{1}{2}\{H,x\},
\end{equation}
which looks very similar to the result we obtained for the stress-energy tensor of the shell, equation \eqref{Trrmassless}. To be more precise, let us denote the stress-energy tensor modes in the dual CFT by $8\pi GT^{CFT}_\pm(y_\pm)=T(y_\pm)$ and write $8\pi G T^{rr}_{\text{shell}}=\delta(v)\rho_{\text{shell}}(\phi)/r$. By comparing with \eqref{Trrmassless}, we thus obtain
\begin{equation}
T_\pm^{CFT}(\phi)=\frac{1}{2}\left(\rho_\text{shell}(\phi)+M_{AdS}\right)
\end{equation}
The interpretation is thus as follows: The shell has a certain angular dependent energy density, directly determined by the angular dependence of the energy in the instantaneous quench in the dual CFT that sourced the shell. This energy density then directly determines the stress-energy tensor modes of the CFT after the quench. To be more precise, the energy density of the CFT directly after the instantaneous quench is equal to the energy distribution of the shell plus the energy density of \ads. The energy density in the CFT will then attain a time dependence by splitting up into the right- and left-moving modes $T^{CFT}_+(y_+)$ and $T^{CFT}_-(y_-)$. This is similar to what was obtained in \cite{Bhaseen:2013ypa,Lucas:2015hnv} for a related setup. In higher dimensions we expect similar behaviour, but the two left- and right-moving modes should then also be subjected to dissipative effects. It is not clear how this analysis would work for the massive shells, and we will leave that as an interesting open problem.

\section{Algorithm for solving \eqref{peq} and \eqref{nueq}}\label{numericssec}
In this section we will explain how to solve the discrete equations \eqref{peq} and \eqref{nueq} in practice. Let us denote $\Phi_i\equiv\phi_{i,i+1}-\psi_i$ and $P_i\equiv p_i\nu_i$. The goal is thus to obtain solutions of \eqref{peq} and \eqref{nueq} for $P_i$ and $\Phi_i$ with periodic boundary conditions (meaning $P_{N+1}=P_1$ and $\Phi_{N+1}=\Phi_1$). The equations \eqref{peq} and \eqref{nueq} can be reformulated as
\begin{equation}
P_{i+1}=\nu_{i+1}-\arctan\left(\frac{\sin(\Phi_i+\psi_i-\psi_{i+1})}{\frac{\sinh\zeta_{i+1}\cos\Phi_i}{\tanh\zeta_i}-\frac{\sinh\zeta_{i+1}\sin\Phi_i}{\sinh\zeta_i\tan(\nu_i+P_i)}-\cos(\Phi_i+\psi_i-\psi_{i+1})\cosh\zeta_{i+1}}\right),\label{peqrec}
\end{equation}
\begin{equation}
\Phi_{i+1}=\arctan\left(\frac{\tan(\Phi_i+\psi_i-\psi_{i+1})(\tan\nu_{i+1}+\tan P_{i+1})}{\tan P_{i+1}-2\cosh\zeta_{i+1}\tan(\Phi_i+\psi_i-\psi_{i+1})\tan P_{i+1}\tan\nu_{i+1}-\tan\nu_{i+1}}\right).\label{Phieqrec}
\end{equation}
Note that now the equations are written such that they can be solved recursively: Given $\Phi_i$ and $P_i$ we can determine $P_{i+1}$, and then given $\Phi_i$ and $P_{i+1}$ we can compute $\Phi_{i+1}$. However, we have to make sure that the periodic boundary conditions $\Phi_{N+1}=\Phi_1$ and $P_{N+1}=P_1$ are satisfied. We will start by considering some special cases where the problem has reflection symmetry in some axis.

\subsection{Axis of symmetry through one of the particles}
We will first assume that there is a reflection symmetry that passes through particle 1. This implies that the wedge for this particle is symmetric, i.e. $p_1=0$, which will simplify the computation. Assuming then that we know $\Phi_1$, we can from \eqref{peqrec} compute $p_2$. Thereafter we can use \eqref{Phieqrec} to compute $\Phi_2$. The algorithm can then be continued to obtain all values of $\Phi_i$ and $P_i$. When we have moved one lap around the circle and come back to $\psi_1$ however, the periodic boundary conditions will generically not be satisfied for an arbitrary guess of $\Phi_1$. To obtain the correct value of $\Phi_1$ we thus use a Newton-Rhapson solver (or other root finding algorithms) to search for the correct value of $\Phi_1$ such that the periodic boundary condition is satisfied. Note that this does not obviously imply that the periodic boundary condition of $P_i$ is satisfied, and indeed there are sometimes spurious solutions where $\Phi_i$ is periodic but not $P_i$. Thus we must now choose the right solution for $\Phi_i$ such that $p_{N+1}=0$ and it seems that there will always exist such a solution. This is the technique used for finding the parameters for Fig. \ref{3particlessym_cs}.
\subsection{Axis of symmetry between two particles}
The other possible case one can consider that has an axis of symmetry, is when the axis of symmetry lies in between two particles, taken conveniently as particles $N$ and $1$. In this case, the symmetry restriction will then instead determine the value of $\Phi_{0}\equiv\phi_{N,1}-\psi_N$. Thus we start instead by guessing the value of $p_1$. From $\Phi_0$ and $p_1$ we then compute $\Phi_2$, and then we can continue this algorithm and compute all values of $P_i$ and $\Phi_i$ recursively. We can then use a root finder to make sure that $P_i$ is continuous, but again we are not guaranteed that $\Phi_i$ is periodic. Thus we must again choose the correct solution of $P_1$ such that also $\Phi_N=\Phi_0$, and such a solution always seems to exist. This is the method used to obtain the parameters in Fig. \ref{4particles}.
\subsection{No symmetry restrictions}
In the case without any symmetry restrictions, the procedure is in practice more complicated although coneptually the same. In that case we first have to guess both $p_1$ and $\Phi_1$, recursively construct all other $p_i$ and $\Phi_i$ and then use a two-dimensional root finder to make sure that $p_1=p_{N+1}$ and $\Phi_1=\Phi_{N+1}$. This is the method used when computing the parameters for Fig. \ref{3particlesgen}.

\section{Summary and conclusions}\label{summarysec}
We have studied collisions of massive pointlike particles in three-dimensional gravity with negative cosmological constant. It was found that we can construct spacetimes with an arbitrary number of particles with any masses and at any angles, colliding to form either a new massive particle or a BTZ black hole. We also took the limit of an infinite number of particles and constructed a new class of spacetimes with a massive thin shell collapsing to a black hole. These thin shell spacetimes are very non-trivial, since both the equation for the location of the shell as well as the mass density depends on the angular coordinate. Previously only rotationally symmetric massive thin shells had been studied in the literature (non-rotationally symmetric {\it massless} shells were studied in \cite{Lindgren:2015fum} using similar methods as in this paper). We then analyzed these thin shell spacetimes using the junction formalism of general relativity, and computed the stress-energy tensor of the shell. We obtain perfect agreement with what is expected by adding up the stress-energy contribution of each individual particle, which is a very non-trivial check on the techniques developed in this paper. We also investigated the massless limit, obtaining agreement with the result in \cite{Lindgren:2015fum}. For the massless shells, we also showed, by doing a coordinate transformation that makes the coordinates continuous at the boundary, that the resulting spacetimes should not really be interpreted as a static black hole (or static conical singularity). Rather, they should be interpreted as spacetimes that have many non-trivial Virasoro modes excited, packaged into the stress-energy tensor modes $T^{CFT}_\pm$ of the dual CFT, with the only constraint that the functions $T^{CFT}_+(x)$ and $T^{CFT}_-(x)$ are equal.\\
\linebreak
The thin shell spacetimes that we constructed in this paper are pressureless, as expected since they are composed of radially falling pointlike particles. However, for the homogenous case it is easy to construct massive shells that have non-zero pressure (see for instance \cite{Erdmenger:2012xu},\cite{Keranen:2015fqa}). It would be interesting to try to build thin shells with pressure by adding up an infinite number of pointlike particles, which then would not follow radial geodesics. The deviation of the particles trajectories from radial geodesics should then induce both pressure and momentum density in the shell's stress-energy tensor. It would also be interesting to investigate if it is possible to build thick shells, by adding up an infinite number of pointlike particles that are not constrained to a single surface. Such constructions could result in non-rotationally symmetric generalizations of the general \ads-Vaidya spacetimes (i.e. not the thin shell special case) and analouges spacetimes with massive shells. We will leave these interesting, but possibly daunting, questions for future work.\\
\linebreak
Another interesting problem is what the properties of the CFT dual of these spacetimes are. The massless shells can be interpreted as corresponding to an instantaneous perturbation in the dual CFT, but the interpretation of massive shells is less understood (but see \cite{Keranen:2014zoa} for some discussions). Studying the dual CFT of these spacetimes, and computing interesting quantities using the AdS/CFT correspondence such as two-point functions and entanglement entropies (which has only been done in the rotationally symmetric case \cite{Balasubramanian:2011ur,Ziogas:2015aja}) will be the top priority for future work.

\section{Acknowledgements}
This work was supported in part by the Belgian Federal Science Policy Office through the Interuniversity Attraction Pole P7/37, by FWO-Vlaanderen through project G020714N, and by the Vrije Universiteit Brussel through the Strategic Research Program ``High-Energy Physics''. The author is supported by a PhD fellowship from the Research Foundation Flanders (FWO); his work was also partially supported by the ERC Advanced Grant ``SyDuGraM", by IISN-Belgium (convention 4.4514.08) and by the ``Communaut\'e Fran\c{c}aise de Belgique" through the ARC program. The author thanks Pujian Mao for some useful discussions.

\appendix
\section{Derivation of equations \eqref{coshatT} and \eqref{coshX}}\label{usefulrel}
We will start by deriving the relation 
\begin{equation}
\cos\hat T=\cosh\hat X=\cos\Phi\cos T+\sin\Phi\sin T\cosh Z,\label{coshatTX}
\end{equation}
where it is implicit that $\hat T$ and $\hat X$ are only defined in the case of a formation of a conical singularity or a black hole, respectively. This follows from \eqref{tanhhatZ} and \eqref{tanhatTcoshhatZ}, since we have
\begin{align}
\frac{1}{\cos^2\hat T}&=1+\tan^2\hat T=1+\frac{(\tan T\cosh Z-\tan\Phi)^2}{(1+\tan\Phi\tan T\cosh Z)^2}\frac{1}{\cosh^2 \hat Z}\\
&=1+\frac{(\tan T\cosh Z\cos\Phi-\sin\Phi)^2-\tan^2T\sinh^2Z}{(\cos\Phi+\sin\Phi\tan T\cosh Z)^2}\\
&=\frac{1+\tan^2 T}{(\cos\Phi+\sin\Phi\tan T\cosh Z)^2}\\
&=\frac{1}{(\cos\Phi\cos T+\sin\Phi\sin T\cosh Z)^2}.\\
\end{align}
Thus we can conclude that 
\begin{equation}
\cos\hat T=\pm(\cos\Phi\cos T+\sin\Phi\sin T\cosh Z).
\end{equation}
A very similar computation shows that in the case of a formation of a black hole, we have
\begin{equation}
\cosh\hat X=\pm(\cos\Phi\cos T+\sin\Phi\sin T\cosh Z).
\end{equation}
The difficult part when deriving this relation is  to determine the sign, and we will argue that we should always pick the plus sign although we will not be completely rigorous. First of all, in the homogeneous case where $\Phi=T=\hat T=\hat X=0$, we must have $\cos\hat T=\cos\hat X=1$. In the inhomogeneous case, we will only be interested in solutions that can be continuously deformed to the homogeneous case, and it is thus reasonable to expect that we should always choose the plus sign. To make this argument a bit more solid, we will now argue that $\sin\Phi$ and $\sin T$ will always have the same sign. This would mean that the right hand side of \eqref{coshatTX} is always (strictly) positive, and when continuously deforming a homogeneous solution to a non-homogeneous one, the plus sign must be kept. The only caveat that remains is then for solutions where $\Phi$ or $T$ would be very large such that $\cos \Phi$ or $\cos T$ can change sign, but we have plenty of numerical evidence that $\Phi$ or $T$ are never that large.\\
\linebreak
To see that $\Phi$ and $T$ always have the same sign, we will start by looking at the differential equations \eqref{odeP} and \eqref{odeT}. It is easy to see that we have $T(\phi_0)=0$ if and only if $\Phi(\phi_0)=0$, at some point $\phi_0$. Moreover, it can be shown that generically $T'$ and $\Phi'$ are non-zero at these points, and that we can only have $T'(\phi_0)=0$ if and only if $\Phi'(\phi_0)=0$. This indicates that $T$ and $\Phi$ flip signs at the same points. Now consider solutions that are small deviations from a homogeneous solution (so that we only keep terms linear in $\tan \Phi$ and $\tan T$). It can be easily shown that these satisfy $\tan T/\tan \Phi=\rho>0$. Thus for perturbative solutions, $\tan \Phi$ and $\tan T$ always have the same sign, and together with the fact that for non-perturbative solutions $\tan\Phi$ and $\tan T$ flip sign at the same points, one can easily convince oneself that $\Phi$ and $T$ always have the same sign.

\section{Derivation of equations \eqref{cs_anglemap} and \eqref{bh_anglemap}}\label{anglemap}
%
In this section we will prove the mapping between the angular coordinates across the shell, equations \eqref{cs_anglemap} and \eqref{bh_anglemap}. These are computed as the sum of the opening angles of the static wedges. Thus in the conical singularity case, we have
\begin{equation}
\hat\phi=\hat{\phi}_0+\sum_{0\leq i<j} 2\nu_{i,i+1}+O(1/N),
\end{equation}
when $\hat\phi$ is inside wedge $c_{j,j+1}^\text{static}$, while in the black hole case\footnote{To be able to do the conical singularity and the black hole case simultaneously, we use $\hat\phi$ to also denote the quantity that was called $\hat y$ in Section \ref{collisionsec}}
\begin{equation}
\hat\phi=\hat{\phi}_0+\sum_{0\leq i<j} 2\mu_{i,i+1},
\end{equation}
when $\hat\phi$ is inside wedge $c_{j,j+1}^\text{static}$. Let us define (see equation \eqref{gammadiff})
\begin{align}
\Delta \Gamma&\equiv\lim_{n\rightarrow\infty} \frac{1}{d\phi}\left[\tan\Gamma_+^{i,i+1}-\tan\Gamma_-^{i,i+1} \right]\nonumber\\
&=\frac{\cos^2T(\tan T\cosh\cZ)'-2\cosh\cZ\rho+\cos^2T+\sin^2T\cosh^2\cZ)}{(\cos T\cos\Phi+\sin T\cosh\cZ\sin\Phi)^2},
\end{align}
where $'$ is derivative with respect to $\phi$. By equating equation \eqref{gammadiff} with either \eqref{gammadiffcs} or \eqref{gammadiffbh} we obtain in the limit that
\begin{equation}
\hat\phi=\hat{\phi}_0+\int_0^\phi \Delta \Gamma d\phi_1\times\bigg\{ \begin{array}{ll} \frac{\cos^2\hat T}{\cosh \hat Z},&\text{Point particle}\\-\frac{\cosh^2\hat X}{\sinh \hat Z},&\text{Black hole}\\\end{array}
\end{equation}
After using \eqref{odeT} to substitute $T'$ we obtain
\begin{equation}
\Delta \Gamma=\frac{(\cos T-\frac{\sin T}{\sinh\cZ}\cZ')(\cos T-\cosh Z \sin T \cot\Phi)}{(\cos T\cos\Phi+\sin T\cosh\cZ\sin\Phi)^2\cosh \hat Z}
\end{equation}
%
%
Now using equations \eqref{coshatTX} we obtain
\begin{equation}
\hat\phi=\hat{\phi}_0+\int_0^\phi\frac{(\cos T-\frac{\sin T}{\sinh\cZ}\cZ')(\cos T-\cosh Z \sin T \cot\Phi)}{\cosh \hat Z}d\phi,
\end{equation}
in the case of formation of a conical singularity, and 
\begin{equation}
\hat\phi=\hat{\phi}_0-\int_0^\phi\frac{(\cos T-\frac{\sin T}{\sinh\cZ}\cZ')(\cos T-\cosh Z \sin T \cot\Phi)}{\sinh \hat Z}d\phi,
\end{equation}
in the case of formation of a black hole. $\hat\phi$ is still increasing despite the suspicious sign, since $\hat Z<0$. We can simplify this even further by using equations \eqref{tanhatTcoshhatZ} and \eqref{tanhXsinhhatZ} to get rid of $\hat Z$. This will result in the equations \eqref{cs_anglemap} and \eqref{bh_anglemap}.

\section{Derivation of the induced metric}\label{app_ind_metric}
In this section we will prove equations \eqref{indmetric} and \eqref{indmetric_bh}, and we will start with the conical singularity case. We will focus on the inside patch, the outside is completely analogous. Let us restate the problem: We want to compute the induced metric on a surface given by the equation
\begin{equation}
\tanh\chi=-\tanh Z(\phi)\sin t,
\end{equation}
in the spacetime with metric
\begin{equation}
ds^2=-\cosh^2\chi dt^2+d\chi^2+\sinh^2\chi d\phi^2.
\end{equation}
We will use the proper time $\tau$ and the angular coordinate $\phi$ to parametrize the geometry on this surface. We will restate some useful relations involving the proper time which are also found in Section \ref{geodesicssec}. They are
\begin{equation}
\sinh\chi=-\sinh Z\sin \tau,\label{ur1}
\end{equation}
\begin{equation}
\cosh\chi=\frac{\cos \tau}{\cos t}=\sqrt{\cos^2\tau+\sin^2\tau\cosh^2Z},\label{ur2}
\end{equation}
\begin{equation}
\tan t=\cosh Z\tan \tau.\label{ur3}
\end{equation}
The induced metric is computed as
\begin{equation}
ds^2=\gamma_{\tau\tau}d\tau^2+2\gamma_{\tau\phi}d\tau d\phi+\gamma_{\phi\phi}d\phi^2,
\end{equation}
where
\begin{equation}
\begin{array}{cc}
 \gamma_{\tau\tau}=&-\cosh^2\chi\dot{t}^2+\dot{\chi}^2\\
 \gamma_{\tau \phi}=&-\cosh^2\chi\dot{t}t'+\dot{\chi}\chi'\\
 \gamma_{\phi\phi}=&-\cosh^2\chi(t')^2+(\chi')^2+\sinh^2\chi\\
\end{array}
\end{equation}
Here $\cdot$ is derivative with respect to $\tau$ and $'$ is derivative with respect to $\phi$. From \eqref{ur1}, \eqref{ur2} and \eqref{ur3} we can derive that
\begin{equation}
\dot\chi=-\sinh Z\cos t,\label{chidot}
\end{equation}
\begin{equation}
\dot t=\frac{\cosh Z}{\cosh^2\chi},\label{tdot}
\end{equation}
\begin{equation}
\chi'=-\frac{\cosh Z\sin \tau}{\cosh\chi}Z',\label{chiprime}
\end{equation}
\begin{equation}
t'=\frac{\sinh Z\sin \tau\cos t}{\cosh\chi}Z'.\label{tprime}
\end{equation}
What we want to show is thus $\gamma_{\tau\tau}=-1$, $\gamma_{\tau\phi}=0$ and $\gamma_{\phi\phi}=\sin^2\tau(\sinh^2Z+(\partial_\phi Z)^2)$. The outside patch works analogously.\\

{\bf $\mathbf{\gamma_{\tau\tau}}$:} We already know that $\gamma_{\tau\tau}=-1$ must hold, since we know that $\tau$ is the proper time, but for completeness we will prove this explicitly here anyway. By using the relations \eqref{ur2}, \eqref{chidot} and \eqref{tdot} we obtain
\begin{align}
\gamma_{\tau\tau}&= -\cosh^2\chi\dot{t}^2+\dot{\chi}^2=-\frac{\cosh^2Z}{\cosh^2\chi}+\sinh^2Z\cos^2t\nonumber\\
&=\frac{1}{\cosh^2\chi}\left(\underbrace{-\cosh^2Z+\sinh^2Z\cos^2\tau}_{=-\cosh^2Z\sin^2\tau-\cos^2\tau}\right)\nonumber\\
&=-1.
\end{align}

{\bf $\mathbf{\gamma_{\phi\phi}}$:} It is also straightforward to compute $\gamma_{\phi\phi}$. We obtain by using \eqref{ur1}, \eqref{ur2}, \eqref{chiprime} and \eqref{tprime} that
\begin{align}
\gamma_{\phi\phi}&=-\cosh^2\chi(t')^2+(\chi')^2+\sinh^2\chi\nonumber\\
&=\sin^2\tau\left(-\sinh^2Z\cos^2t(Z')^2+\frac{\cosh^2Z(Z')^2}{\cosh^2\chi}\right)+\sinh^2\chi\nonumber\\
&=\frac{\sin^2\tau}{\cosh^2\chi}\left(-\sinh^2Z\cos^2\tau+\cosh^2Z\right)(Z')^2+\sin^2\tau\sinh^2Z\nonumber\\
&=\sin^2\tau(\sinh^2Z+(Z')^2).
\end{align}

{\bf $\mathbf{\gamma_{\tau\phi}}$:} From \eqref{chidot}-\eqref{tprime} it directly follows that $\gamma_{\tau\phi}=0$.\\
\linebreak
Now let us consider the black hole case. We thus consider a metric on the form
\begin{equation}
ds^2=-\sinh^2\beta d\sigma^2+d\beta^2+\cosh^2\beta dy^2.
\end{equation}
In this spacetime our shell is specified by a function $B(y)$ by the equation $\coth\beta=\coth B\cosh\sigma$ and we want to compute the induced metric on this shell. We use $y$ and the proper time $\tau$ as our intrinsic coordinates on the shell. The induced metric is computed as
\begin{equation}
ds^2=\gamma_{\tau\tau}d\tau^2+2\gamma_{\tau\phi}d\tau dy+\gamma_{\phi\phi}dy^2.
\end{equation}
where
\begin{equation}
\begin{array}{cc}
 \gamma_{\tau\tau}=&-\sinh^2\beta\dot{\sigma}^2+\dot{\beta}^2,\\
 \gamma_{\tau\phi}=&-\sinh^2\beta\dot{\sigma}\sigma'+\dot{\beta}\beta',\\
 \gamma_{\phi\phi}=&-\sinh^2\beta(\sigma')^2+(\beta')^2+\cosh^2\beta.
\end{array}
\end{equation}
Now $'$ means derivative with respect to $y$. The equation for the shell can be described in terms of the proper time by
\begin{equation}
\cosh\beta=-\cosh B\sin\tau,\label{ur4}
\end{equation}
\begin{equation}
\sinh \beta=\frac{\cos\tau}{\sinh\sigma}=\sqrt{\sin^2\tau\sinh^2B-\cos^2\tau},\label{ur5}
\end{equation}
\begin{equation}
\coth\sigma=-\sinh B\tan\tau.\label{ur6}
\end{equation}
The derivatives are
\begin{equation}
\dot\beta=-\cosh B\sinh \sigma,\label{betadot}
\end{equation}
\begin{equation}
\dot \sigma=\frac{\sinh B}{\sinh^2\beta},\label{sigmadot}
\end{equation}
\begin{equation}
\beta'=-\frac{\sinh B\sin \tau}{\sinh\chi}B',\label{betaprime}
\end{equation}
\begin{equation}
\sigma'=\frac{\cosh B\sin \tau\sinh \sigma}{\sinh\beta}B'.\label{sigmaprime}
\end{equation}
The computations are very similar to the conical singularity case.\\

{\bf $\mathbf{\gamma_{\tau\tau}}$:} By using the relations \eqref{ur5}, \eqref{betadot} and \eqref{sigmadot} we obtain
\begin{align}
\gamma_{\tau\tau}&= -\sinh^2\beta\dot{\sigma}^2+\dot{\beta}^2=-\frac{\sinh^2B}{\sinh^2\beta}+\cosh^2B\sinh^2\sigma\nonumber\\
&=\frac{1}{\sinh^2\beta}\left(\underbrace{-\sinh^2B+\cosh^2B\cos^2\tau}_{=-\sinh^2B\sin^2\tau+\cos^2\tau}\right)\nonumber\\
&=-1.
\end{align}

{\bf $\mathbf{\gamma_{yy}}$:} It is also straightforward to compute $\gamma_{\phi\phi}$. We obtain by using \eqref{ur4}, \eqref{ur5}, \eqref{betaprime} and \eqref{sigmaprime} that
\begin{align}
\gamma_{yy}&=-\sinh^2\beta(\sigma')^2+(\beta')^2+\cosh^2\beta\nonumber\\
&=\sin^2\tau\left(-\cosh^2B\sinh^2\sigma(B')^2+\frac{\sinh^2B(B')^2}{\sinh^2\beta}\right)+\cosh^2\beta\nonumber\\
&=\frac{\sin^2\tau}{\sinh^2\beta}\left(-\cosh^2B\cos^2\tau+\sinh^2B\right)(B')^2+\sin^2\tau\cosh^2B\nonumber\\
&=\sin^2\tau(\cosh^2B+(B')^2).
\end{align}

{\bf $\mathbf{\gamma_{\tau y}}$:} From \eqref{betadot}-\eqref{sigmaprime} it directly follows that $\gamma_{\tau y}=0$.

\section{Consistency condition on the induced metric}\label{app_cons_cond}
In this section we will prove equation \eqref{indmetric_consistency}. This equation states that the relation between $R$ and $\bar{R}$ (or equivalently between $Z$ and $\pc Z$ in the conical singularity case, and between $Z$ and $B$ in the black hole case) is consistent with how $\phi$ is related to $\bar \phi$ when crossing the shell, in the sense that the induced metric are the same in both patches. The ($d\bar \phi^2$ part of the) metric outside the shell can be written as
\begin{equation}
\bar{\gamma}_{\bar\phi\bar\phi}d\bar\phi^2=(\bar{R}^2+\frac{(\partial_{\bar\phi} R)^2}{\bar{f}(\bar R)})d\bar\phi^2=\left(\bar{R}^2\left(\frac{d\bar\phi}{d\phi}\right)^2+\frac{(\partial_{\phi} R)^2}{\bar{f}(\bar R)}\right)d\phi^2\equiv\bar{\gamma}_{\phi\phi}d\phi^2.
\end{equation}
From the inside of the shell the ($d \phi^2$ part of the) induced metric is $\gamma_{\phi\phi}d\phi^2=(\sinh^2Z+(\partial_{\phi} Z)^2)d\phi^2$, and thus what we want to show is that $\bar{\gamma}_{\phi\phi}=\gamma_{\phi\phi}$.\\
\linebreak
We will start by proving the useful relation \eqref{barRrel2}, namely that
\begin{equation}
\frac{\sqrt{\bar{f}(\bar{R})}}{\bar R}=\frac{1}{\sinh Z}\left(\cosh Z\cos T-\cot \Phi\sin T\right).\label{barRrel}
\end{equation}
Note also that the above left hand side is equal to $\coth \pc Z$ in the conical singularity case, and equal to $\tanh B$ in the black hole case. In the conical singularity case, it can be obtained by starting with equations \eqref{Zprel1} and \eqref{Zprel2}. By solving for $\coth \pc Z$ we immediately obtain
\begin{equation}
\frac{\coth \pc Z}{\cos \hat{T}}=\frac{\cosh Z\cosh\hat Z-\sinh Z\sinh\hat Z\cos \Phi}{\sinh Z\cosh\hat Z\cos\Phi-\sinh\hat Z\cosh Z}.
\end{equation}
Now we can use \eqref{sinhatTcoshhatZ} and \eqref{sinhatTsinhhatZ} to eliminate $\hat Z$, and then using \eqref{coshatT} to eliminate $\cos\hat T$, the result \eqref{barRrel} follows. In the case of a formation of a black hole the same relation can be proved. By comparing the embedding equations \eqref{adscoord} and \eqref{bhcoord} at the point $\tau=-\pi/2$ we can obtain $\sinh B=\sinh \pc Z\cos\pc\Phi$ and $\cosh B\cosh\hat X=\cosh \pc Z$. Now By using \eqref{bhcoshZp} and \eqref{bhcoshZ} we obtain
\begin{equation}
\frac{\tanh B}{\cosh \hat X}=\tanh \pc Z\cos\pc\Phi=\frac{\cosh\hat Z\sinh Z\cos\Phi-\sinh\hat Z\cosh Z}{\cosh Z\cosh\hat Z-\sinh Z\sinh\hat Z\cos\Phi}.
\end{equation}
We then use \eqref{coshX}, \eqref{sinhhatXsinhhatZ} and \eqref{sinhhatXcoshhatZ} to eliminate $\hat Z$ and $\hat X$, and then equation \eqref{barRrel} follows.\\
\linebreak
The next step is to obtain an expression for $\partial_\phi R$. This can be obtained by taking the derivative of \eqref{barRrel} with respect to $\phi$. This yields
\begin{align}
\frac{\partial_\phi \bar{R}}{\sqrt{\bar f(\bar R)}}\left(1-\frac{\bar f(\bar R)}{\bar{R}^2}\right)=&\partial_\phi T\frac{1}{\sinh Z}\left(\cosh Z\sin T+\cos T\cot \Phi\right)+\partial_\phi\Phi\frac{\sin T}{\sinh Z\sin^2\Phi}\nonumber\\
&+\partial_\phi Z\frac{1}{\sinh^2Z}(-\cos T+\cosh Z\sin T\cot \Phi).
\end{align}
Now we can use \eqref{odeP} and \eqref{odeT} to eliminate $\partial_\phi \Phi$ and $\partial_\phi T$, and \eqref{barRrel} to substitute $\bar{f}/\bar{R}^2$. After simplifying the result, we end up with the simple expression
\begin{equation}
\frac{\partial_\phi \bar{R}}{\sqrt{\bar f(\bar R)}}=\partial_\phi Z\cos T+\sinh Z\sin T.\label{partialR}
\end{equation}
Now, equations \eqref{cs_anglemap} and \eqref{ZpZ}, or \eqref{bh_anglemap} and \eqref{BZ}, both imply that
\begin{equation}
\frac{d\bar{\phi}}{d\phi}=\frac{1}{\bar{R}}\left(\sinh Z\cos T-\sin T\partial_\phi Z\right).
\end{equation}
The remainder of the proof is now straightforward. Starting with $\bar{\gamma}_{\phi\phi}$, we obtain
\begin{align}
\bar{\gamma}_{\phi\phi}&=\bar{R}^2\left(\frac{d\bar\phi}{d\phi}\right)^2+\frac{(\partial_{\phi} R)^2}{\bar{f}(\bar R)}\nonumber\\
&=\left(\sinh Z\cos T-\sin T\partial_\phi Z\right)^2+(\partial_\phi Z\cos T+\sinh Z\sin T)^2\nonumber\\
&=\sinh^2Z+\left(\partial_\phi Z\right)^2\nonumber\\
&=\gamma_{\phi\phi}.\label{finalgammagamma}
\end{align}
This completes the proof that the induced metric is the same from both sides of the shell.

\section{The stress-energy tensor for the shell}\label{app_shellT}
In this appendix we will outline how to prove equation \eqref{Stautaupp}, namely we will evaluate the $\tau\tau$ component of the stress-energy tensor of the shell in the case where the shell is obtained as a limit of pointlike particles. Just as in Appendix \ref{app_cons_cond} we put the black hole and conical singularity on the same footing by using $\bar{R}(\phi)$ to characterize the shell in the outside patch, but we will now use $Z(\phi)$ inside the shell to simplify some of the expressions. We will use $'$ to denote derivatives with respect to $\phi$.\\
\linebreak
The energy density is then given by
\begin{equation}
8\pi G S_{\tau\tau}=\frac{\bar{K}_{\phi\phi}-K_{\phi\phi}}{\sin^2\tau(R^2+\frac{(R')^2}{f(R)})},\label{SmS}
\end{equation}
The extrinsic curvatures $K_{\phi\phi}$ and $\bar{K}_{\phi\phi}$ inside respectively outside the shell are defined by \eqref{Kij}, and can be computed using a symbolic manipulation software. The result is
\begin{equation}
\frac{K_{\phi\phi}}{\sin \tau}\equiv\frac{\sinh^2Z\cosh Z-\sinh Z Z''+2\cosh Z(Z')^2}{(\sinh^2Z+(Z')^2)^\frac{1}{2}},
\end{equation}
and 
\begin{align}
\frac{\bar{K}_{\phi\phi}}{\sin\tau}\equiv&\Bigg[(\sinh^2Z+(Z')^2)^2\frac{\sqrt{\bar{f}}}{\bar R}-(\sinh^2Z+(Z')^2)\frac{(\bar R')^2}{\sqrt{\bar f}\bar{R}}+\sinh Z\cosh Z Z'\frac{\bar R'}{\sqrt{\bar f}}\\
&-(\sinh^2Z+(Z')^2)\left(\frac{\bar R''}{\sqrt{\bar f}}-\frac{\bar R}{\sqrt{\bar f}^3}(\bar R')^2\right)+Z'Z''\frac{\bar R'}{\sqrt{\bar f}}\Bigg]\\
&\times \frac{1}{(\sinh^2Z+(Z')^2)^\frac{1}{2}(\sinh^2Z+(Z')^2-\frac{(\bar R')^2}{\bar f})^{\frac{1}{2}}},
\end{align}
The goal is to write this in terms of $\rho$, by using the differential equations \eqref{odeP} and \eqref{odeT} to eliminate the derivatives. From \eqref{partialR} we obtain that
\begin{equation}
\frac{\bar R''}{\sqrt{\bar f}}-\frac{\bar R}{\sqrt{\bar f}^3}(\bar R')^2=Z''\cos T+Z'\cosh Z\sin T+T'(\sinh Z\cos T-Z'\sin T).\label{partial2R}
\end{equation}
$T'$ can be eliminated using \eqref{odeT}, which states that
\begin{equation}
T'=\rho-\sin T\cos T\cot \Phi-\cosh Z\sin^2T-\frac{Z'\sin T}{\sinh Z}(\cosh Z\cos T-\sin T\cot\Phi).\label{eqTp}
\end{equation}
Note also that $\sqrt{\sinh^2Z+(Z')^2-\frac{(\bar R')^2}{\bar f}}=\sinh Z\cos T-Z'\sin T$ (see equation \eqref{finalgammagamma}). By using \eqref{barRrel}, \eqref{partialR}, \eqref{partial2R} and \eqref{eqTp} and then simplifying the expression using for example Mathematica, we end up with the remarkably simple result
\begin{equation}
8\pi G S_{\tau\tau}\sin\tau=-\frac{\rho}{\sqrt{\sinh^2Z+(Z')^2}},
\end{equation}
which is the same as \eqref{Stautaupp} after the substitution $\sinh Z=R$.

%
\section{Derivation of equation \eqref{peq}}\label{peqapp}
In this section we will outline how to derive equation \eqref{peq}, which is a consistency condition imposed on the parameters of the wedges such that the geometry after the collision is consistent. The goal is to find a constraint that will make sure that the intersection $I_{i-1,i}$ is mapped to $I_{i,i+1}$ via the holonomy of particle $i$. Recall that $I_{i,i+1}$ is the intersection between the surface $w_+^i$ and $w_-^{i+1}$, and is a radial geodesic at angle $\phi_{i,i+1}$. We will assume that $\phi_{i-1,i}$ and $\phi_{i,i+1}$ are known, and then find a constraint on the value of $p_i$ in terms of these angles. The holonomy of particle $i$ is given by
\begin{equation}
\bu_i=\cos\nu_i+\sin\nu_i\cosh\zeta_i\gamma_0-\sin\nu_i\sinh\zeta_i\cos\psi_i\gamma_1-\sin\nu_i\sinh\zeta_i\sin\psi_i\gamma_2.
\end{equation}
$I_{i-1,i}$ will be parametrized as
\begin{equation}
I_{i-1,i}=\cosh\pc\chi\cos\pc t+\cosh\pc\chi\sin\pc t\gamma_0+\sinh\pc\chi\cos\phi_{i-1,i}\gamma_1+\sinh\pc\chi\sin\phi_{i-1,i}\gamma_2,
\end{equation}
and $I_{i,i+1}$ as
\begin{equation}
I_{i,i+1}=\cosh\chi\cos t+\cosh\chi\sin t\gamma_0+\sinh\chi\cos\phi_{i,i+1}\gamma_1+\sinh\chi\sin\phi_{i,i+1}\gamma_2.
\end{equation}
Now computing $\bu_i^{-1}I_{i,i+1}\bu_i=I_{i-1,i}$, and extracting the coefficients of $\bm{1}$ and $\gamma_i$, gives the equations
\begin{subequations}
\begin{align}
\cos \pc t\cosh\pc\chi=&\cos t\cosh\chi,\label{eq1}\\
\sin \pc t\cosh\pc\chi=&2\sin^2\nu_i\cosh\zeta_i\sinh\zeta_i\cos(\phi_{i,i+1}-\psi_i)\sinh\chi\nonumber\\
 &-2\sin\nu_i\cos\nu_i\sinh\zeta_i\sin(\phi_{i,i+1}-\psi_i)\sinh\chi\nonumber\\
 &+2\sin^2\nu_i\sin t\sinh^2\zeta\cosh\chi+\sin t\cosh\chi,\label{eq2}\\
\cos\phi_{i-1,i}\sinh\pc\chi=&-2\sin^2\nu_i\sin\psi_i\sinh^2\zeta_i\sin(\phi_{i,i+1}-\psi_i)\sinh\chi\nonumber\\
&-2\sin^2\nu_i\cos\psi_i\cosh\zeta_i\sinh\zeta_i\sin t\cosh\chi\nonumber\\
&-2\sin^2\nu_i\cos\phi_{i,i+1}\cosh^2\zeta_i\sinh\chi\nonumber\\
&+2\sin\nu_i\cos\nu_i\cosh\zeta_i\sin\phi_{i,i+1}\sinh\chi\nonumber\\
&+2\sin\nu_i\cos\nu_i\sin\psi_i\sinh\zeta_i\sin t\cosh\chi+\cos\phi_{i,i+1}\sinh\chi,\label{eq3}\\
\sin\phi_{i-1,i}\sinh\pc\chi=&-2\sin^2\nu_i\sin\psi_i\sinh^2\zeta_i\cos(\phi_{i,i+1}-\psi_i)\sinh\chi\nonumber\\
&-2\sin^2\nu_i\sin\psi_i\sinh\zeta_i\cosh\zeta_i\sin t\cosh\chi\nonumber\\
&-2\sin\nu_i\cos\nu_i\sinh\zeta_i\cos\psi_i\sin t\cosh\chi\nonumber\\
&-2\sin\nu_i\cos\nu_i\cos\phi_{i,i+1}\sinh\chi\cosh\zeta_i\nonumber\\
&-2\sin^2\nu_i\sin\phi_{i,i+1}\sinh\chi+\sin\phi_{i,i+1}\sinh\chi.\label{eq4}
\end{align}
\end{subequations}
Since the intersections $I_{i-1,i}$ and $I_{i,i+1}$ are located on $w_-^{i}$ and $w_+^i$ respectively, we also have the equations
\begin{subequations}
\begin{align}
\tanh\pc\chi\sin(-\phi_{i-1,i} +\Gamma_-^i+\psi_i)&=-\tanh\zeta_i \sin\Gamma_-^i \sin \pc t,\nonumber\\
\tanh\chi\sin(-\phi_{i,i+1} +\Gamma_+^i+\psi_i)&=-\tanh\zeta_i \sin\Gamma_+^i \sin t,\label{weqapp}
\end{align}
\end{subequations}
where
\begin{equation}
\tan\Gamma_\pm^i =\pm\tan((1 \pm p_i)\nu_i) \cosh\zeta_i.
\end{equation}
Now we will find a value of $p_i$ such that the above equations are solved. This can be done by first solving for $\sinh\pc\chi$ in \eqref{eq4}, and then substituting that into \eqref{eq3}. By also using \eqref{weqapp} to eliminate $\cosh\chi$ and $\cosh\pc\chi$, we can solve for $p_i$ as
\begin{align}
  &\tan(p_i\nu_i)=\nonumber\\
  &\frac{\sin(\phi_{i-1,i}+\phi_{i,i+1}-2\psi_i)\sin\nu_i}{\sin\nu_i\cosh\zeta_i\cos(\phi_{i-1,i}+\phi_{i,i+1}-2\psi_i)+\cos\nu_i\sin(\phi_{i,i+1}-\phi_{i-1,i})-\cosh\zeta_i\sin\nu_i\cos(\phi_{i,i+1}-\phi_{i-1,i})},
\end{align}
which is equivalent to \eqref{peq}. It can now be checked explicitly that also \eqref{eq2} is satisfied, and then we know from the embedding equation \eqref{embedding_eq} that \eqref{eq1} must be satisfied as well (to be more precise, from \eqref{embedding_eq} we obtain that \eqref{eq1} is satisfied up to a sign, and this sign determines in what interval we choose $\pc t$).

\section{Summary of notation}\label{notationsapp}
In this appendix we summarize our notation and all the different variables and coordinate systems that have been defined in this paper.
{\renewcommand{\arraystretch}{1.2}
\begin{longtable}{|p{2cm}|p{2cm}|p{10cm}|}
 \hline
 {\bf Quantity}&{\bf Discrete analog}&{\bf Explanation}\\\hline\hline
 $\chi,t,\phi$&$-$&Coordinates used to describe \ads, in the original spacetime where the initial wedges are cut out.\\\hline
 $\tilde\chi,\tilde t,\tilde\phi$&$-$& Coordinates that will generically describe \ads after a boost has been applied, and in a collision process (when a conical singularity forms) they describe the coordinates where the resulting wedges take the form of static circular sectors in \ads.\\\hline
 $\tilde\chi,\tilde t,\hat\phi$&$-$&Coordinates describing the spacetime after the static wedges (after a collision process when a conical singularity forms) have been pushed together and the angular coordinate is now continuous (but the range is not 0 to $2\pi$).\\\hline
 $\beta,\sigma,y$&$-$&Coordinates that describe a BTZ black hole with unit mass, given by the metric \eqref{betabhcoord}. After a collision process where a black hole forms, the final wedges will be mapped to static circular sectors in this geometry. \\\hline
 $\beta,\sigma,\hat y$&$-$&Coordinates used after the static wedges (in the case when a black hole forms) have been pushed together and $\hat y$ is now continuously parametrizing the whole spacetime (but note that the range is not 0 to $2\pi$).\\\hline
 $r,t,\phi$&$-$&Coordinates after transforming the AdS part of the spacetime to the metric \eqref{rtpmetric}, by the transformation $\sinh\chi=r$.\\\hline
 $\bar{r},\bar{t},\bar{\phi}$&$-$&Coordinates after bringing the final geometry to the form \eqref{rtpbarmetric}, by the transformations given by either \eqref{bartransf_cs} (in the conical singularity case) or \eqref{bartransf_bh} (in the black hole case).\\\hline
 $\rho$&$2\nu_i$&The mass density of the particles, or more precisely the rest mass density per unit angle (in units of $1/8\pi G$) or deficit angle density per unit angle.\\\hline
 $-$&$\psi_i$&Angular location of the particles in the $(\chi,t,\phi)$ coordinates.\\\hline
 $Z$&$\zeta_i$&The radial location of the particles at proper time $\tau=-\pi/2$, in the coordinates $(\chi,t,\phi)$ (or equivalently the boost parameter of the particle in those coordinates).\\\hline
 $\pc{Z}$&$\pc\zeta_i$&The radial location of the particles at proper time $\tau=-\pi/2$, in the coordinates $(\pc\chi,\pc t,\pc\phi)$ (or equivalently the boost parameter of the particle in those coordinates).\\\hline
 $B$&$-$&The radial location of the particles at proper time $\tau=-\pi/2$, in the coordinates $(\beta,\sigma,y)$.\\\hline
 $R$&$-$&The radial location of the particles at proper time $\tau=-\pi/2$, in the coordinates $(r,t,\phi)$.\\\hline
 $\bar R$&$-$&The radial location of the particles at proper time $\tau=-\pi/2$, in the coordinates $(\bar r,\bar t,\bar\phi)$.\\\hline
 $\hat Z$&$\zeta_{i,i+1}$&This is the boost parameter required to bring the final wedges to a static coordinate system (in the case of formation of conical singularity). In the case of formation of a black hole, this boost is an intermediate step when tansforming the wedges to static wedges (circular sectors in the BTZ black hole metric with unit mass).\\\hline
 $T$&$p_i\nu_i$ &The parameter associated to a particle's wedge that specifies its deviation from being a symmetric wedge.\\\hline
 $\Phi$&$\phi_{i,i+1}-\psi_i$ &The angle of the radial geodesic of the resulting object that is between wedge $i$ and $i+1$, minus the anglular location of particle $i$.\\\hline
 $-$&$\nu_{i,i+1}$&Half the opening angle of a final wedge after it has been transformed to static coordinates (in the case when a conical singularity forms).\\\hline
 $-$&$\mu_{i,i+1}$&Half the opening angle of a final wedge after it has been transformed to static coordinates (in the case when a black hole forms).\\\hline
 $\hat T$&$p_{i,i+1}\nu_{i,i+1}$&The parameter associated to a final wedge after a collision, that specifies the deviation from it being a symmetric wedge (in the case when a conical singularity forms).\\\hline
 $\hat X$&$\xi_{i,i+1}$&The parameter associated to a final wedge after a collision, that specifies the deviation from it being a symmetric wedge (in the case when a black hole forms).\\\hline
 $-$&$w^i_\pm$&The two surfaces that border one of the wedges associated to a particle.\\\hline
 $-$&$w^{i,i+1}_{\pm}$&The two surfaces that border one of the final wedges after the collision. Note that $w^{i,i+1}_+=w^{i+1}_-$ and $w^{i,i+1}_-=w^{i}_+$.\\\hline
\end{longtable}
}







\end{document}